# Prediction of the Kamchatka July 29, 2025, earthquake by the evolution of low-magnitude seismicity recovered using waveform cross-correlation at IMS seismic arrays

I.O. Kitov

**Abstract**

Waveform cross-correlation (WCC), when used for detection at seismic arrays, allows for a reduction in threshold by an order of magnitude compared to beamforming. An additional benefit of WCC detection is the significant suppression of signals unrelated to the studied area. This effect simplifies and improves the accuracy of local phase association with event hypotheses located near corresponding master events. All the advantages of WCC-based processing are best utilized in the detailed study of low-magnitude seismicity prior to mega-earthquakes. For the Sea of Okhotsk Mw8.3 earthquake that occurred on May 24, 2013, WCC processing detected the onset and subsequent evolution of intensive, low-magnitude seismicity during the five days leading up to the mainshock. The Kamchatka earthquake on July 29, 2025, is analyzed using a similar WCC-based processing extended by a detailed study of the recurrence curves of the cross-correlation bulletin (XSEL). The transition to the coseismic phase was marked by a progressive increase in the intensity of seismic activity with growing magnitude. These XSEL events fell below the detection level of the International Data Centre, which utilizes the same data stream from the International Monitoring System.

## Introduction

Earthquake prediction is a major task in geophysics. Seismicity is generated by the Earth's internal dynamics and its interaction with extraterrestrial objects. Seismology provides information on the elastic and inelastic properties of the Earth's interior, allowing for the estimation of the distribution of principal geophysical parameters and their evolution over various time scales. Larger earthquakes are one of many mechanisms for the Earth's internal energy release. Magnitude 8.0+ earthquakes directly illustrate plate tectonics and demonstrate the relative motion of plates. The tremendous elastic energy generated by such earthquakes is a threat to human beings. Long-term forecasting of the largest earthquakes is based on the assumption of their repeatability with periods observed in the past. At this point, threats to the population and infrastructure can be mitigated through building codes and general awareness. Immediate, partly *post-factum* prediction is based on the possibility of transferring information faster than seismic waves propagate. Having a few seconds' warning before the destructive wave arrives can be very helpful in avoiding life-threatening risks.

Short-term prediction is crucial for reducing all types of risks and threats. It should involve continuous monitoring based on a set of reliable, measurable parameters indicating the current state of potential catastrophic earthquake zones. An alert should be issued when one or more parameters reach their corresponding threshold values, which are well-estimated based on previous experience and historical data.

In [Kitov, 2026b], it was shown that the waveform cross-correlation (WCC) method applied to seismic arrays of the International Monitoring System (IMS) of the Provisional Technical Secretariat (PTS) of the Comprehensive Nuclear-Test-Ban Treaty Organization (CTBTO) in Vienna, Austria, allowed for finding a large number of low-magnitude events before the May 24, 2013, Sea of Okhotsk earthquake (M24). It had coordinates 54.89°N, 153.31°E, Mw(USGS)=8.3, $m_b$(IDC)=6.27, and a hypocenter depth of 604 km. The depth of this event and its position within the subducting Pacific plate effectively separate it from any other seismic activity. The International Data Centre (IDC) of the PTS CTBTO has reported ~100 seismic events in this zone since 2001. This makes the Sea of Okhotsk earthquake a laboratory case for WCC testing. It was found that the flux of low-magnitude events jumped by a factor of ~5 approximately three days before the M24 mainshock, and the evolution of the event sizes indicates the possibility of being related to the earthquake preparation process.

Shallow seismicity near the Kamchatka Peninsula is very active, and the earthquake preparation process has to be recovered and separated from the background process. This is the next challenge for the study of earthquake preparation processes with potential prediction capabilities, all based on the synergy of the WCC and seismic arrays of the IMS as the most sensitive tools in global and regional seismic observations. Seismic arrays were proposed as a measure to reduce the detection threshold for underground nuclear test monitoring [*e.g.*, Sultanov, 1996]. The concept of a phased antenna for seismic waves was tested in the early 1960s in many countries and demonstrated excellent results. The IMS includes more than 30 array stations distributed across continents. Station aperture varies from a few hundred meters to over 40 kilometers. Theoretically, for the case of stochastic and non-additive noise, the gain of an array station is proportional to the square root of the number of individual sensors. The array aperture and the number of individual sensors depend on the target seismic phase and are tuned to the number of wavelengths, *i.e.*, the wave apparent velocity and frequency band [Schweitzer *et al*., 2012]. For low-velocity and high-frequency local phases, the aperture can be less than 1 kilometer. For regional P-wave phases with velocities of 6.0 to 8.5 km/s and frequencies up to 10 Hz, the optimal array aperture is between 1 and 5 km. For teleseismic phases, an aperture of 15 or more kilometers can be the best choice.

The relative array station sensitivity to various phases is important for assessing network performance relative to a given source location. For the Kamchatka region, IMS arrays for regional observations (PETK, PDYAR, MJAR, and USRK) dominate in events at distances below 40° including the subduction zones near Kamchatka and Japan. Figure 1 shows the IMS stations with available data for this study. IMS arrays between 30° and 60° (ZALV, KURK, MKAR, BVAR, ARCES, FINES, SPITS, LZDM, HILR, NVAR, and PDAR) contribute to a larger portion of events in the IDC bulletin for Kamchatka and its surroundings. There are also several large-aperture arrays beyond 60°: AKASG, NOA, EKA, and WRA, which are sensitive to events worldwide. The distance and azimuth distribution of the IMS arrays is favorable for the task of the weak signals detection. This is not a common case for seismic regions with potentially catastrophic earthquakes. For example, South America lacks regional IMS arrays, making a study similar to that of the Sea of Okhotsk unfeasible there.

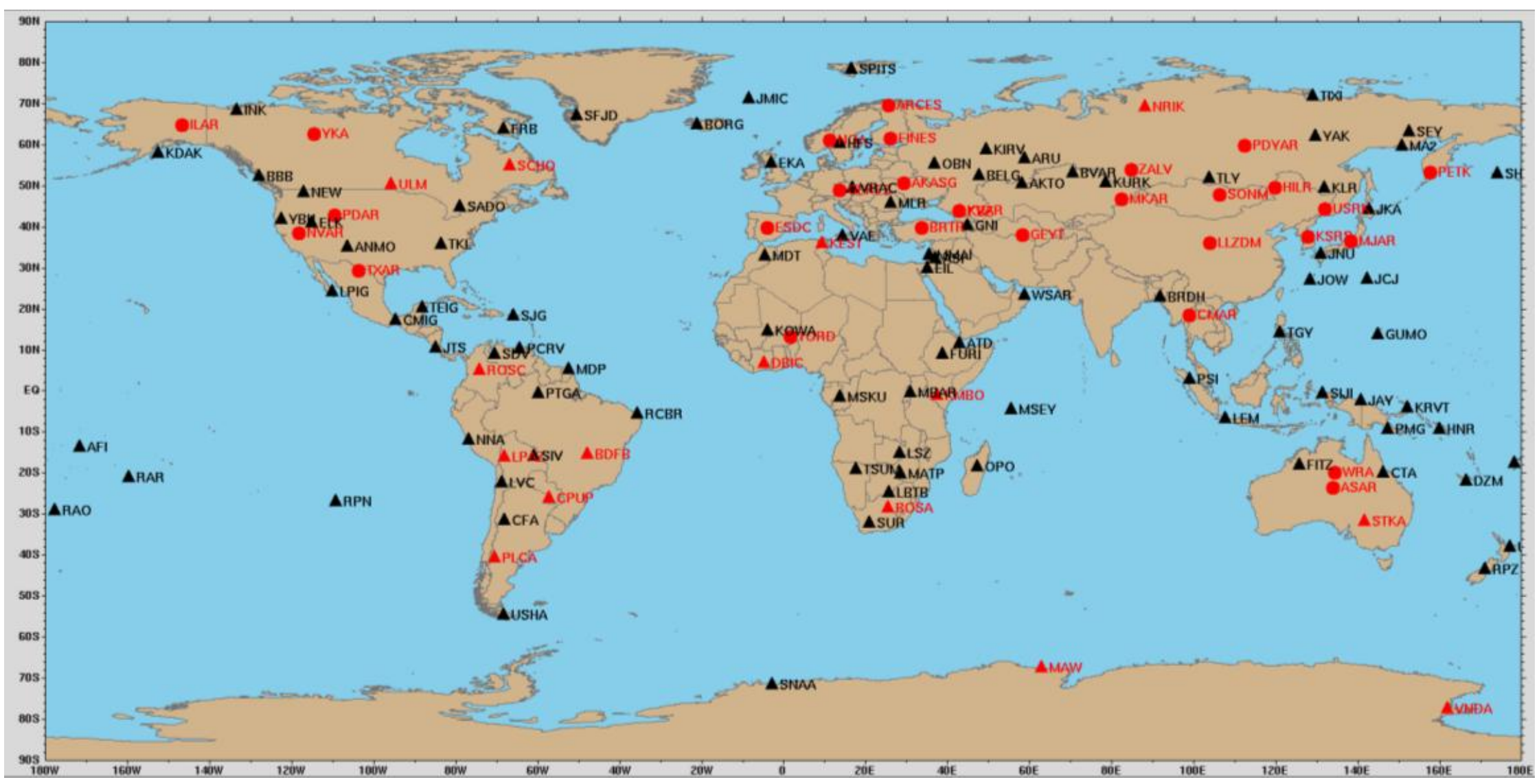

Figure 1. IMS seismic stations with data available for the current analysis. Red – primary stations. Black – auxiliary stations. Circles – array stations. Triangles – 3-C stations.

Waveform cross-correlation is a relatively new method of seismic data processing that uses high-quality digital records and historical templates. Digital seismology has been developing since the early 1960s. The WCC-based detection method was well-known long before the era of digital seismology as a core technique of the matched filter [Turin, 1960]. The first attempts to apply WCC to data from 3-C stations were reported in the early 1990s [Israelsson, 1990; Joswig, 1990; Joswig and Schulte-Theis, 1993]. Broader attention to WCC was sparked by the possibility of improving event location accuracy by one to two orders of magnitude [Schaff and Richards, 2003; Gibbons *et al*., 2007; Waldhauser and Schaff, 2008; Selby, 2010]. The accuracy of seismic magnitude estimates was also improved [Schaff and Richards, 2014; Bobrov *et al*., 2014]. A major benefit of WCC is its capability to reduce the detection threshold by an order of magnitude, leading to significant improvements in the completeness, consistency, and quality of seismic catalogs and bulletins [Schaff, Richards, 2004]. Array stations have been growing in number and data quality since the first experiments in the 1960s, but the joint use of arrays and WCC started only in the 2000s after the IMS was installed and made available to the broader seismological community [Gibbons, Ringdal, 2004, 2006].

Following the approach developed by Kitov [2026ab], here we apply the same WCC-based methods to the IMS arrays and several near-regional 3-C stations to study the preparation process of the M8.8 Kamchatka July 29, 2025, earthquake (J29). There is also an M7.4 event that occurred on July 20, 2025 (J20) with its hypocenter close to that of the J29 mainshock.

The task to detect signals and events below the threshold established by the IDC for the purpose of global monitoring of underground nuclear tests is challenging. Regional seismic networks, on the other hand, have much lower thresholds despite the fact that they are using only 3-C stations. In remote regions without regional networks, the arrays of the IMS provide a much lower detection threshold, and the IDC dominates in global bulletins. In regions like the subduction zone of the Kamchatka Peninsula [Kitov, 2026a] and the Sea of Okhotsk [Kitov, 2026b], where regional networks are onshore and the distance to the closest stations is several hundred kilometers, the IMS arrays and the IDC successfully compete with regional bulletins during times of low seismicity. For the highest seismicity episodes, the arrays demonstrate better results as ambient noise literally blinds regional 3-C stations, while remote 3-C stations are inferior to remote arrays by a factor of 4 or more.

The WCC approach allows for detecting much smaller events and can play a crucial role during quiet periods. It is better at locating these small events using regional and teleseismic networks with an azimuth gap lower than that of a one-sided regional network. This new task involves creating event hypotheses that were not detected by any other methods, and identifying signals that are not visible below the noise level. In physics, there is no requirement for the matched filter [Turin, 1960] to only detect signals that are visible to analysts or operators. In seismology, legacy observations were traditionally made by analysts, and this tradition has been maintained throughout the digital era. The new approach is currently under development and requires a detailed description and discussion.

## Data and WCC methods

***IDC processing: an example of automatic processing and the source of data***

The IMS transmits continuous data to the IDC. An extended quality check of the raw data starts the automatic processing. It consists of three stages needed to include late seismic data from primary stations, data segments requested from auxiliary stations, and infrasound data for joint seismic-acoustic analysis. The final outcome of automatic processing, Standard Event List 3 (SEL3), provides initial event hypotheses for the interactive review conducted by IDC analysts. At the interactive stage, a Reviewed Event Bulletin (REB) is produced for further operations related to the CTBT mandate, but these operations do not change the seismological content of the REB. All automatic and interactive stages of the IDC seismic data processing are described in [Coyne *et al*., 2012].

For the WCC-based methods of detection and phase association, automatic IDC processing serves as the basic scheme for mandatory stages. The official IDC bulletin provides a quality reference for the cross-correlation automatic bulletin (XSEL), since they are both based on the very same IMS data set. The differences in processing are not only in algorithms and parameters, but also in objectives. The IDC has to follow the CTBT mandate and separates seismic stations into two groups: primary and auxiliary. The former define the statistical significance of the event hypotheses, and the latter are only used to improve the estimates of the principal event parameters such as hypocenter, origin time, and magnitudes. For the purpose of the Kamchatka earthquake study and other civil applications of the WCC-based methods, there is no difference between primary and auxiliary IMS stations. Their input to statistics is defined by their historical participation rates in the events listed in the IDC bulletin within the region under investigation. The IDC has to consider each event hypothesis as it is. The objective is to find all the events of concern to the CTBT, and not to miss any.

IDC detections with calculated principal characteristics - arrival time, azimuth, slowness, amplitude, period, *inter alia* - are promoted to the detection list (IDCX). At the station processing stage (StaPro), the initial phase names are assigned to the IDCX detections using their apparent velocities and after their grouping for a single-station location. There are many rejected detections in the IDCX according to their parameters. They are converted into noise phases (N-phases), which can participate but are intentionally excluded from the following phase association process. The detections corresponding to regular seismic phases with the full set of estimated parameters are used in the Global Association (GA) process to create a set of event hypotheses. These hypotheses are then passed on to the SEL3.

The core of the GA is a location algorithm searching for potential sources to associate the available phases in the statistically optimal way. At the final GA stage, the "event definition criteria" (EDC) are evaluated. They are based on the event defining parameters of the associated phases: arrival time, station-event azimuth, and scalar slowness. This procedure evaluates the statistical significance ("minimum standards for the amount and type of information required for events included in the bulletins") of the event hypotheses using fixed weights of the measured defining parameters. For example, the weight of the defining arrival time of the first P-wave at an array station and at a 3-C station is 1.0. The weights of azimuth and slowness are 0.4 at arrays and 0.2 at 3-C stations. The secondary (not the first P) phases have lower weights and are counted in only when the first P is defining. For automatic processing, the EDC control the total number of created event hypotheses. For the interactive review, the EDC require much higher statistical significance for event hypotheses in the final IDC bulletins. These EDC have to balance the rate of valid and false (*e.g.*, not matching the EDC, but potentially physically reliable) events. For the REB, the minimum requirement is at least 3 associated primary stations, two of them arrays with the defining arrival time, azimuth, and slowness. The third station can be a 3-C one, and the arrival time can be the only defining parameter to add weight to the event statistics.

The REB is just one of many possible versions that can be created using the same data from the IMS. One of the constraints of the REB is the human resources available for the interactive analysis. The number of event hypotheses in the SEL3 should not exceed the IDC human capacity to process all of them in accordance with the standard procedure. After the Sumatra M9.1 earthquake occurred on December 26, 2004, interactive analysis was impossible due to an unprecedented number of SEL3 seeds and an indigestible number of phases that needed to be considered by IDC analysts.

The statistical approach to the creation of event hypotheses is common among seismological agencies and varies depending on their objectives. WCC processing also follows the same design of automatic processing, with natural differences in the defining parameters and station/phase weight definitions related to the results of cross-correlation with high-quality signals generated by historical events characterized by accurate locations and magnitude estimates. The requirement of at least 3 associated stations is adopted in the WCC phase association to make it possible to directly compare the XSEL to the REB and achieve high statistical significance for the WCC-created event hypotheses.

The generic approach to the event definition criteria of the IDC, as based on adjustable station/phase weights, allows for flexibility in responding to any potential change in the seismic monitoring network. The EDC can also be tuned to the requirements of bulletin reliability and/or sensitivity. Such changes, if implemented, may result in significant shifts in the reliability, completeness, consistency, and accuracy of the final bulletin, not to mention the demand for human resources. The EDC re-optimization in response to possible changes in the requirements of the CTBT, in a world of quick progress in science, technology and international relations, has to be tested accordingly. The WCC processing with multiple sets of EDC proposed in [Kitov, 2026b] and further developed and tested in this paper could be used for EDC testing at the IDC.

***WCC-based detection***

By construction, any recurrence curve demonstrates that seismic events in a given region are repeatable in locations and magnitudes, and thus have to generate similar signals observed at the same stations. A WCC detector searches for signals similar to a given waveform template selected from the historical records on the same station. The level of similarity is usually defined by the cross-correlation coefficient (*CC*). This measure is not appropriate for detection purposes, however, since *CC* strongly depends on the spectral content and length of the template and the sought signals. In the low-frequency band, the *CC* value is usually higher than for a high-frequency representation of the same actual signal. But the *CC* amplitudes for ambient noise are also increasing respectively. Longer templates generate lower *CC* values as seismic signals are prone to various kinds of dispersion, and the similarity declines quickly with the distance between the sources of templates and sought signals. Detection by the maximum |*CC*| value from a set of time series obtained for various filters and template lengths would be highly biased.

In addition, ambient seismic noise can contain signals similar to the sought one. The signals from the immediate aftershocks of the Kamchatka 2025 earthquake are a good example, as the signals associated with the events created by the IDC are very similar in shape and amplitude to a multitude of unassociated signals. These coherent signals create the ambient noise where WCC has to find valid signals fully coherent to the “noise signals”. Both the beamforming and WCC techniques usually fail to distinguish between identical signals separated by a few seconds, and this decision has to be made by IDC analysts. The Sumatra 2004 case showed that the human resources available at the IDC are not adequate for such a task.

The matched filter uses WCC as a quantitative similarity estimator. It requires the noise to be stochastic in order to maximize the signal-to-noise ratio, SNRcc, for the *CC*-trace. Stochastic noise conditions are necessary for the matched filter to operate as an optimal detector. For a sought signal, the actual properties of the ambient noise can vary from realizations very close to the stochastic case to situations where the signals creating the noise are almost fully coherent with the sought signal. Therefore, the effectiveness of WCC detection depends on three signals: the template, the sought signal, and the noise. The latter signal is subject to significant changes in amplitude of the components coherent with the template and sought signals. In a model case, the fully stochastic noise generated by a computer can be approximately 10 times larger in amplitude than the sought signals, and still the SNRcc would be above the detection threshold of 3.5 [Adushkin *et al*., 2025a, Kitov, 2026c].

The signal-to-noise ratio technique in the matched filter applied to the *CC* time series is more efficient than a generic *CC*-value detector. It becomes even more efficient when extended by the addition of stochastic noise to the filtered waveforms before the *CC* calculation. For example, it allows for finding aftershocks occurring within minutes from the J29 mainshock and hidden in the extremely high noise coherent to the sought signals at seismic stations [Kitov *et al*., 2026]. It also helps to find low-magnitude events in the quiet seismic zone before the J29 mainshock [Kitov, 2026a], which can be related to the asperity that had stopped the J20 fault propagation.

The SNRcc values are calculated on a *CC* time series according to the same algorithm as the standard SNR of the IDC [Coyne *et al*., 2012]. Short-term (STA) and long-term (LTA) average (RMS or absolute) amplitudes represent signal and noise, respectively. The length of the STA running window is usually adjusted to the expected signal rise-time and ranges from 0.5 s to 1.5 s (0.6 s is used in this study). The LTA window can be adjusted to the specific noise properties at a given station, but values between 40 s and 90 s (60 s is used) produce very similar results in most cases.

The STA/LTA ratio is defined as the SNRcc value. It is used for signal detection, meaning finding a time count when the average signal amplitude is larger than the noise amplitude by a certain factor, also known as the detection threshold. The STA running window is positioned half its length ahead of the LTA window. When the SNRcc threshold is reached, the LTA is frozen and does not change for a period approximately equal to the template length or the cross-correlation window width (CCWW). This condition prevents signal energy from affecting the LTA estimates.

The standard SNR detector is triggered when a signal arrives at a sensor and the energy flux or power (STA) exceeds the average noise level plus a few (detection threshold) standard deviations. As the signal amplitude and power increase over time after the arrival, the SNR also grows with the signal amplitude. The peak amplitude used for magnitude estimation and the maximum SNR are typically reached from a fraction of a second (*e.g*., for underground explosions) to tens of seconds (*e.g.,* for shallow earthquakes) after the arrival. The IDC uses a 5.5 s segment after the arrival to measure the peak amplitude and SNR because its sources of concern are explosions. The WCC detection procedure aims to cover a variety of potential signal properties such as rise-time, amplitude, length, sequence of regular phases, spectral content, among others. This ensures that all ranges of these parameters are considered to avoid missing low-amplitude signals that are crucial for various scientific studies and technical applications.

For a *CC* time series, the detection threshold does not define the time of signal (energy) arrival. It indicates the start of correlation growth between the running template and the data segment of the same length. The start of *CC* growth can be related to the sought signal or induced by a random realization of a noise segment coherent with the template. The actual arrival of a sought signal corresponds to a time point when it synchronizes the most with the template. At this moment, SNRcc has to reach its maximum value. Therefore, the time count with the maximum SNRcc is the arrival time. The difference between the trigger time related to the SNRcc threshold and the arrival time can reach tens of seconds. Special measures for the SNRcc trace quality check are implemented to protect the WCC detection procedure from the false alarms related to the noise realizations. Such false detections can mask valid signals between the trigger and arrival times.

In standard IDC processing, various regular phases have to be detected and identified by their properties. This is important for creation of an event hypothesis for a unique source not related to any other sources worldwide. A sequence of regional phases $P_g$, $P_n$, $S_n$, $L_g$, $R_g$ allows for a single station event location and characterization. It is used in automatic processing to create a seed event for the subsequent processing stages, both automatic and interactive. The depth sensitive secondary phases, such as pP and sP, allow for accurate hypocenter location with low depth uncertainty estimates.

The WCC-based processing uses the full waveform containing all phases of interest [Bobrov *et al*., 2014]. There is no need for separation of regular phases as they can be a significant source of information on the similarity between a high-quality template signal and the below-the-noise-level sought signal. The first P-phase is a mandatory part of any template. This rule is based on the observational fact that only P-phases are usually detectable from low-magnitude sources at teleseismic distances. It is used in the EDC adopted by the IDC – no secondary phase is counted in without the first P-wave associated with the same source. The secondary phases can be included in the waveform templates depending on their potential input to the *CC* obtained from the long-term practice and extended testing. The CCWW can cover a range from 4 seconds for impulse P-phases produced by underground explosions and deep earthquakes, up to 200 or more seconds for near-regional wavetrains of regular phases ranging from Pg/Pn to Rg. For the widest templates, only the longest *CC* time series has to be calculated with the shorter CCWWs retrieved from it.

The *CC* time series is calculated using the standard definition of the cross-correlation coefficient as a dot product of two (demeaned) vectors:

$$CC(t) = \mathbf{d}(t)\cdot\mathbf{m} \,/\, \|\mathbf{d}\|\cdot\|\mathbf{m}\|,$$

where **d(t)** is the running data vector at time *t*, and **m** is the template vector. For an array station, individual *CC*-traces are calculated, $CC_i(t)$, where *i* is the channel index from 1 to N. Then, the $CC_i(t)$ values are averaged for a given *t* over all channels to obtain the average *CC*(*t*) value. There are no beamforming time shifts applied, as the template and sought signals are synchronized in time over all individual sensors. The same procedure is applicable to 3-C stations. For a one-channel case, no averaging is applied.

Our objective is to study the evolution of seismic process before, during, and after two major earthquakes that occurred at 06:28:14 on July 20 and at 23:24:48 on July 29, 2025. There were just a few earthquakes detected by the IDC during the two weeks before the J20, and approximately 600 aftershocks between July 20 and July 29. During July 30, there more than 700 aftershocks of the J29 M8.8 mega-earthquake were detected. As of May 2026, there have been several thousand aftershocks of the J29. The post-seismic activity will continue. In Figure 2a, signals from the J20 mainshock and its immediate aftershocks over a 90-minute period are shown at IMS stations PETK (epicentral distance ~5°; station-event azimuth ~70°-120°), PDYAR (R~30°; seaz~90°), MKAR (R~48°; seaz~50°), and WRA (R~75°; seaz~10°). There was an earthquake at 06:02:49 ($m_b$=4.95), which is also seen at all the stations. This could be a foreshock related to the earthquake preparation process. There were three major aftershocks during the first hour. The aftershock at 6:49:00 had a larger magnitude, $m_b$=5.69 (IDC) than the J20 mainshock ($m_b$=5.39) and higher amplitudes at all stations. The second (07:07:41; $m_b$=5.96) and third (07:22:57; $m_b$=5.86) major aftershocks were compatible by amplitude with that of the J20 mainshock. There were many smaller aftershocks as well as those not seen in Figure 2 within the coda-waves of the major aftershocks during the first hour.

At PETK (upper trace), the J20 creates a long wavetrain with the peak observed tens of seconds after the arrival in the signal filtered (3rd order, BP) between 0.6 Hz and 4.5 Hz. The coda-wave is a mixture of primary P-phases from different parts of the earthquake fault and secondary phases from the mainshock, including surface waves likely responsible for the peak amplitude. All these phases are coherent with the corresponding phases in a potential sought signal if the J20 or any other events close to it were taken as a master event (ME).

At teleseismic stations, the peak amplitudes are observed in the first P-wave, but the length of the coda-wave is even large than that at PETK. This is a result of various velocities of the secondary phases, and their mixture expands in time. The secondary phases attenuate much faster in amplitude and, in relative terms, the coda-wave is not as prominent at WRA as at PDYAR. To detect small aftershocks in such prominent coda-waves is a problem not only for 3-C stations but also for the beamforming and WCC at arrays. At the same time, to validate the advantages of the WCC approach, detection of the smallest REB events together with additional XSEL events with the same statistical properties (as per EDC) is needed.

a)

b)

Figure 2. Waveforms for a) 2025201 (90 min) and b) 2025211 (full day) at stations PETK, PDYAR, MKAR, and WRA. All original records are filtered between 0.6 Hz and 4.5 Hz.

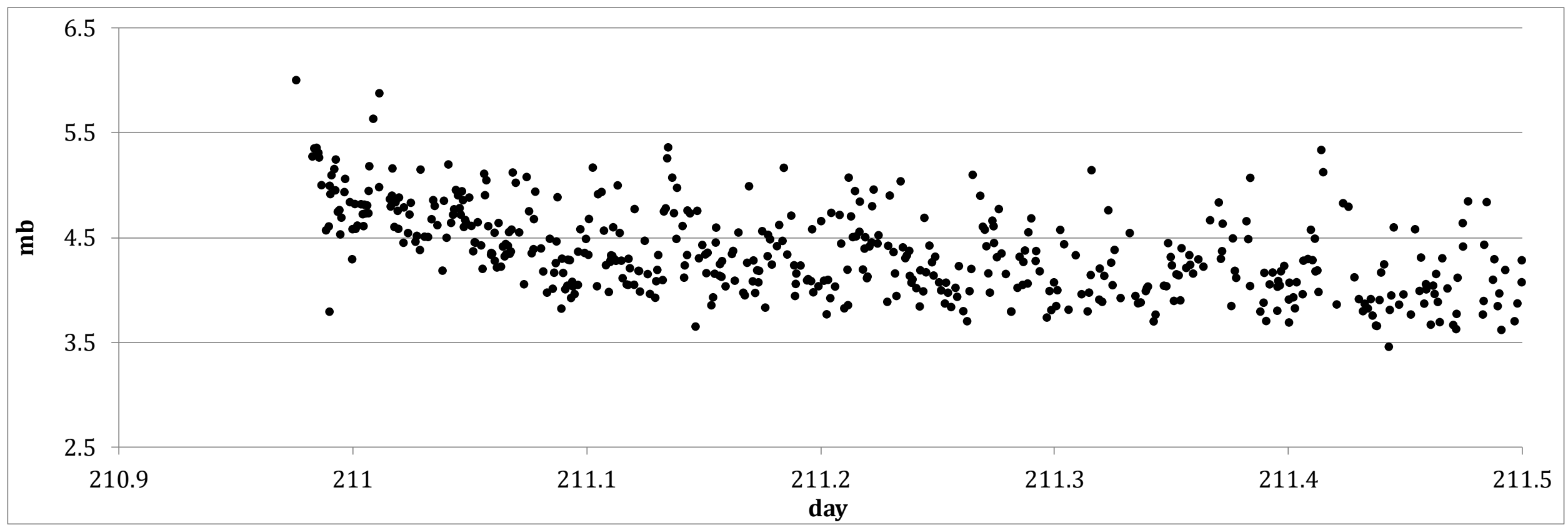

Figure 3. The period from the end of 29 July to noon on 30 July, 2025. The deficit of dozens of events with $m_b$ between 4.0 and 4.5 is clear during the first hour after the mainshock.

In Figure 2b, data from the same stations are shown for July 30. The number of visible aftershocks is very large, and the others listed in the REB have amplitudes too low to be seen among the mainshock and the largest aftershocks. The noise level is very high during the first hour of July 30. There are also ~30 minutes of high noise on July 29 after the mainshock and no aftershocks reported within the next 10 minutes before 23:34:49 ($m_b$=5.28). Some of the aftershocks within this 10-minute interval were recovered using the WCC with the addition of stochastic noise [Kitov *et al*., 2026]. Detection of relatively large aftershocks ($m_b$ from 3.5 to 5.0) was suppressed during the first hour after the J29 mainshock as Figure 3 shows.

The templates to find smaller seismic events before the J20 and the J29, as well as after them can be taken from the historical IDC records, including the earthquakes available on the day of processing. Three earthquakes with close epicenters but different depths are listed in Table 1 as examples of MEs for the WCC processing, with the waveforms shown in Figure 4. The focal depth is responsible for the differences between the unfiltered waveforms observed at PETK (Figure 4a). The deep earthquake generates a more impulsive signal and the templates for the two near-surface ones would have a low-amplitude segment before the peak, as better seen in the filtered (3$^{rd}$ order, BP, 2.0-5.0 Hz) version of the waveforms in Figure 4b. These specific features are decisive for the difference in WCC detection of the near-surface and deeper sources as these three MEs demonstrate.

A reasonable spacing between the MEs for signals at teleseismic distances ranges from 70 km to 120 km. The spatial resolution is defined by the slowness of the phases, which is low at distances above 30°–60° (0.07–0.08 s/km). Shallow earthquakes are very similar for a given geological structure. The cross-correlation exercise with earthquakes from the southern to the northern part of the Mid-Atlantic Ridge showed high similarity of the waveforms throughout the entire length [Bobrov et al., 2017]. The depth sensitivity requires a smaller spacing due to its effect on the waveform shape. For routine WCC processing at the global level, a set of MEs was prepared within 40-km-thick layers from 0 km to 700 km. The depth of a sought event is fixed to the depth of the corresponding ME, since the depth sensitivity of the first P-wave is very low. When focused on the event depth, WCC can be extended by templates of the depth-sensitive phases together with the first P in order to improve the depth estimates. In any case, the change in signal shape with depth provides a good initial selection of the most relevant ME.

Table 1. Three high quality MEs for further comparison of detection performance

| REB ORID | jdate | Origin time | Lat, deg | Lon, deg | Nass | mb | depth, km |
|---|---|---|---|---|---|---|---|
| 23694976 | 2023068 | 02:06:51 | 51.66 | 159.63 | 46 | 5.0 | 0.0 |
| 23689063 | 2023067 | 16:10:51 | 51.62 | 159.58 | 54 | 4.9 | 12.0 |
| 23637715 | 2023056 | 06:25:22 | 51.69 | 158.43 | 73 | 4.5 | 66.8 |

a)

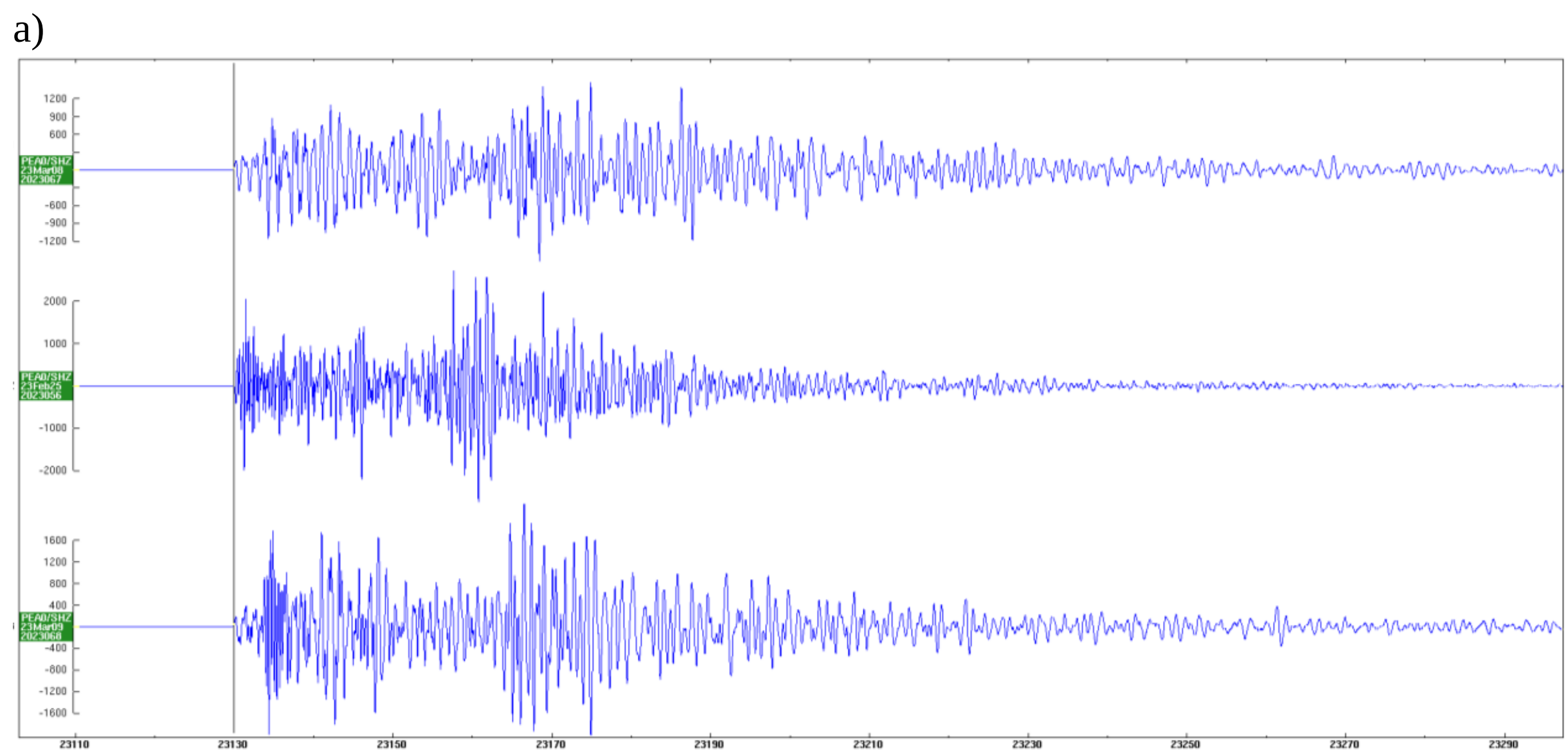

b)

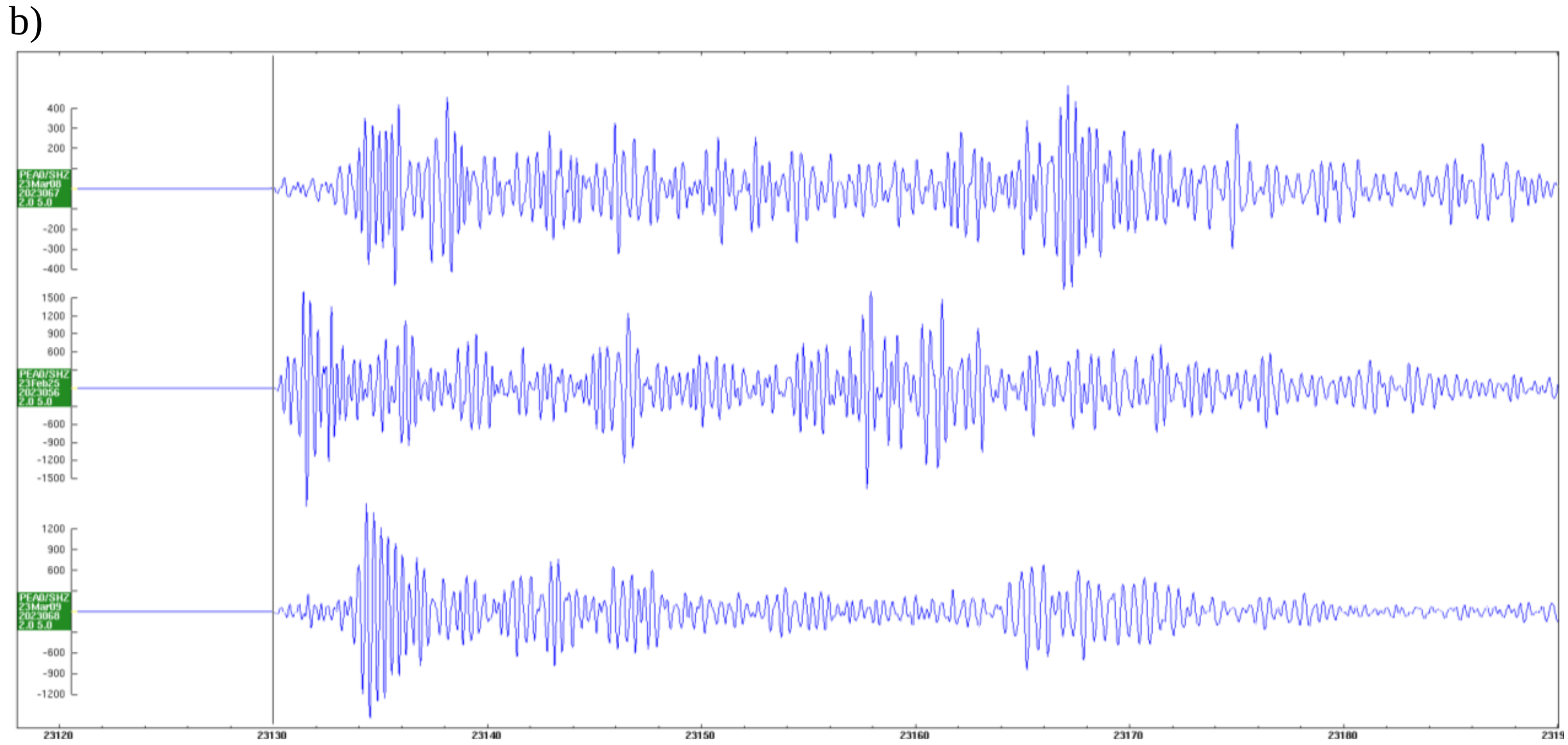

Figure 4. a) Unfiltered signals from three MEs in Table 1 at station PETK. b) Same signals. 3-rd order, 2-5 Hz BP filter is applied.

In Figure 5, the evolution of the signal with distance is shown for the deep ME in Table 1. The signals are filtered and demonstrate their high-frequency nature in the first P-wave. The pre-signal noise is not seen at PETK relative to the signal ($SNR_{IDC}$~10000), with the peak amplitude reached in ~1-2 s after the arrival. At PDYAR, the peak amplitude is observed in ~4 seconds after the arrival; the signal is also long and specific. The coda-wave is much higher than the noise at PDYAR ($SNR_{IDC}$~33) and at MKAR ($SNR_{IDC}$~115), where the signal peaks at the same time as at PETK. At WRA, the first arrival is impulsive and the coda-wave is well above the pre-signal noise ($SNR_{IDC}$~65). All four signals are specific and have high signal-to-noise ratios. Both features are crucial for effective WCC performance.

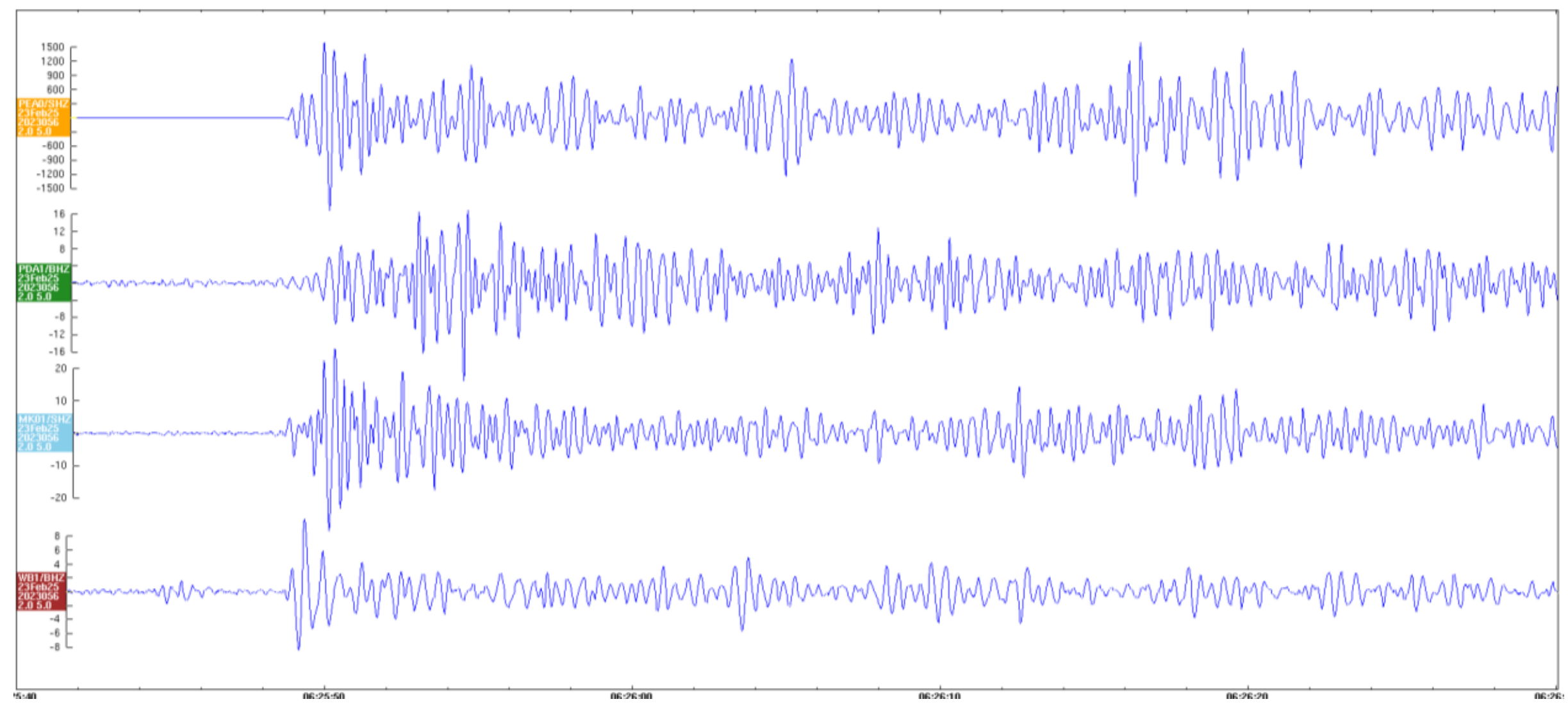

Figure 5. Templates of the ME on February 25, 2023 (jdate: 2023056) at four IMS array stations: PETK, PDYAR, MKAR, and WRA.

The WCC detection process not only finds signals similar to the templates and reduces the detection thresholds at the IMS array stations, as a realization of the matched filter detector, but also suppresses signals arriving from different azimuths and distances. The latter property refers to the change in relative arrival times of secondary phases with different apparent velocities.  The WCC changes the meaning of valid/false detection for the following phase association. For beamforming, the average daily rate of detections varies around some tuned level [Saragiotis, Kitov, 2020], with the deviations depending on the global seismicity. For example, the Sumatra 2004 and the Kamchatka 2025 earthquakes generated an extremely large number of detections at almost all IMS stations, except those in the core shadow zone where only secondary phases were observed. The presence of numerous detections near the detection threshold, with inaccurate estimates of the principal parameters and their tolerances in automatic processing leads to multiple phase misinterpretations (*e.g.*, PP used as P) and misassociations (P wave from a different source). Numerous weaker events in the SEL3, with some of them promoted to the REB, often lose many, or sometimes all, associated phases in the interactive analysis. Analysts have to find or generate new ones in order to create the REB event hypotheses.

Mega-earthquakes do not occur often, and the daily seismicity at a representative IMS station is defined by events at teleseismic distances with magnitudes between 3.5 and 5.5 as well as events at regional distances with lower magnitudes. Signals from these events are effectively suppressed by the WCC detector except for those which are close to MEs from a given set, if any. As a result, there are no detections to be misinterpreted or disassociated as such. There are days with just a few signals detected by the WCC for a given ME. This number depends on the seismic activity around the ME. The detection threshold can be adjusted to the current level of activity, however. This operation would increase the number of valid detections and likely increase the probability of false detections related to noise components slightly coherent with the template.

There is a task of tuning a WCC detector to an optimal balance between valid detections and false alarms, where the valid detections are those associated with the final set of event hypotheses generated by one ME. All detections made by the WCC are assumed to be associated with (found and not found) events close in space to the corresponding ME. Therefore, the station-event angles and apparent velocities of the regular phases related to the detections are practically the same as for the parent event, with minor adjustments for location discrepancies. For the EDC adopted by the IDC, this is an important source of information for the statistical significance of the REB events with 3 to 6 associated P-phases. For a larger number of associated first P-phases, only arrival times are used. For a WCC detection, in some cases it is impossible to visually confirm a signal below the noise level, even if its signal-to-noise ratio at the *CC*-trace, SNRcc, is high and it could potentially contribute to creating a hypothesis for the event. Analysts' work with such detections is challenging, and hypotheses generated by the WCC often get dismissed due to a lack of visual verification. Therefore, the WCC detector tuning has to rely on associating phases with reliable event hypotheses.

One can formally use statistical significance for an event hypothesis created by the association of the WCC-generated phases. The EDC use the same approach and tested the SEL3 and REB statistical significance in GSETT 3 experiment [Ringdal, 1994]. Statistical significance depends on the probabilities of the detected phase being associated with a source, characterized by its location and origin time, compared to the probabilities of randomly generated detections with SNRcc values distributed by an exponential law being associated with the same source. Thus, the distribution of SNRcc values determines the statistical significance and power of the WCC-based hypotheses. The WCC detection list includes signals with SNRcc above a predetermined threshold, as well as those satisfying several additional criteria, with the primary criterion being the standard SNR threshold. By adjusting this threshold, you can control the detection rate. The standard SNR estimates are taken in a short interval after the *CC* detection time and in many cases present the peak SNR of the ambient noise not related to the signal below the noise level. An acceptable rate is ~1500 detections per day, with hourly variations of ±30–50% possible for various specific conditions. Most of these detections on quiet days, characterized by the absence of events in the REB in the area of study, are likely to be below the IDC detection threshold. In certain cases, such detections can be due to WCC's side sensitivity to intense sources like the Tohoku earthquake. Spontaneous signals above the threshold are also possible due to data issues such as zeros and spikes. Quality controls can eliminate such cases by imposing strict requirements on the number of consecutive SNRcc estimates above the threshold. One can also use adaptive thresholds to adjust for the actual event flux to the background noise levels. These adaptive thresholds are useful for tuning the WCC during periods of low and high seismicity.

During WCC processing at a high threshold value, say 100, a quasi-uniform SNRcc distribution can be generated for a given ME. Under these circumstances, detections do not occur, and all SNRcc estimates represent noise within the LTA time series without interruptions. During quiet times without detected events, we expect that random signal generation will create an SNRcc frequency distribution with a quasi-exponential roll-off. Any departure from this baseline scenario indicates the presence of actual detections, as the template reacts to similar signals within the data set. The point at which this departure occurs can serve as a detection criterion.

Over an hour processing interval, the number of SNRcc values should be 3600 times the sampling rate at a given station. IMS stations have sample rates ranging from 20 to 100 per second, with 40 Hz being most common among the IMS arrays and 3-C stations. Figures 6 to 8 show several SNRcc frequency distributions for the same four IMS stations as in Figure 5. The same three MEs from Table 1 are used to evaluate potential variations in individual performance. Two threshold sets are used in the WCC processing: those used in actual processing from 3.1 (PETK, PDYAR, and MKAR) to 3.3 (WRA), and a detection-prohibiting value of 100 for all four stations. In Figure 6, the results are presented for a quiet day on 2023334, with no REB events between the start of 2023333 and the end of 2023335. A six-hour period is selected between 12:00 and 18:00. The SNRcc distributions for the threshold of 100 are crucial for understanding the WCC processing. At all the four stations, the corresponding curves for all the three MEs demonstrate a practically exponential roll-off from the peak between ~1.0 (WRA) and ~1.2 (PETK, MKAR) to a maximum SNRcc of 4 to 5 at PDYAR, MKAR, and WRA. At PETK, the maximum SNRcc is ~6, but the exponential roll-off is observed up to ~4. This behavior is expected for a stochastic distribution. At WRA, the exponential decay constant for the third ME in Table 1 is -3.0. This value and the range between the actual threshold (3.3) and the maximum SNRcc of ~4.5 can be used for the modeling of random noise for further use as a benchmark for the evaluation of the statistical significance of the XSEL events. For example, the requirement of SNRcc above 4.5 would reject any random detection.

For the actual thresholds, the SNRcc frequency distributions in Figure 6 demonstrate the presence of detections with “frozen” LTA values by a sharp peak near 3.1. When STA/LTA reaches the threshold level, the following SNRcc values should grow with time to the maximum value if they are related to a sought signal. If the STA/LTA ratios immediately after the threshold are not related to the sought signals, sporadically producing a few large values and then dropping back below the threshold, the LTA returns to the “calculated” mode. The maximum SNRcc value is above 10 at PETK and PDYAR, where the longest templates are used. These detections belong to low-magnitude events generating weak signals. The corner magnitude in a recurrence curve does not indicate the absence of events with lower magnitudes but rather the inability of the observational system to detect them following its internal rules. Low-magnitude events always occur in the actual seismic process, and their number grows exponentially with decreasing magnitude. The WCC processing finds some of them, and the SNRcc values of the corresponding signals are above the detection thresholds.

The differences in the MEs’ performance for the working detection thresholds are clear in Figure 6. The deep ME is the most sensitive at PDYAR and WRA. The first ME in Table 1 is very effective at PETK, and the second ME is the best at MKAR. These estimates are only for the quiet day when the events are below the IDC detection threshold. For the days of higher activity, the relative performance can change. Figure 7 shows the results for the period between 00:00 and 06:00 on July 20, 2025. This is also a relatively quiet period without REB events, but preceding the J20 earthquake. At first glance, it seems to be even more “quiet”, with lower

SNRcc curves at all stations. For example, there is no large SNRcc generated by the third ME at WRA. All the curves are closer to the exponential distribution. This can be an important feature related to the earthquake preparation process.

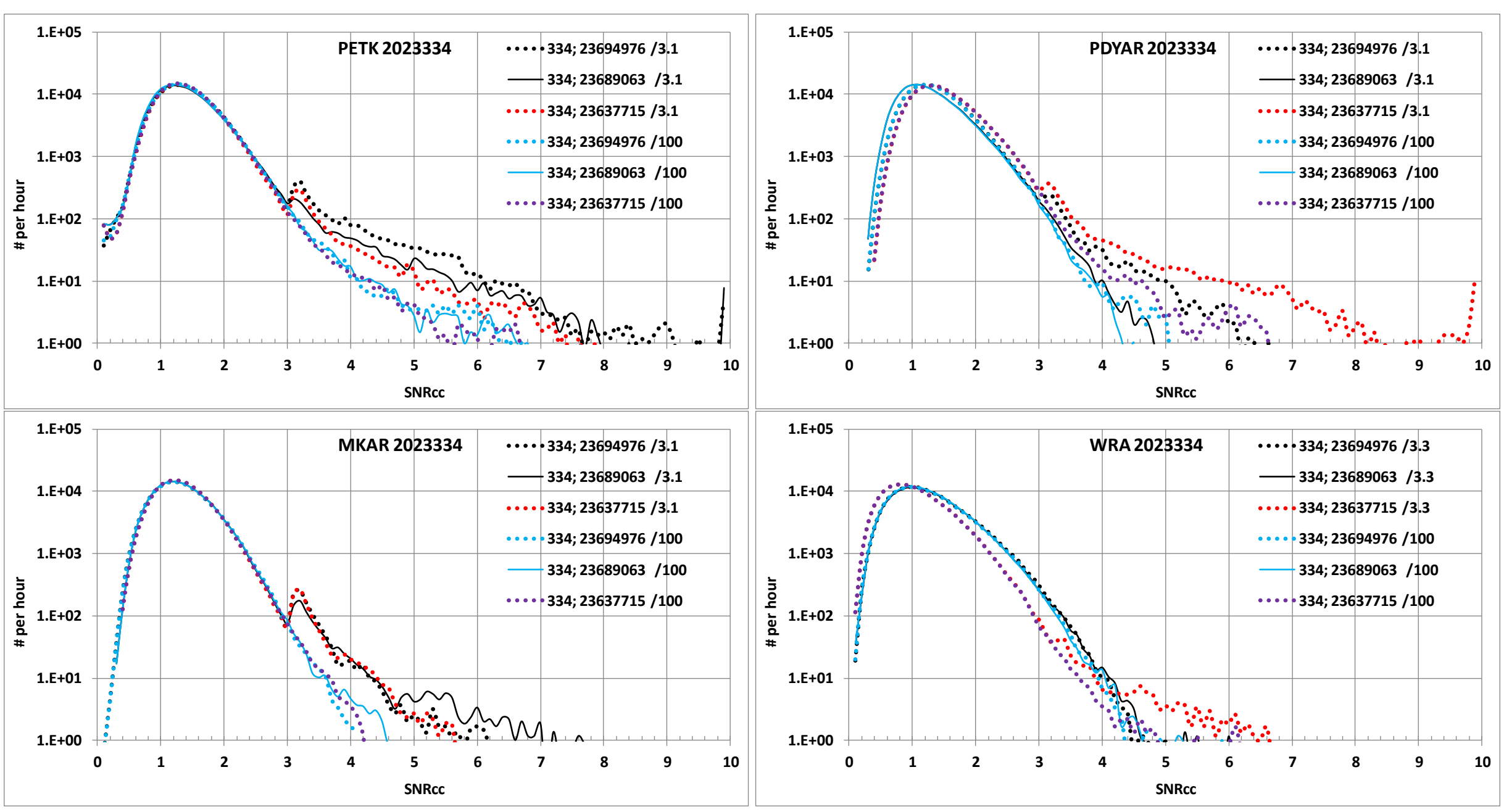

Figure 6. Frequency distribution (0.1 bin width) of SNRcc at PETK, PDYAR, MKAR, and WRA for a quiet day of 2023334. Three MEs are shown with two detection thresholds: working 3.1 (3.3 for WRA) and a detection-prohibiting of 100. The distributions vary with station and ME.

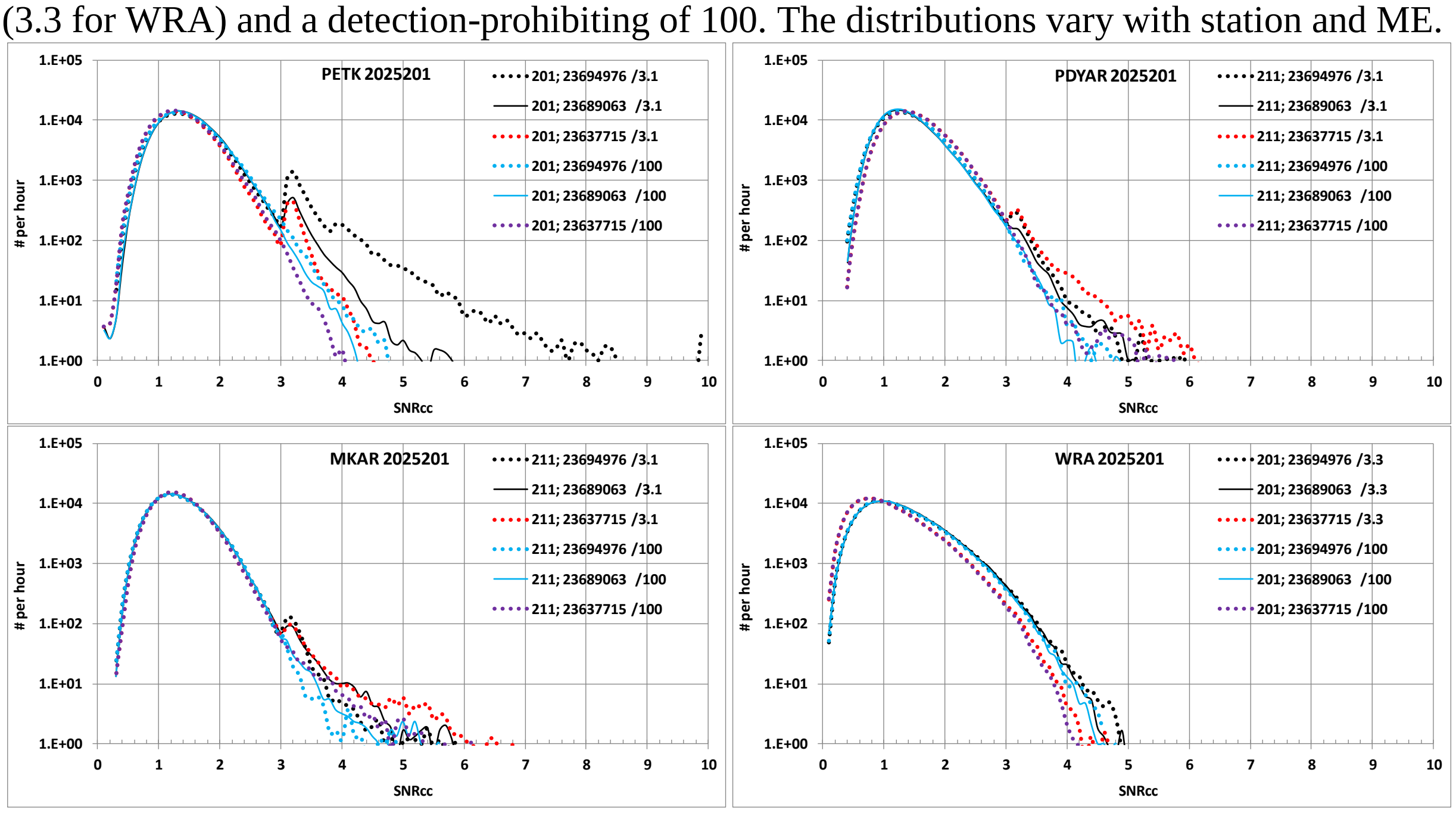

Figure 7. Same as in Figure 6 for the 6 hours before the J20 earthquake. The rates of detections are lower than for the quiet day 2023334. Almost no detections at remote station WRA.

The SNRcc frequency distributions obtained for the period from 00:00 to 06:00 on July 30 are shown in Figure 8. This is the period with the highest seismic activity likely ever detected by the IDC. At PETK and PDYAR, the curves for the working threshold show heavy tails and a significant number of detections above SNRcc ~10. For the small-aperture array MKAR, the highest SNRcc is ~9, while the large-aperture array WRA generates a large number of detections with SNRcc>10. The first ME is the most sensitive at PETK and PDYAR, and almost as sensitive as the third ME at MKAR and WRA. The curves for the threshold of 100 also demonstrate heavy tails. This observation reveals that the detected signals have sharp arrivals with larger amplitudes, as depicted in Figure 2b. The pre-signal LTA includes some elevated *CC* values, but the peak SNRcc is much higher than the surrounding SNRcc estimates.

The number of SNRcc values above the working threshold is very high at all stations but only a few (approximately 60 to 100 per hour) are selected as detections according to the WCC procedure. This detection rate may not be suitable for tasks requiring very high seismicity recovery. Special measures are needed to reduce the distance between subsequent detections in the phase association processing. These tasks are beyond the scope of the current study, which is focused on signals below the noise level that can be detected by WCC before the mainshocks. In this study, aftershocks of major earthquakes serve as a source for the recurrence curves obtained for time intervals of various lengths - from days to the entire observational period since 2001.

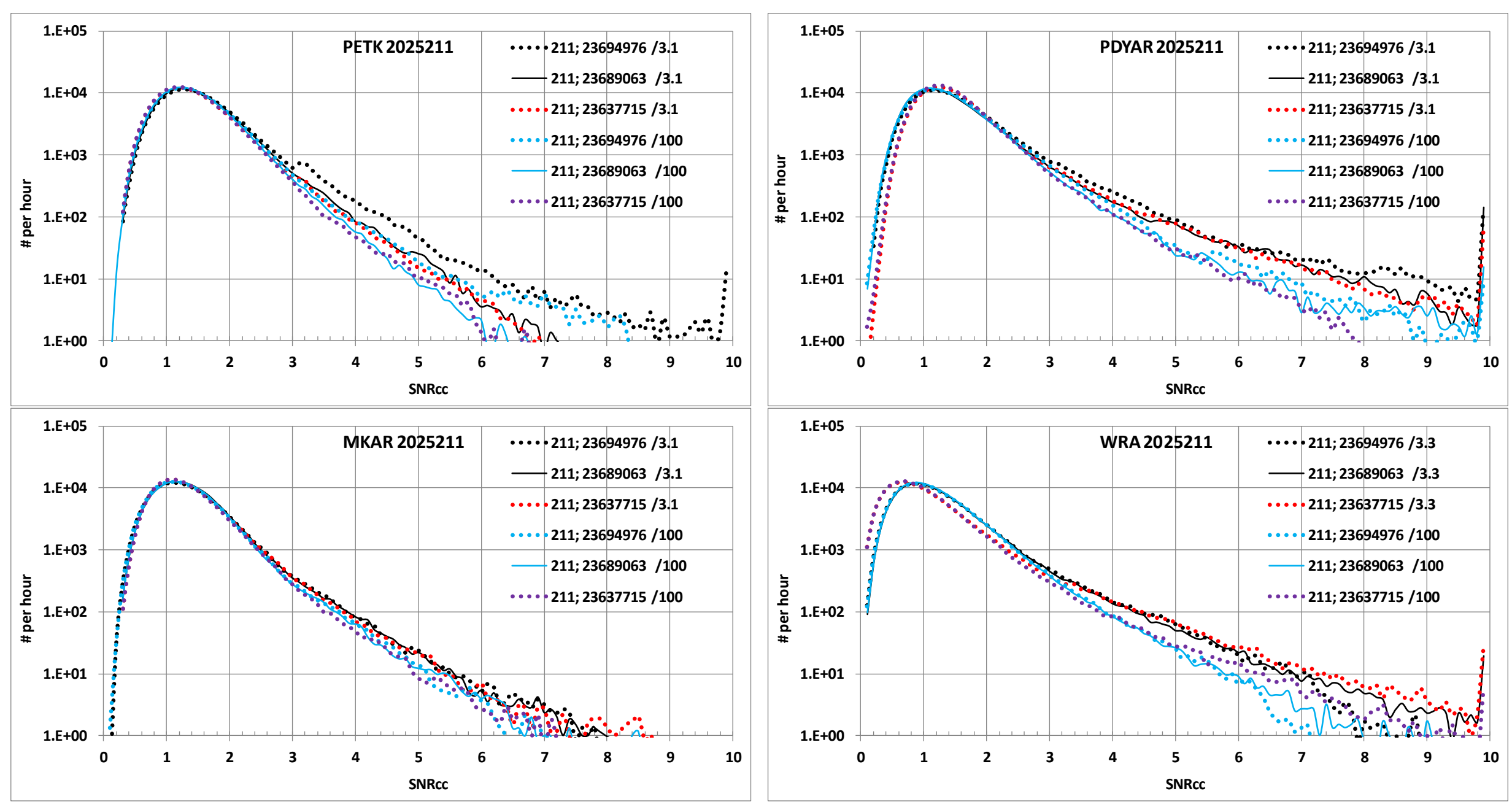

Figure 8. Same as in Figure 6 for the most seismically active period between 0:00 and 6:00 a.m., 2025211. The detection rates are high at all stations. Station PETK may suffer significant elevation in the coherent noise amplitude and lower SNRcc.

The earthquake preparation process includes a growth in the size of the sources emitting the sought signals. Simultaneously, the ambient noise at recording stations progressively includes these signals, which rise in amplitude, from all the sources in the area of the mainshock preparation. When using beamforming as a detection method, this noise, coherent with the detected signals, can mask smaller events at the array stations, as suppression is based on destructive interference. The same effect occurs with WCC-based techniques, but it is possible to improve array resolution by adding randomly generated time series to the real data. In a previous study, we used this method and identified several aftershocks that were missed by IDC within the first 10 minutes following the J29 mainshock [Kitov *et al.*, 2026].

Detection thresholds at individual IMS stations were tuned to the total number of detections in numerous tests and studies of different cases. Phase association processing relies on the detection quality and relevance to the corresponding ME, which is selected from the REB. There are some places where only poor REB events are available with a low number of associated stations and low-quality templates. The tuned detection threshold must provide detection rates within a narrow range, but the detections have to be reliable not to bias the final XSEL. Automatic bulletins always face this dilemma between number and quality. The SEL3 of the IDC contains up to 60% of events that are rejected at the interactive stage. A few percentage points of the SEL3 are located at distances within 90°-180° from their final REB solutions.

Sometimes, the SEL3 seed events lose all associated phases and are replaced by analysts with different ones.

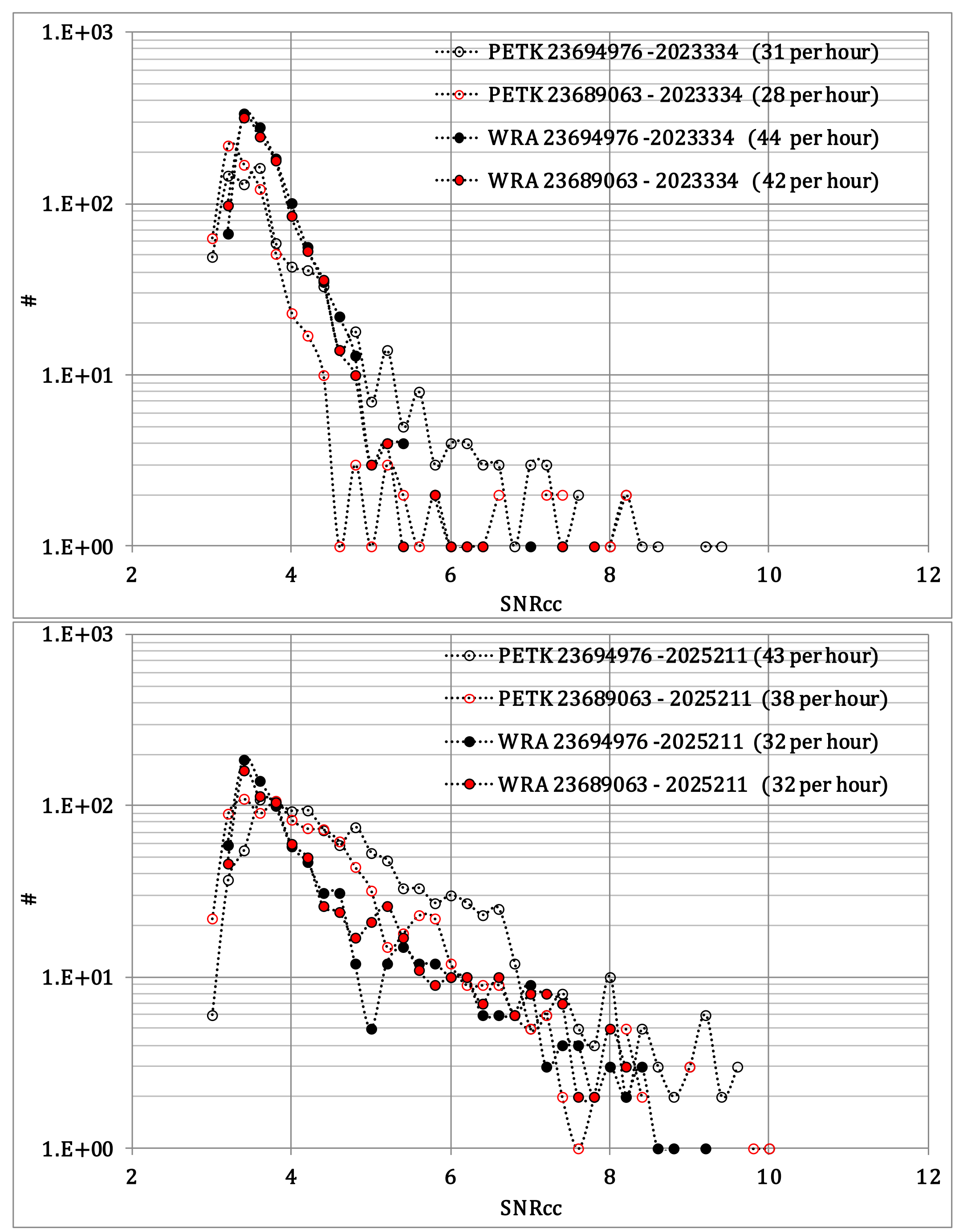

Figure 9. Frequency distribution of the SNRcc for all detections made by MEs with orid=23694976 and orid=23689063 at stations PETK and WRA during the whole days of 2023334 (quiet) and 2025211 (most active seismicity).

Frequency distributions of SNRcc values for all detections produced by the first (23694976) and second (23689063) ME in Table 1 for the days of the lowest (2023334) and the highest (2025211) activity at stations PETK and WRA are shown in Figure 9. The rates range between 28 (PETK on 2023334) and 44 (WRA on 2023334) detections per hour. These rates are below the target value of 60 per hour for average seismicity with IDC-detected events. On the quiet day, the low rate is likely related to a lack of events, even the smallest ones. On July 29, 2025, the activity was very high as were the ambient noise amplitude and coherence with the sought signals. As a result, the detection rate was also lower than average. For the best 100 MEs used in routine WCC processing, some principal statistical parameters are presented in Table 2. The average detection rate varies from 28±13 (WRA, 2023334) to 42±7 (PETK, 2025211). The maximum rate among all 100 MEs was 59 detections per hour, which is close to the target rate, with a minimum of 7 per hour related to a relatively poor ME in a very quiet part of the studied area. A rate of 30 to 40 per hour results in a time spacing between subsequent detections of about 1.5 to 2 minutes. For a random detection distribution with the same average spacing, it would be difficult for several stations to be associated with a unique source with a fixed location.

Table 2. Statistical information on the performance of the best 100 ME on 2023334 and 2025211.

| Station | WRA | | PETK | |
|---|---|---|---|---|
| Day | 2023334 | 2025211 | 2023334 | 2025211 |
| mean | 28 | 31 | 33 | 42 |
| st dev | 13 | 5 | 6 | 7 |
| max | 56 | 49 | 50 | 59 |
| min | 7 | 16 | 20 | 7 |

The quality of an XSEL bulletin should be verified through interactive analysis or by comparing it to a high-quality bulletin from the same area. The REB can serve as a reference for the XSEL in the studied region, allowing a direct arrival-time comparison between associated signals at the same IMS station in two bulletins. According to the procedure adopted by the IDC, it is only necessary to have one common arrival, *i.e.*, two similar signals within the allowed retiming window (±10 seconds), in order to match an XSEL and REB event. This means that these events are related to the same physical source.

***Local association***

The WCC arrival list consists of individual lists for each of the used MEs. These individual lists do not intersect but may contain arrivals corresponding to the same physical signals. By design, all templates of a given ME are only able to find signals from other events within a small footprint around its epicenter. For sought events within 1/4 of the signal wavelength, *CC* can be between 0.95 and 0.99 if the noise is negligible [Arrowsmith *et al*., 2006; Baisch *et al*., 2008]. Such a situation was observed for the signals from the DPRK announced underground tests [Selby, 2010; Kitov *et al*., 2025]. For very high *CC* and SNRcc values, the accuracy of the arrival time estimate for the sought signal can be as high as 0.001 to 0.005 seconds. As a result, the RMS travel time residuals for a sought event located using the WCC arrivals can be of the same magnitude of 0.005 to 0.01 seconds. In terms of the spatial location accuracy, it corresponds to 40-80 meters for an apparent wave velocity of 8 km/s, such as the $P_n$-phase. The accuracy of the relative location can be very high, but the absolute location accuracy of a sought event is practically the same as for the corresponding ME. When an ME is located 100 km from its true position, any sought event will also be located with the same error. This effect has to be taken into account when the set of MEs is selected from the REB or any other bulletin. The ME density has to compensate for their potential location errors. The confidence ellipses provide valuable information, but they also serve as estimates of location precision. The relative location of an event hypothesis using all associated phases (stations) and confined to a small area around an ME is called Local Association (LA). The LA version used in this study was described in [Bobrov *et al*., 2014; Bobrov *et al*., 2016ab; Kitov and Sanina, 2025ab].

The *CC* and SNRcc values rapidly fall with the sought event's distance from a given ME and with the amplitude of the sought signal relative to the noise level. For an array station, a virtual event location moved across the LA grid shown in Figure 10 can change the scalar slowness and azimuth of the arriving plane wave, potentially enhancing or suppressing the multichannel cross-correlation with the template. Changes in the actual event location can also impact the source function of the sought event, and differences in propagation paths may distort the shape of the sought signal compared to that of the ME.

The search area for a given ME is defined by the degradation of correlation with distance and the presence of different MEs. In Figure 10, the search area is defined by the distribution of nodes and the size of the mesh cell. There are 16 nodes in both the horizontal and vertical directions between the center and the circle. With a step of 6 km, R=96 km. The number and length of steps can vary, and rectangular cells can be replaced by any appropriate shape to allow for tight and nearly even coverage. To prevent edge effects, if the best event hypothesis is located beyond 0.9R, it is rejected, and adjacent MEs have to be responsible for it. This means that the search radii of neighboring MEs have to overlap to fully cover the studied area.

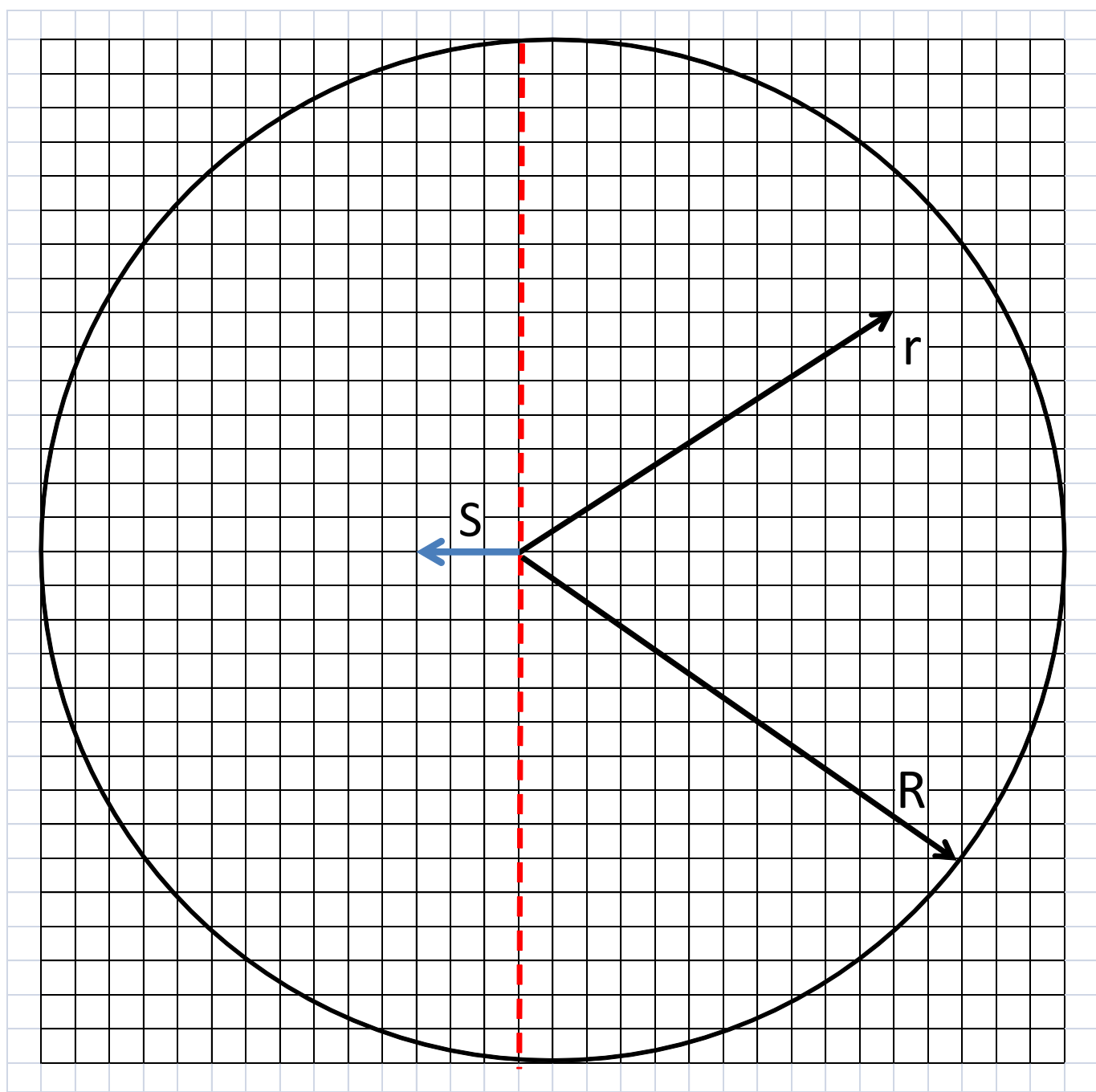

Figure 10. Virtual location grid. **r** - vector from the ME in the center to a grid node. **S** - slowness vector of the plane wave (dashed line) arriving at the station. The circle with radius R limits the distance to the used nodes. For the closest stations, the node position can change the shape of the generated signal and the vector slowness.

A valid event hypothesis is defined by its location and origin time. The origin time for a specific location is determined by the arrival time of the associated phase and the event-station travel time. Arrival time is always a measured parameter. Travel time can be either theoretical, as in the IDC's GA processing, or empirical, as in the WCC's LA processing. The GA method seeks the best absolute locations of sources globally and involves a large number of permutations between virtual sources and their signals. The LA focuses on a smaller sequence of signals most likely generated by sources near the corresponding ME location. To estimate the travel time for a signal, the GA uses source hypotheses in various regions and travel-time tables corresponding to various phases, as the phase name can be altered in the GA. The LA has a small footprint with virtual source locations, extremely (0.001 s) accurate empirical travel times taken from the REB (or any other bulletin for different networks), and fixed phase name. Therefore, a virtual event origin time at the (0,0) node can be calculated as the WCC arrival times less the empirical ME-station travel times independent of the status – associated/not associated. The arrival times obtained by the WCC detector at all stations can be reduced to their corresponding origin times at the ME location once and for all. For the other nodes, the travel times have to be corrected for the vector to the node, **r**, and the slowness vector of the signal at the related station (Figure 10). This is a very accurate time correction, as the node vector is predefined and the slowness vector for a given station does not change much between the nodes of a small grid search area compared to the ME-station distances.

For a valid event hypothesis, the origin times for three or more associated signals (stations) have to be within a predefined tolerance. At the IDC, the allowed travel (origin) time residual relative to the theoretical time is 2 s for the first P-phases, and 2.2 s for the $P_g$- and $P_n$-waves. These residuals correspond to the desired statistical significance of the REB events. A minimum of three first P-phases at different stations have to be associated with a valid event. These criteria are utilized in the WCC processing to align with the IDC approach and the REB. Since XSEL arrivals potentially have higher relative arrival-time accuracy, the origin-time tolerance can be lower than for the IDC without sacrificing performance efficiency, as measured by the number of valid (matching REB) events. In contrast, automatic IDC bulletins (*e.g.*, SEL3) allow for a travel-time tolerance much larger than 2 s, reaching up to 5 to 7 seconds. This increases the risk of phase misassociation and misinterpretation, leading to the rejection of over 60% of SEL3 hypotheses.

The WCC processing, while automatic, strictly adheres to the full list of REB requirements, with the exception of the distinction between primary and auxiliary stations. It is worth noting that there are many physically viable event hypotheses created below the EDC during the interactive review, but rejected during the automatic EDC compliance stage to meet the CTBT requirements for event quality and monitoring threshold distribution. These rejected events could be helpful to tune the WCC processing parameters, but their creation is sporadic.

Three or more arrivals with origin times within a ±2 s interval from their average value are not a guarantee of a reliable event hypothesis. There are many arrivals at various stations, and the event hypotheses have an additional degree of freedom - varying location. The REB provides valuable information on IMS station performance for historical events in any seismic region. Some stations contribute to almost all REB events in specific areas, such as IMS stations WRA and ASAR to events within Australia. Therefore, an REB event hypothesis within Australia without these stations, when they are operational, would be rather false even though other parameters may match the EDC. There are also IMS stations contributing only to the largest events in this region or not contributing at all. The difference between the most sensitive and ineffective stations can be described by their participation rates in the REB events. These rates are calculated for each IMS station in regions with repeating seismic events. For example, the first P-phases detected by PETK have the highest probability of being associated with events in the studied area and can be characterized by a station weight (or probability) of 1.0. Table 3 lists the weights of the top 30 stations.

To create a statistically significant event hypothesis with three associated stations, two of which must be from the top four, one can establish a minimum event weight of 2.65. This will result in only a few valid three-station configurations. When more than three stations are associated, their weights are combined, and the 2.65 threshold can be matched for many configurations with two stations from the top four participating. For events with more than five associated stations, having one from the top four is typically sufficient, as the association of a station adds statistical significance due to its probability of being associated compared to random chance. Station PETK is not needed for such cases, as evidenced by the REB listing events without PETK for the aftershocks on July 29. Noise can mask the appropriate signals and teleseismic stations are sometimes more efficient.

The threshold value of the STA/LTA ratio on a *CC* trace is a powerful parameter for the detection rate. It also impacts the maximum SNRcc value, as demonstrated in Figures 6 through 8. Figure 9 shows that on a quiet day like 2023334, when the detection algorithm may generate more noise phases, the maximum SNRcc is constrained to an exponential decay trend with only a few signals with SNRcc greater than 5. Station WRA has several signals with SNRcc above 5 and lower weight. It is beneficial for event hypotheses with 5 or more associated stations. Station PETK has ~30 signals with SNRcc>6 and ~60 with SNRcc>5 for the ME with orid= 23694976 during the day of 2023334. This is less than 2 and 3 signals per hour, respectively. Therefore, the maximum SNRcc is a critical parameter for the statistical significance of a hypothesis, in addition to its weight. On quiet days, the SNRcc values are also higher on average than for many other stations, including WRA. Requiring SNRcc>5 for station association in 3- to 5-station events will enhance the statistical significance of these hypotheses. The probability of random association is exceedingly low when considering the detection rates at the top stations and the overall statistics for the best 100 MEs (Table 2).

A similar approach to statistical significance related to the probabilities of detections above predefined SNRcc thresholds is the sum of SNRcc values for all associated stations. For 3- to 7-station event hypotheses, this sum assesses the credibility of association for all signals together, as well as any individual signal before it is associated. The threshold for the SNRcc sum has to rely on the statistics of detections for a given region, considering factors such as the distance-azimuth distribution of IMS stations, the history of seismicity, and the quality of MEs, as these may vary.

Table 3. Stations' weights for the event hypotheses creation.

| Station | weight | Station | weight | Station | weight |
|---|---|---|---|---|---|
| PETK | 1.00 | **JKA*** | 0.70 | ARCES | 0.60 |
| PDYAR | 0.90 | MJAR | 0.69 | USRK | 0.56 |
| KURK | 0.86 | PDAR | 0.68 | **SEY** | 0.55 |
| MKAR | 0.85 | WRA | 0.67 | **SHEM** | 0.55 |
| TXAR | 0.83 | NOA | 0.66 | SONM | 0.52 |
| FINES | 0.80 | SPITS | 0.65 | ZALV | 0.52 |
| AKASG | 0.74 | NVAR | 0.63 | **KLR** | 0.50 |
| ASAR | 0.73 | YKA | 0.63 | **MA2** | 0.50 |
| BVAR | 0.72 | KSRS | 0.62 | **TLY** | 0.50 |
| ILAR | 0.71 | CMAR | 0.61 | **YAK** | 0.50 |

*3-C stations are highlighted bold

In summarizing the LA setting for the current study, we can formulate several simple rules to adjust the principal parameters to the desired magnitude range:

1) The number of associated stations is 3 or more. Two-station events are not considered, but a minimum of 4 stations is used in some cases of many stations with the weights near 1.0.

2) The origin-time tolerance can vary from 0.1 s to the tolerance adopted in the SEL3. The WCC detection algorithm allows consecutive arrivals to be very close to each other. Therefore, a detection list at a given station has to be checked for the spacing between consecutive arrival times for the same ME. There should be only one arrival at the same station within an association window of twice the tolerance. When data are read in, the next detection closer than the doubled tolerance plus 1 second is removed. This operation results in a decreasing number of detections available for association with an increasing origin time tolerance. It may have a negative impact on the LA efficiency for periods of very high activity with very close arrivals.

3) The minimum total event weight can vary from 1.5 (3 stations with 0.5 weight) to 2.76 (3 top stations).

4) The weight of one of the best stations to be associated with any 3- to 5-station events is between 0.7 (top 11 stations) and 1.0. A lower weight is not supported by the REB for Kamchatka, but in remote regions such as mid-oceanic ridges and transform faults in the Indian Ocean where stations have low participation rates, a weight of 0.6 may be considered reasonable.

5) The minimum sum of SNRcc values for 3-, 4-, and 5-station events is calculated by adding the 3, 4, and 5 lowest detection thresholds. The sum step for larger events is determined by the minimum detection threshold of the next station to be associated. There is no limit to the maximum level. In practice, a sum of 30 for a 3-station solution makes it almost impossible to create events in routine processing. For the DPRK tests, the SNRcc is above 10 for many stations for any explosion as an ME.

6) The lowest possible value for the minimum SNRcc at one of the top stations from 3) is the detection threshold. This minimum can be set at any level depending on the studied case.
7) The grid radius can be a few hundred meters, such as for the DPRK test site. For an event occurring in remote regions, like the French nuclear test site at Mururoa Atoll in the South Pacific Ocean, a radius of 100 or more kilometers is appropriate for the far teleseismic stations since location algorithm minimizes travel time residuals rather than distances. The number of steps is tuned to the time resolution related to the radius and associated phases from Pg to PKPbc.

There are other LA parameters that allow for fine-tuning to specific cases of seismicity, such as the aftershock sequence of the J29. These parameters remain unchanged throughout all calculations after being tuned in the test calculations.

***Conflict resolution***

By design, each ME generates a detection list, and then the LA creates a set of event hypotheses with a part of all available detections associated with them. Not all detections can be associated since the sensitivity and resolution of the involved stations are not uniform and the best stations may find signals not visible to other stations. Another issue is the generation of detections and creation of hypotheses by adjacent MEs. For example, signals from the six DPRK underground explosions were detected by many IMS stations. When used as an ME, any of them could find all of the others at the stations where it had waveform templates. In turn, for any DPRK test, all the other five can detect signals at the IMS stations with corresponding templates. It is worth noting that the WCC detector, using templates from the larger DPRK tests, can find signals from the low-yield DPRK-1 test at stations where it has no arrivals in the REB [Kitov *et al*., 2025a; Adushkin *et al*., 2025b]. For the XSEL, only one event hypothesis from the five available is allowed. The event weight is the most important characteristic of the statistical significance of the related hypothesis. One can use the station weights for the DPRK test site and estimate the total weights for the five competing hypotheses. Let's assume for this specific example that the event weights created by the DPRK-4 and DPRK-5 explosions are the same. Then the next deciding parameter is the number of associated stations (Nass). This parameter is useful when Nass is high and the lists of associated stations include those with low weights. When the event weights are equal, Nass can differ by one or two stations and resolve the conflict. There is a possibility that the weights and Nass are equal between competing events. Then the final decision is made by the RMS origin-time residual. These values are always different, and the decision is unambiguous. The conflict is resolved. As a result, the best event hypothesis is promoted to the final XSEL and four excellent event hypotheses have to be rejected with their associated phases moved to the list of unassociated. In practice, the MEs are not equal and some of them may create more XSEL hypotheses than the others despite very close quality of template and the same list of stations. For the DPRK test site, the fifth explosion is likely the best for the place [Kitov *et al*., 2026].

There are tens of thousands of MEs in the global WCC processing, with an average spacing of approximately 1.3° and some places with higher ME density. The depth bins for the ME distribution are 40 km thick, and two ME can be almost collocated and have a depth difference of a few kilometers if they are in different depth bins. This situation is close to the DPRK setting of MEs. Several neighboring MEs can find the same physical source of seismic emission by their respective templates at the same stations. Competition for the same physical signals creates conflicts at stations. These conflicts are resolved unambiguously between all pairs of MEs at all related stations according to the WCC Conflict Resolution (CR) procedure. The final XSEL includes unique solutions only.

The principal difference between the DPRK case and the routine WCC processing with thousands of MEs covering a very large area is in the probability of two competing phase not belonging to the same physical source, making the conflict random. The CR makes its choice since the arrival-time difference between the signals is within its responsibility, but one or two rejected phases should not reject the entire event hypothesis when its other associated phases at different stations are not in any conflict. Therefore, an event hypothesis losing one or more arrivals has to be checked against the EDC related to the WCC processing. If it does not pass the test, it has to be rejected. Otherwise, it can be promoted to the final XSEL when no other conflicts have to be resolved. The number of lost stations and the corresponding lost weight could be used to retain the event quality by introducing corresponding tolerances. These are important parameters for the highest seismic activity periods when competition for detections among MEs is enormous. In this study, these parameters are retained at the level tuned to the highest activity after the major events – an event hypothesis can lose up to 12 associated stations and a total weight of 6.0. It does not harm the WCC processing of average activity between the major events as the number of conflicts falls fast with the decreasing number of event hypotheses.

***Seismicity of the Kamchatka region as per the IDC***

The Pacific plate subducting under the Kamchatka Peninsula creates active tectonic forces converted into high seismic activity with extremely large events, like the Kamchatka 1952 (Mw=8.8-9.0) and 2025 (Mw=8.8) earthquakes. As of December 2025, the IDC catalog lists ~27,000 events within the studied area: 45°N - 65°N, 145°E - 165°E. Figure 11 shows the evolution of seismicity since the start of IDC processing in January 2001. In addition to the July 20 and 29 earthquakes, there were several high-magnitude events expressed by aftershock density, as shown in Figure 12. The most prominent ones occurred on November 15, 2006, May 19, 2013, and December 20, 2018. There are approximately 7-year long gaps between the major events. The locations of the 2006 and 2018 events are far from the other four events, with the 2013 and 2025 earthquakes being almost collocated. The J20 can be considered as a foreshock of the July 29 earthquake, with the event on September 18, 2025, event being its aftershock. The J20 could also be treated as a failed attempt of the J29 [Kitov *et al*., 2026]. The study area was selected to include the larger aftershock sequences of the neighboring large-scale events. The preparation of the J29 can depend on the processes in the adjacent, seemingly independent, structures. This possibility should be studied in the overall approach to the earthquake preparation process recovered using waveform cross-correlation.

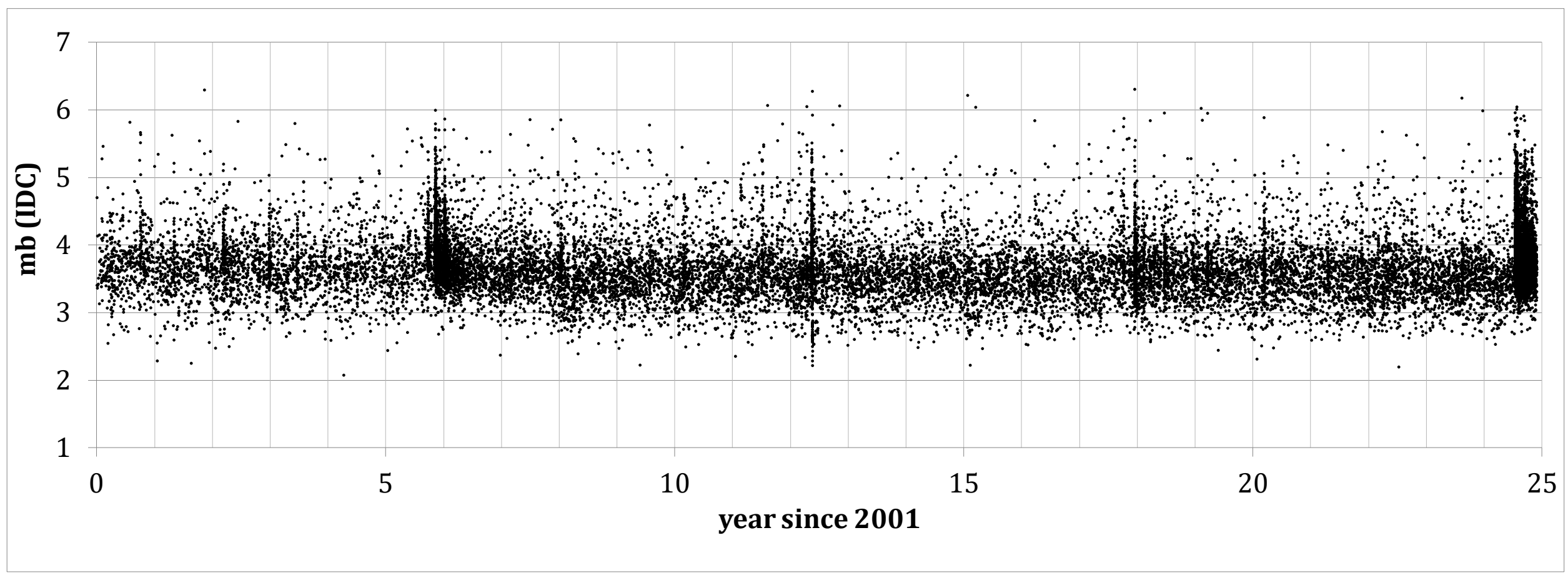

Figure 11. Body wave magnitude $m_b$(IDC) as a function of time since the IDC start in 2001. Minimum event magnitude is 2.0 [Coyne *et al*., 2012]. Maximum magnitude is of 6.30.

The J29 event is the largest observed by the IMS stations and processed by the IDC for the studied area. The events in Figure 11 are represented by their $m_b$ values since the IDC calculates only body- and surface-wave (Ms) magnitudes. The latter is not available for the low-magnitude events. The difference between $m_b$ and Mw is increasing with Mw for the events above magnitude ~6.0 because the body wave magnitude scale saturats when the corner frequency of the event spectrum falls below the lower frequency of the $m_b$-band 0.8 Hz - 4.5 Hz.

This magnitude scale difference is not a problem, since the events of interest have magnitudes below the corner magnitude of the recurrence curve for the Kamchatka region and are mainly not detected by the IMS/IDC standard procedure. The Mw estimates are not available for these events.

The seismicity map in Figure 12 shows the abundance of potential sources of high-quality templates for the WCC processing. There are aftershocks of the J29 not included in this map, because these events occurred after December 2025. They may prove to be the most efficient MEs. The WCC has been on a learning curve since the start of IDC processing in 2001. For other regions with seismic networks, observational periods may vary in length, but the availability of MEs has to increase in both number and geographical coverage.

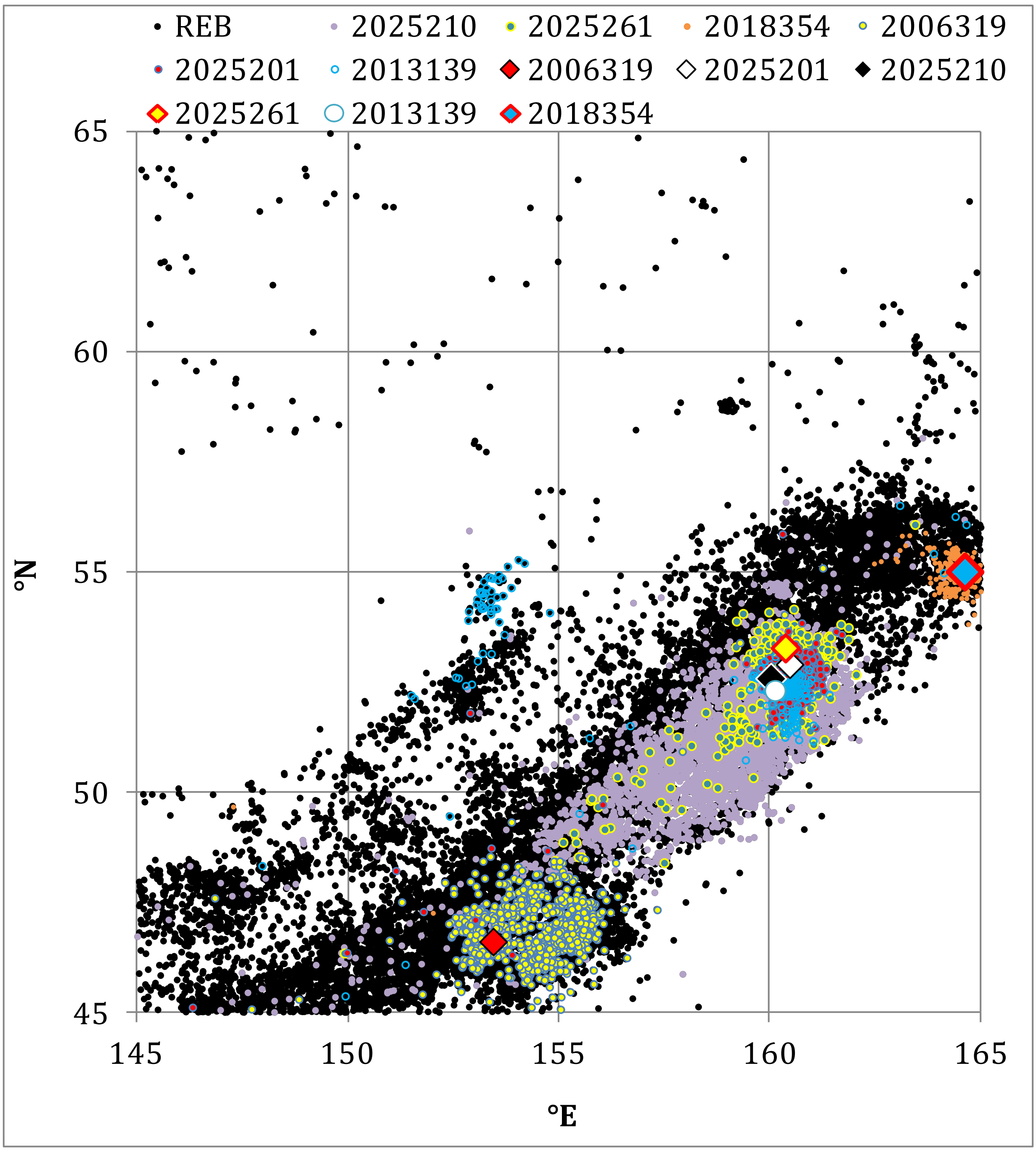

Figure 12. Seismic activity within the studied region as per IDC. Aftershock sequences of the 2006319, 2013139, 2018354, 2025201, 2025210, and 2025261 are highlighted.

***Master events***

In seismic applications, waveform cross-correlation is based on templates corresponding to regular signals detected at the stations associated with seismic events. The IMS seismic network uses arrays and 3-C stations to provide data to the IDC. For a 3-C station, one or all three components can be used as templates. All three components are useful for near-regional signals from low-magnitude sources. At larger distances, the ambient noise on the horizontal components is usually higher than the signals from almost all sources. Since all three components are recorded by the same instrument, they are synchronized in time, and the arrival times at the three components of the sought signals are also simultaneous. The difference between 3-C signals from the sought sources and other, unrelated sources arises from the differences in the apparent velocities of the secondary phases and their azimuth- and distance-dependent attenuation and dispersion.

The IMS array stations allow for a significant reduction in the detection threshold compared to collocated 3-C stations, and are much more sensitive to the azimuth and scalar slowness of the primary and secondary phases. This makes them a major contributor to the REB events, also because they provide a higher statistical significance to the associated phases due to lower tolerances related to the estimates of arrival time, azimuth, and slowness conducted by beamforming [Schweitzer *et al*., 2012]. As a waveform template, a multichannel signal at an array station makes the matched filter detector more powerful due to additional signal specificity. The performance of a WCC-based detector, as an implementation of the matched filter detector, depends on the specificity of the template and sought signals, but also on the properties of the ambient noise. These conditions make the selection of the master events with the associated waveform templates a crucial task for the completeness, consistency, and accuracy of resulting bulletin.

The most efficient way to select master events (MEs) is to conduct a preliminary study of the level of cross-correlation between REB events within an area where the MEs may compete for the same signals, as described in the CR procedure. The main task of the selection procedure is to identify those REB events that have matched the largest number of neighbors using standard WCC processing. The REB is the best source of meta-information on the locations and magnitudes of all events and their associated phases. The estimates of the signal-to-noise ratio (SNR) for the associated phases are important for the quality checking of the corresponding signals.

The latest update of the ME set was completed in 2023, but the current available computing resources are insufficient to repeat it with the new data. There were approximately 1000 MEs in the region for a prototype routine WCC processing at the IDC, which was halted in 2024. For this study, the best 100 out of the 1000 were selected. This reduction in number was primarily due to computer limitations (one i9 processor with 24 cores). The feasibility study status also allows for a reduction in the number of MEs, as the task is to uncover the earthquake preparation processes for further full-scale investigation, including in different regions.

The upper panel in Figure 13 shows the geographical distribution of the best 100 MEs (yellow circles) across the region, covering the entire depth range from 0 km to 700 km. The slightly increased spacing between these MEs compared to the 1000 MEs case is partly compensated by their quality, as demonstrated by four years of test routine WCC processing in a prototype IDC pipeline. This represents a basic set of the WCC processing. There was a significant change in the IMS seismic network related to observations in the Kamchatka region – a new sensitive array station, PDYAR, was installed in 2023 to replace the old IMS 3-C station PDY. Since it did not participate in the WCC selection procedure, it is absent from the 100 best MEs set. To address this deficit, a set of 195 MEs was generated, all containing PDYAR. These PDYAR events are more widespread across the region. They are shown by red circled in Figure 13. In total, there are 295 MEs used in this study.

The basic assumption is that the whole region has the potential to impact the J29 preparation process. Seismic activity in both the upper and lower latitudes of the area may reveal specific features of this process. An extension of this assumption is to test each potentially independent part separately using the MEs related to these parts. Therefore, the entire set of 295 MEs was processed by the WCC detector. The LA was applied to various subsets, including the whole set. The differences in the resulting XSELs were revealed and interpreted in terms of the potential influence of various parts on the J29 preparation, as described by the evolution of low-magnitude events.

Several key geographical parts of the entire seismic activity within the region can be identified from Figure 12. Aftershocks of the J20 earthquake are distributed across a relatively small spot that includes the J29 hypocenter and many of its aftershocks, including those on September 18, as well as the epicenter of the 2013 earthquake and its aftershocks. The inability of the J20 faulting process to follow the path of the J29 fault is likely related to an asperity in its way. This spot also includes the 2013 earthquake and its aftershocks, making it the most seismically active spot in the region based on event density, deserving special consideration. This spot is referred to as the “Asperity” and includes 49 MEs, as shown in the lower panel of Figure 13. In contrast, the other 245 MEs define the area referred to as “Out of asperity”. One out of 50 events within the “Asperity” area is excluded by its depth and is also not included into any of the subsets except the PDYAR set, to which it belongs. The performance of these MEs is a significant factor in understanding elastic energy accumulation and related seismic processes beyond the most active spot. Another important area is the southern rim of the region. Here, the seismic process is likely decoupled from the influence of the J29 earthquake, as indicated by the spread of its aftershocks. On the other hand, it may act as a solid wall or boundary where elastic energy is not released in the aftershock process. This area is referred to as the “Rim PDYAR”, as only MEs from the PDYAR set are selected.

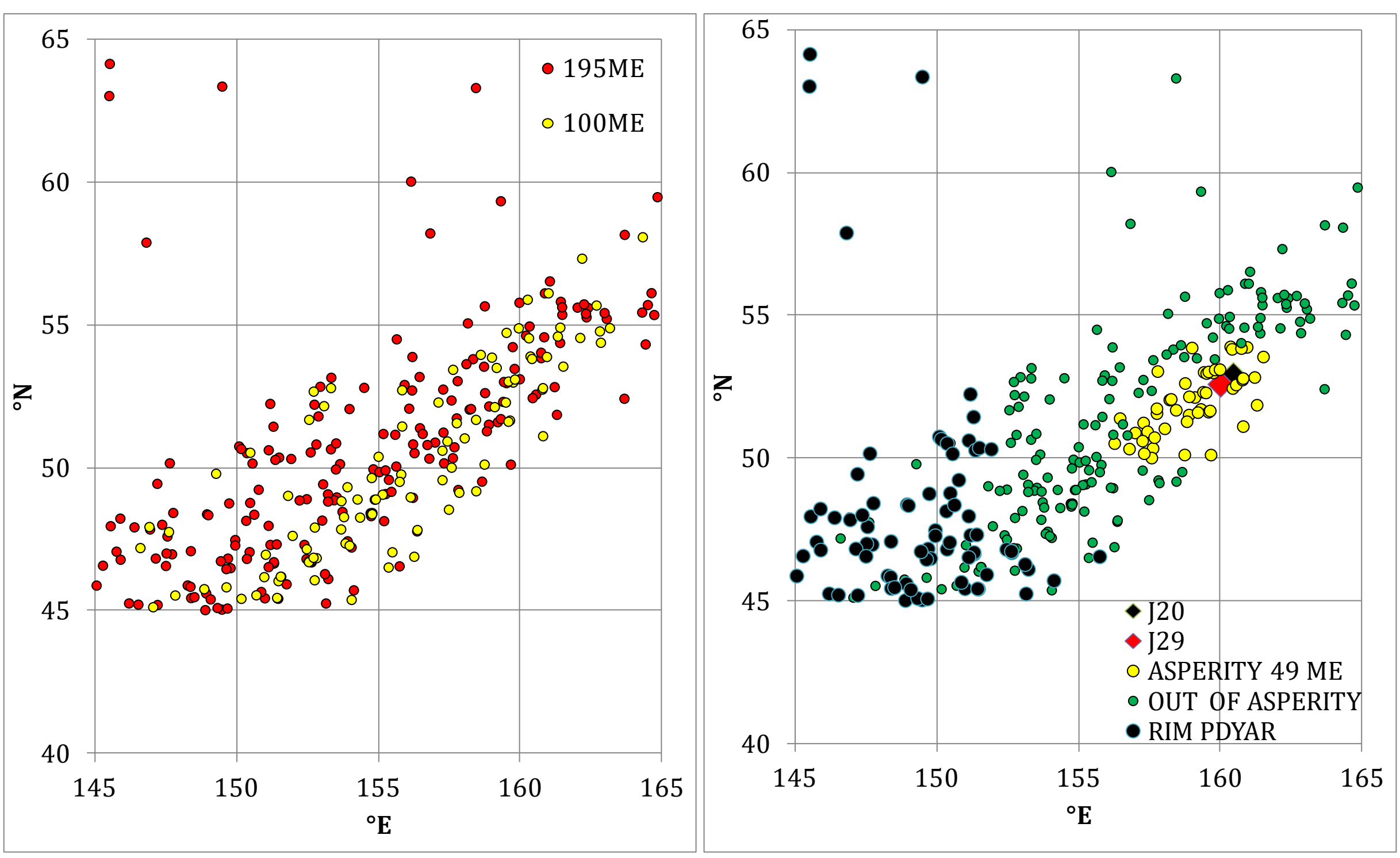

Figure 13. Two representations of 295 MEs used in the analysis of seismicity between July 6 and August 3, 2025. The upper panel includes the best 100 MEs and 195 PDYAR MEs. The lower panel shows geographical excerpts of MEs from the full set. Asperity related 49 MEs yellow circles). Out of asperity 245 MEs (green circles). 79 MEs with PDYAR located on the rim of the studied area (black circles over green ones). The J20 and J29 epicenters are also shown.

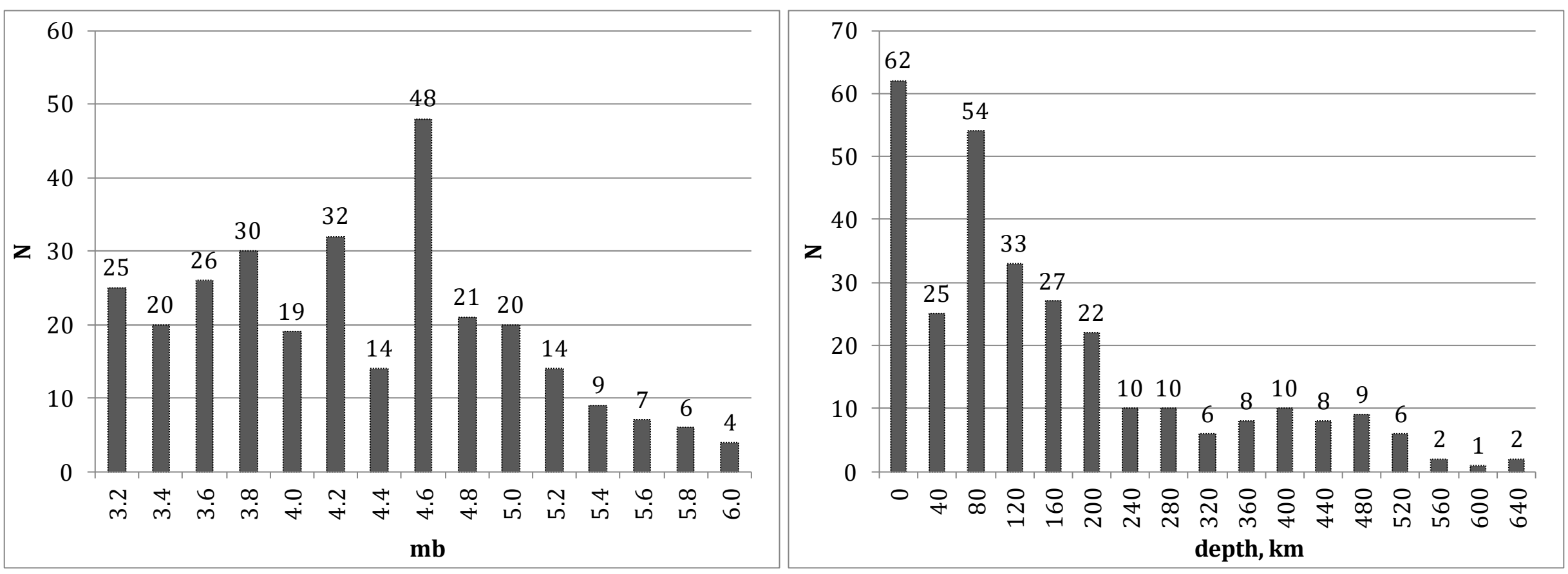

Figure 14. Frequency distributions of $m_b$ and focal depth for the 295 MEs.

The frequency distributions of the 295 MEs over $m_b$ and depth are shown in Figure 14. The deep MEs have lower magnitudes as they are corrected for the source position, and some deep events also have lower Nass, as the deep seismic activity is weak and the ME choice is scarce. Each and every ME, together with all associated templates, follows the same detection and local association processing stages. There are 295 individual detection lists and corresponding XSELs. Each individual XSEL can be compared to the REB before the CR stage, where some event hypotheses can be suppressed. Therefore, the effectiveness of one ME can be assessed at personal and collective levels.

***Varying the statistical significance threshold for XSEL events***

The detection lists obtained for thresholds adjusted to the purposes of finding ultra-weak signals contain signals with very low standard SNR values. Sometimes these signals are below the noise level. The SNR threshold is used to reject some of the detections that have appropriate SNRcc but too low-amplitude actual signals. As it works in the gray zone where valid and false detections are in almost equal proportions, its tuning was a mandatory task in the preparations to routine WCC processing on a prototype pipeline. The tuning results are used in this study. Therefore, the SNR filter is applied because it demonstrated an important positive outcome for the rate of associated detections.

Since the WCC-detected signals are mainly in the gray zone, the LA process is able to associate only a part of all detections with event hypotheses. These hypotheses are defined by the level of statistical significance based on two principal parameters: the probabilities of association of the corresponding stations with REB events in the region, and the non-random the SNRcc values of the detections. The lower the allowed statistical significance, the larger the number of created XSEL event hypotheses. To understand the effect of the defining parameters in the LA on the generation of XSEL events and their statistical properties, one can introduce a set of thresholds and calculate the corresponding XSELs. The numbers of XSEL events and their distributions over the event weights are then compared to the known bulletins for confirmation or rejection. The REB is the main source of such information.

There are several examples of WCC processing at the IDC with the XSEL passing the interactive review stage, which proved that the REB misses from 50% to 100% of REB-ready events. These events must be included in the REB according to the CTBT requirements. All events that could be found in the IMS data have to be found. The REB-ready events that were not found by the IDC but were detected by the WCC are the smallest and most relevant to the CTBT verification regime. When assessing WCC processing, it is important to consider the REB as a reliable source of information for larger events. However, for smaller events it has to be corrected by a factor of 2 to 3.

A tentative set of LA parameters meeting the requirements for the weakest events study was first introduced by Kitov [2026b] and tested in the recovery of the seismic process before the Sea of Okhotsk earthquake that occurred on May 24, 2013. There were no REB events for more than one year before this major earthquake. The WCC allowed for finding a sequence of weak earthquakes with an increasing magnitude right before the mainshock. Here, this set is accepted as successful for the purpose of weak seismicity recovery and analysis of the earthquake preparation process. The Kamchatka 2025 earthquakes had a history of seismicity detected by the IDC before the J20 and J29 major events. This allowed for assessing the performance of the set of LA thresholds and estimating their properties through the recurrence curves they produce.

The REB and other bulletins cannot find the events that are most important for understanding the earthquake preparation process [Schaff *et al.,* 2025]. The WCC approach focuses on events below the corner magnitude of the recurrence curves, between 2.0 and 3.0-3.4. The signals from these events are not detectable by analysts even on the best array stations with waveforms optimally filtered and beamed to the source hypocenter. Therefore, the set of LA thresholds was established without any reference to the REB or other bulletins. Instead, the thresholds were set from scratch, drawing from previous experience with XSELs. Due to numerous defining parameters in the LA processing, conducting a grid search through all of them with only one basic WCC detection list for the Sea of Okhotsk, containing a few hundred XSEL events at best, would not be feasible.

The approach was simplified by the introduction of two LA versions, each with different sets of defining parameters and six cases per version allowing for the fine-tuning of statistical significance. These two versions needed to be sufficiently distinct in a statistical sense to avoid overlap between the cases. The strict version of the LA parameters has to be close to the corner magnitude of the recurrence curve for the region in order to provide a smooth transition from the XSEL to the REB. This version should not generate any XSEL events during quiet seismic periods, similar to the REB. The following LA parameters are used for this task: 1) the number of associated stations is 3. 2) The origin-time tolerance is a case dependent parameter. 3) The minimum total event weight is 2.5. 4) The weight of one of the best stations to be associated with any 3- to 5-station events is 0.855. 5) The minimum sums of SNRcc values for 3-, 4-, and 5-station events are 15.0, 18.5, and 22.0, respectively. 6) The lowest possible value for the minimum SNRcc at one of the top stations from 3) is 5.0. 7) The grid radius is 48 km, defined by 12 steps of 4 km. The event hypotheses beyond a radius of 43.2 km are rejected. This radius can be less than half the distance between neighboring MEs, but it is important to increase statistical significance by reducing the flexibility in location.

On the opposite side of the XSEL sensitivity is the weak LA version. The defining parameters were guesstimated using the experience with LA processing in a number of previous studies and from the SNRcc curves in Figures 6 through 8, as well as similar curves for other involved stations. The parameters for the weak LA version are as follows: 1) the number of associated stations is 3. 2) The origin-time tolerance is a case dependent parameter. 3) The minimum total event weight is set at 1.8. 4) The weight of one of the top stations to be associated with any 3- to 5-station events is 0.80. 5) The minimum sums of SNRcc values for 3-, 4-, and 5-station events are 14.0, 17.5, and 21.0, respectively. 6) The lowest possible value for the minimum SNRcc at one of the top stations from 3) is 4.5. 7) The grid radius is 90 km, defined by 15 steps of 6 km. The event hypotheses beyond a radius of 81.0 km are rejected.

There is a wide gap in statistical significance between the weak and strict versions. The task at hand is to accurately measure potentially subtle changes in the fluxes of extremely weak signals and events that may not be detected by other methods. An added flexibility to the versions is provided by changing the origin-time tolerance. This parameter directly impacts statistical significance by eliminating random detections in proportion to the width of the association window, which is twice tolerance. For signals related to the sought seismic activity, probability of rejection increases when narrowing the window, but the dispersion of arrival times relative to the average origin time depends on the quality of the signals defined by their position on the SNRcc curves. The valid signals tend to cluster closer together.

After a brief test, six different origin-time tolerances, Δt, were introduced to determine their impact on the XSELs from the same detection lists: 5.0, 3.0, 2.0, 1.0, 0.5, and 0.25 seconds. These six tolerances create 12 version/case pairs. They can be aligned in a formal order: $v_1c_1$, ..., $v_1c_6$, $v_2c_1$,...,$v_2c_6$, where $v_1c_1$ corresponds to the weak LA version with $\Delta t$ =5.0 s. The shortest $\Delta t$ allows for practically only actual event hypotheses to be created. For detections randomly distributed in time and in SNRcc value, the probability of 3 or more of them at the best stations having origin times in a 0.5 s time window is extremely low, considering the observed detection rates of 30 to 40 per hour. The downside of having a shorter $\Delta t$ is the higher likelihood of missing many weaker but valid XSEL event hypotheses, associating weaker signals with poor arrival-time estimates and larger travel-time residuals. The $v_2c_6$ pair should not be able to detect too many new XSEL events in addition to the REB. Overall, the XSEL events that match the REB events have to be of a higher quality related to the SNRcc values of the associated signals.

The weak LA version with the narrowest origin-time window, $v_1c_6$, has to be focused on the new XSEL events with the best quality below the corner magnitude of the recurrence curve. These events are highly statistically significant, but can also include weaker WCC arrivals. The $v_1c_1$ pair has the highest resolution. It may contain many valid events and likely some false events, with both types formally matching the EDC. For an XSEL obtained with a given version and case pair, a formal statistical threshold can be defined to distinguishing between valid and false events as determined by the LA algorithm. Then, the strict and weak versions for the same case, $\Delta t$, define a specific range of event quality between their corresponding thresholds. The assumption behind introducing two versions and six cases was that the corner magnitude of the recurrence curve of the detected XSEL events depends on these thresholds of statistical significance. The potential influence of source mechanisms, as well as of the noise level at stations around the arrival times of all associated phases, is averaged as for the long-term REB recurrence curve.

Formally, for pairs of $v_ic_j$, we introduce quality boundaries $B^i_j$, where $i$=1,2 represents the version - weak or strict, and $j$=1,...,6 represents the tolerance case. By design, $B^1_j \leq B^2_j$ and $B^i_j \leq B^i_{j+1}$. The quantitative meaning of these boundaries should be revealed by WCC calculations, with a decreasing number of event hypotheses generated as $j$ increases for a given $i$. When the sought seismic activity from a completely quiet state reaches the lower boundary $B^i_j$, the number of weak events grows, while the number of strict events, which must have larger magnitudes and statistical significance, remains constant. The change in the ratio of the event numbers in the weak and strict XSEL versions, $B^1_j/B^2_j$, expresses the change in the seismic process within the range between two version and the same case, $\Delta t$.

When the activity starts to grow from a quiet state, the $B^1_j/B^2_j$ ratio also increases with the growing number of XSEL events above $B^1_j$. The activity may not reach its upper boundary, $B^2_j$, and then may start to fall back to the initial level. In some cases, the activity may reach the upper boundary and move above it, as happens after larger earthquakes. When the activity penetrates above the upper boundary $B^2_j$, the ratio starts to fall because of an increasing number of events above the upper boundary and a constant or decreasing number above the lower boundary. The larger earthquakes generate noise that masks the smaller events. In an aftershock sequence following the largest earthquakes, the boundaries $B^1_j$ and $B^2_j$ can be displaced to almost the same magnitude threshold above the corner magnitude of the REB recurrence curve. At that point, the $B^1_j/B^2_j$ ratio may reach the lowest level, expressing the magnitude distribution of earthquakes according to the Gutenberg-Richter law. With the decaying post-seismic activity, the ratio grows again as the number of events above $B^1_j$ has to decrease rapidly and the number above $B^2_j$ decreases much slower. In conclusion, there are six quality ranges potentially related to magnitude to follow the seismic process before an earthquake. It is assumed that the XSEL events in these ranges span the magnitude range from the corner magnitude down to magnitude 2. The boundaries can be adjusted if they do not fit the assumptions.

***The recurrence curve for the low-magnitude events below the IDC detection threshold***

Recurrence curves for various seismic regions are estimated using long-term observations. They represent both the structure of the energy-emitting media, such as the zones of continental faults or subducting plates, and the external energy flux creating stresses in the zone that results in energy release in earthquakes of different sizes. The external energy flux can be considered almost constant, generated by tectonic processes lasting millions of years. Seismic energy release, as part of the Earth's internal thermal energy dissipation, is a quasi-periodic process at a historical timescale of observations, but changes with the changing geological structures created and destroyed by tectonic forces. The recurrence curves are snapshots of long-term evolution and may change with changing tectonics and geological structures. At the same time, any recurrence curve is a cumulative observation with practically invariant properties, allowing for the prediction of important features such as seismic periodicity, activity, and the largest possible earthquakes.

The short-term spatial and magnitude structure of the seismic process is much more variable than the overall recurrence curve for a given region obtained from decades of continuous observations. This is a well-known effect that poses a challenge to predicting large-scale events that endanger populations and infrastructure in various regions, with the Kamchatka Peninsula being one of the most vulnerable seismic hazard regions on Earth.

The recurrence curves obtained from the IDC catalog for the studied region have to be depth-dependent because the magnitude estimates are subject to depth corrections. For near-surface events (0-40 km), the recurrence curve is likely the most relevant for the current analysis. The IDC has fixed the J29 mainshock depth to the free surface according to a procedure not to report the free-depth solution for events with poor depth estimates [Coyne *et al.*, 2012]. Approximately 16,000 events out of ~26,500 within the area have depth estimates below 40 km, and almost 15,000 of them are fixed on the surface. Figure 15 demonstrates the recurrence curve for these near-surface events and its extrapolation to the lower magnitude range. The number of events missing from the IDC bulletin increases below the corner magnitude.

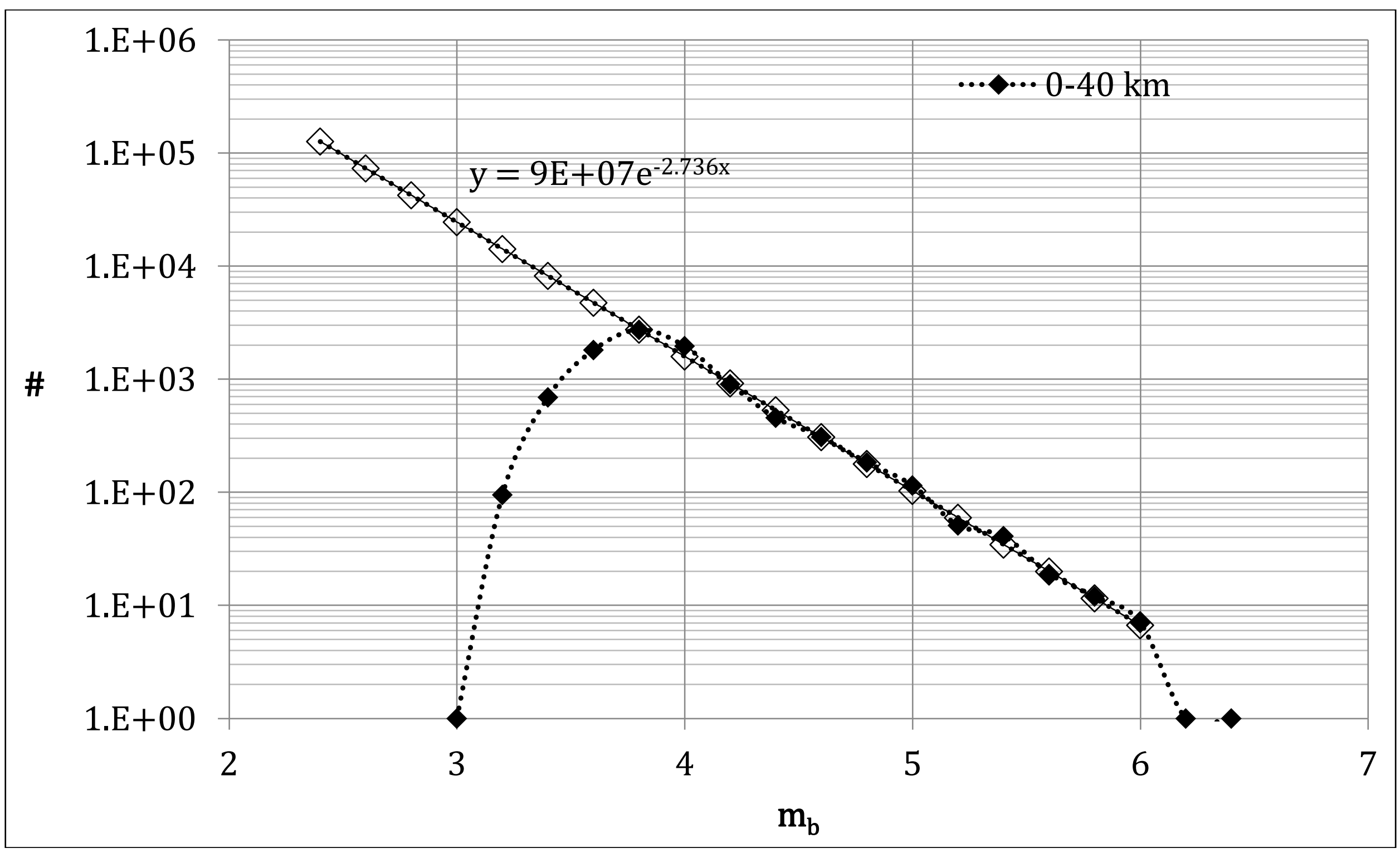

Figure 15. Recurrence curve for the depth range 0 km to 40 km. The approximated recurrence curve is extrapolated to lower magnitudes to show the deficit of event reported by the IDC.

At array stations, the WCC detector allows us to find many of the missed events, as the detection threshold is 5 to 10 times lower than for the standard beamforming. At 3-C stations, the gain is lower but still contributes to improving the bulletin. The actual WCC gain for arrays is difficult to estimate in standard threshold values, since the WCC detector uses a different time series. The WCC processing is accompanied by estimates of the standard SNR following the IDC rules. The arrival time of the WCC detection is determined by a peak in the SNRcc time series that is calculated using the CC-time series. The peak SNR value is taken in a 5.5 s window following the WCC arrival time and then used as the standard SNR. For a valid IDC signal, such a WCC-based SNR is close to the IDC estimates. When the IDC detection is absent, the SNR related to the WCC detection is taken from a series containing noise, and the peak SNR is a poor representative of the WCC signal amplitude for the magnitude estimates. This is a magnitude estimate representing the peak noise value. When using the RMS amplitude in the detection cross-correlation window as an amplitude approximation, the noise magnitude is smoothed but still represents noise rather than the corresponding WCC signal, which is often buried deep within the noise.

To characterize the set of 12 version/case pairs, it is important to obtain recurrence curves with a corner value of statistical significance pertaining to each pair. The statistical significance of XSEL events cannot be calculated directly from the data, nor can the body-wave magnitude estimates. To present the results obtained for the 12 pairs reasonably, one has to find a relative scale that provides a reliable and clear functional dependence. This scale should serve as the magnitude in a standard recurrence curve.

The number of events, N(m), above a specific magnitude larger than the corner one can be estimated by integrating the regression line in Figure 15:

$$f(m) = 2.34\cdot 108\times 10^{(-1.2\cdot m)}\,, \qquad (1)$$

where the decay constant *b*=2.736 is reduced to 1.20 for base 10, and the coefficient $9.0\cdot 10^7$ is corrected for a bin width of 0.2. For m=5.0, the predicted and observed number of events during the entire IDC observation period between January 2001 and December 2025 is 245. When extrapolated, relationship (1) allows for estimating the number of events above any magnitude below the corner value, including those missed in the IDC bulletin for various reasons. For a magnitude interval $[m_1,m_2]$, the number of events for the whole studied period is as follows:

$$dN = N(m_1) - N(m_2) = (2.34\cdot 10^8/2.736)[\exp(-2.736*m_1) - \exp(-2.736*m_2)] \qquad (2)$$

If we neglect the second term for magnitudes above 7, the total number above $m_1$ is as follows:

$$N(m_1) = (2.34\cdot 10^8/2.736)\exp(-2.736*m_1) \qquad (3)$$

Relationship (3) has to be valid for any part of the recurrence curve above the corner magnitude and follow an exponential distribution. A reliable recurrence curve has to include a sufficient number of events in relatively narrow bins of 0.1 to 0.3 magnitude units to reduce the uncertainty of statistical estimates. This requires a longer observation time, even in a seismically active region like the Kamchatka Peninsula with the subducting Pacific plate. Shorter time intervals may not contain enough data, or the data may be biased. For example, the Kamchatka earthquake on July 29, 2025, generated thousands of aftershocks in a relatively small area. In the days following the mainshock, relatively large and smaller aftershocks can be under-represented in the recurrence curve. The highly elevated ambient seismic noise after large events masks the smaller ones. As a result, the decay constant can be shifted to lower values. This is likely the case under consideration, and the recurrence curve in Figure 15 does not necessary have to be the reference for the XSEL bulletin.

***Estimation of magnitude scale for the XSEL events***

The number of events in the 12 version/case pairs forms a sequence. When these pairs were introduced, there was no assumption of a specific magnitude threshold value related to each pair. The only assumption was that the number of events in a given version decreases with the case order number, meaning it decreases with a decreasing origin-time tolerance. The number of events in all cases of the weak version was not always expected to be higher than the numbers in all cases of the strict version. Ideally, the magnitude thresholds for the 12 pairs should form a sequence with a constant increment of Δm. Then, the corresponding numbers of XSEL events would follow an exponential law. To test the actual distribution of the magnitude thresholds in the 12 pairs, one can present the results of the XSEL calculations using the order number of the version/case instead of magnitude and estimate the difference between the ideal exponential curve and the observed curves. The order numbers range from 1 to 6 for the cases of the weak version and from 7 to 12 for the cases of the strict version.

This representation is an important constraint on the technique used to estimate the change in the number of low-magnitude events in the range where no REB events are available. The time evolution of energy release from the smallest cracks to the level of IDC-detectable earthquakes during a rapid transition toward the mainshock is measured in narrow magnitude bins. The wave of events progresses through the bins and passes the stages of increase, peak, and possibly a subsequent decrease in the number per unit time as the wave progresses to larger events when approaching the coseismic phase.

An example of aligning the version/case pairs by their order numbers is shown in Figure 16. It displays a black-dot curve for the XSEL event hypotheses created during the 10-day period after the J20 earthquake, including the day of July 29 but excluding the last hour with the J29 mainshock. The first pair ($v_1c_1$, weak LA version and $\Delta t$=5.0 s) has ~9000 XSEL event hypotheses, and the 12th pair ($v_2c_6$, strict LA version and $\Delta t$=0.25 s) generated ~1600 XSEL hypotheses. The numbers of XSEL events for all pairs are sufficient to generate reliable statistics. This period includes several hundred REB events, which are not enough to create a statistically reliable recurrence curve. The XSELs for these days have hypotheses matching almost all REB events, including those added to the REB by IDC analysts without seed events in the automatic bulletin SEL3.

The sequence of the number of XSEL events the in 12 pairs represents a standard recurrence curve by the number of events above some threshold below the REB corner magnitude. Relationship (3) predicts an exponential decrease in this number for a constant step between the 12 magnitude thresholds. The black-dot recurrence curve in Figure 16 shows that the thresholds for the 12 pairs for the case of the 11-day period with active seismicity are distributed with an approximately constant magnitude increment as $R^2$=0.995 indicates. However, the exponential decay constant, $b$=0.18, is much lower than that for the entire REB recurrence curve in Figure 15.

Therefore, it is important to compare the black-dot XSEL curve in Figure 16 with the REB recurrence curve for the same period. Unfortunately, there are only 633 REB events in the aftershock sequence of the J20 earthquake. This is not enough to obtain a reliable recurrence curve. To improve these statistics, 779 REB events for 25 hours of after the J29 earthquake were added. The recurrence curve with 1412 events is shown in Figure 17. In addition, the recurrence curves for three different periods are shown: 1) the 5-day period between July 30 and August 3, 2025; 2) the aftershocks of J29 between July 29 and December 4, 2025; 3) all REB events in the area between January 1, 2001, and December 4, 2025.

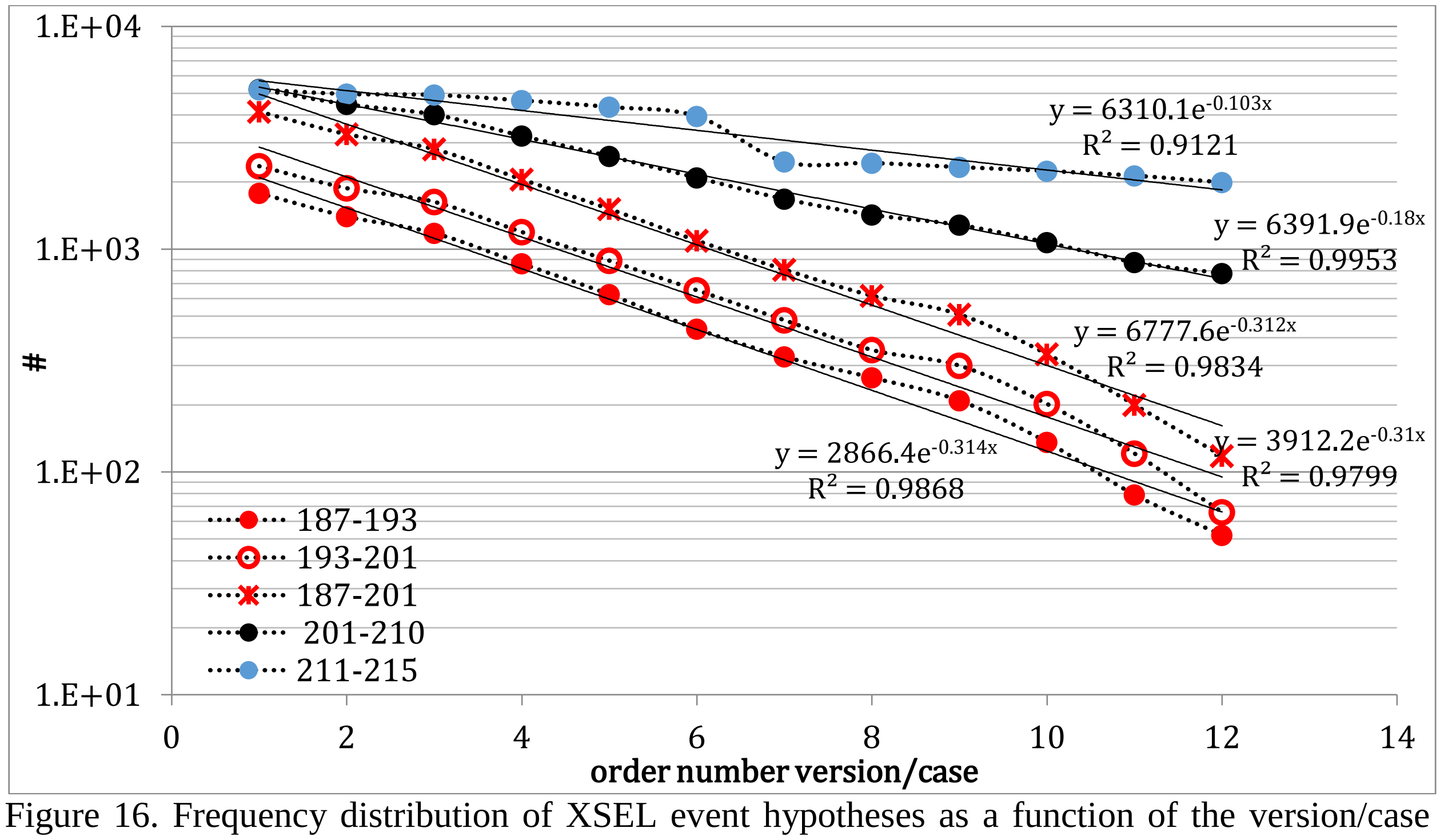

Figure 16. Frequency distribution of XSEL event hypotheses as a function of the version/case order number for various time periods: blue dots represent the period between July 30 and August 3, 2025; black circles represent the period between July 20 (day 201 after the major earthquake) and July 29 (day 210); red signs represent two separate intervals between July 6 (187) and July 20 (201), as well as the entire periods.

These four REB recurrence curves were normalized to the corresponding total number of events. The obtained **pdf**'s reveal significant differences in the decay constant between the frequency distribution of magnitudes in the short period between July 10 and July 29 (*b*=1.04) and that of the entire REB (*b*=2.62). For the period of the most intensive post-seismic activity between July 30 and August 3, 2025, *b*=1.91 and $R^2$=0.993. For the aftershock sequence between July 29 and December 4, 2025, *b*=2.42 and $R^2$=0.996. For the 11-day period characterized by non-peak post-seismic activity, the *b*-value is much lower than for a shorter period of an extremely high aftershock activity generated by the J29 event as well as for longer periods of observations. In all four curves, the REB events from all depths are included.

The XSEL curves, which have a generic x-axis scale in Figure 16, can be converted into the standard $m_b$ form as shown in Figure 18. For example, the XSEL curve for the 2025201-2025210 period with a linear scale from 1 to 12 and *b*=0.18 can be reduced to $m_b$ scale by a simple multiplication factor of 0.1282 resulting in an actual *b*=1.0405. Figure 18 shows an example of such a conversion. Another issue with this conversion is the absolute $m_b$ value for the converted XSEL curve, as the methods to obtain the REB and the XSEL are different and the numbers are not directly compatible. The $m_b$ start point of the XSEL curve with *b*=1.04 is adjustable. In Figure 18, the REB and the converted XSEL curve intersect near the corner magnitude by adjusting the XSEL curve’s start point by 2 units of magnitude and the level by an amplitude factor of 2.5. This adjustment is arbitrary, and the XSEL curve can be moved closer to or further from the REB curve. However, the XSEL trend line extrapolated to larger values has to pass though the measured REB values. For the 2025201-2025210 period, the XSEL curve in Figure 18 covers the magnitude range from 2.1 to 3.5. This is a simple guesstimate of potential improvements in the event catalog using the WCC-based approach.

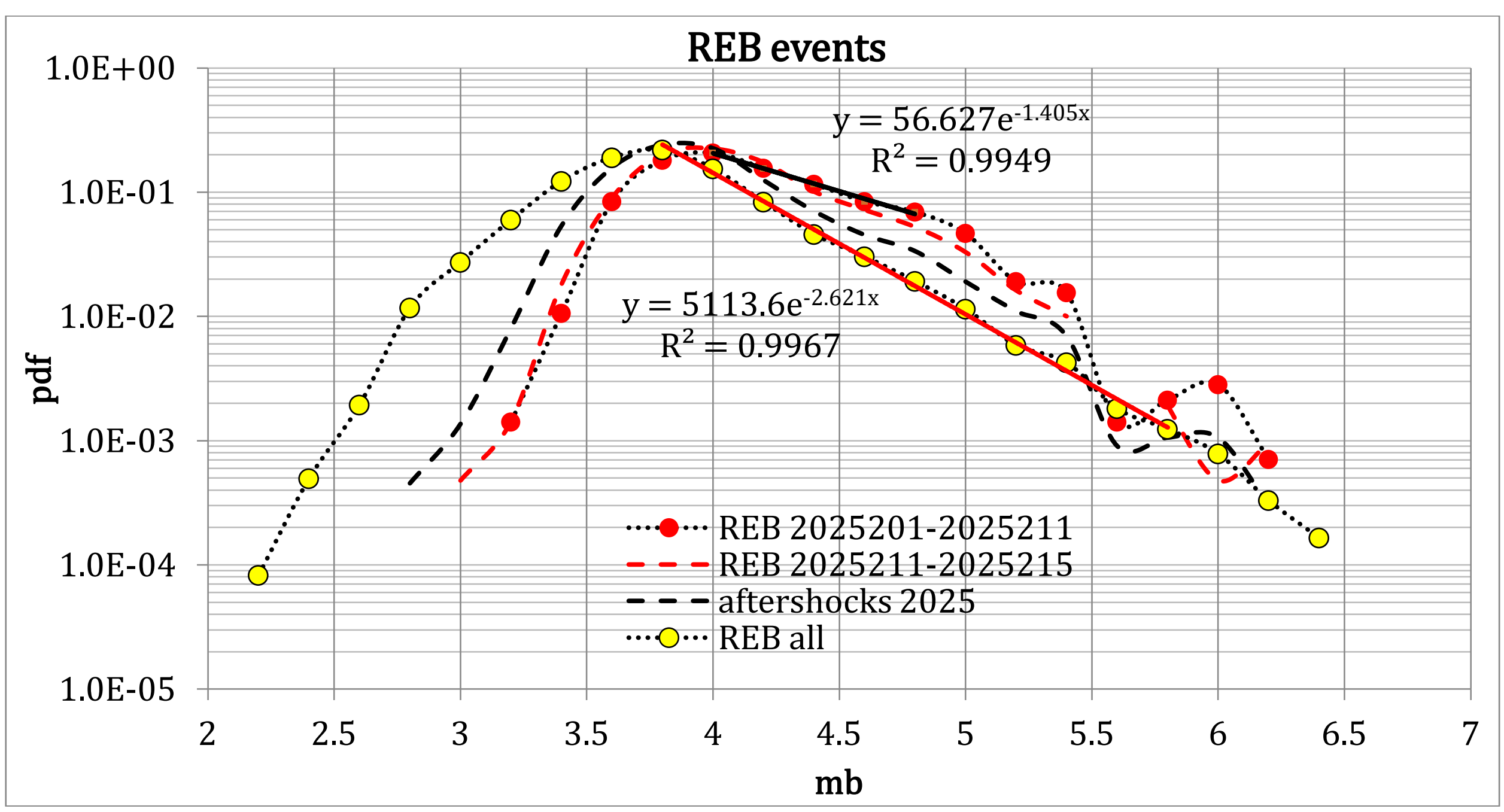

Figure 17. Four **pdf** calculated for the REB events in various periods: for the entire REB for the region with all depths included; for all aftershocks between the July 29 and December 4, 2025 earthquake ($b$=2.42, $R^2$=0.996); for 5 days after the J29 ($b$=1.91, $R^2$=0.993); for the 11-day period between July 20 and July 30, 2025.

The XSEL for the 2025211-2025215 period is shown in Figure 18 to demonstrate that the XSEL scale conversions are case-dependent. In order to match the decay constant and the REB level, the following values were used: a magnitude scale factor of 0.05302, a start point shift of 3.0, and an amplitude factor of 1.5. The high level of post-seismic activity in the first days after the J29 makes it difficult to find low-magnitude events, and the XSEL curve covers only a narrow $m_b$ interval between 3.05 and 3.65. This is a guesstimate as well.

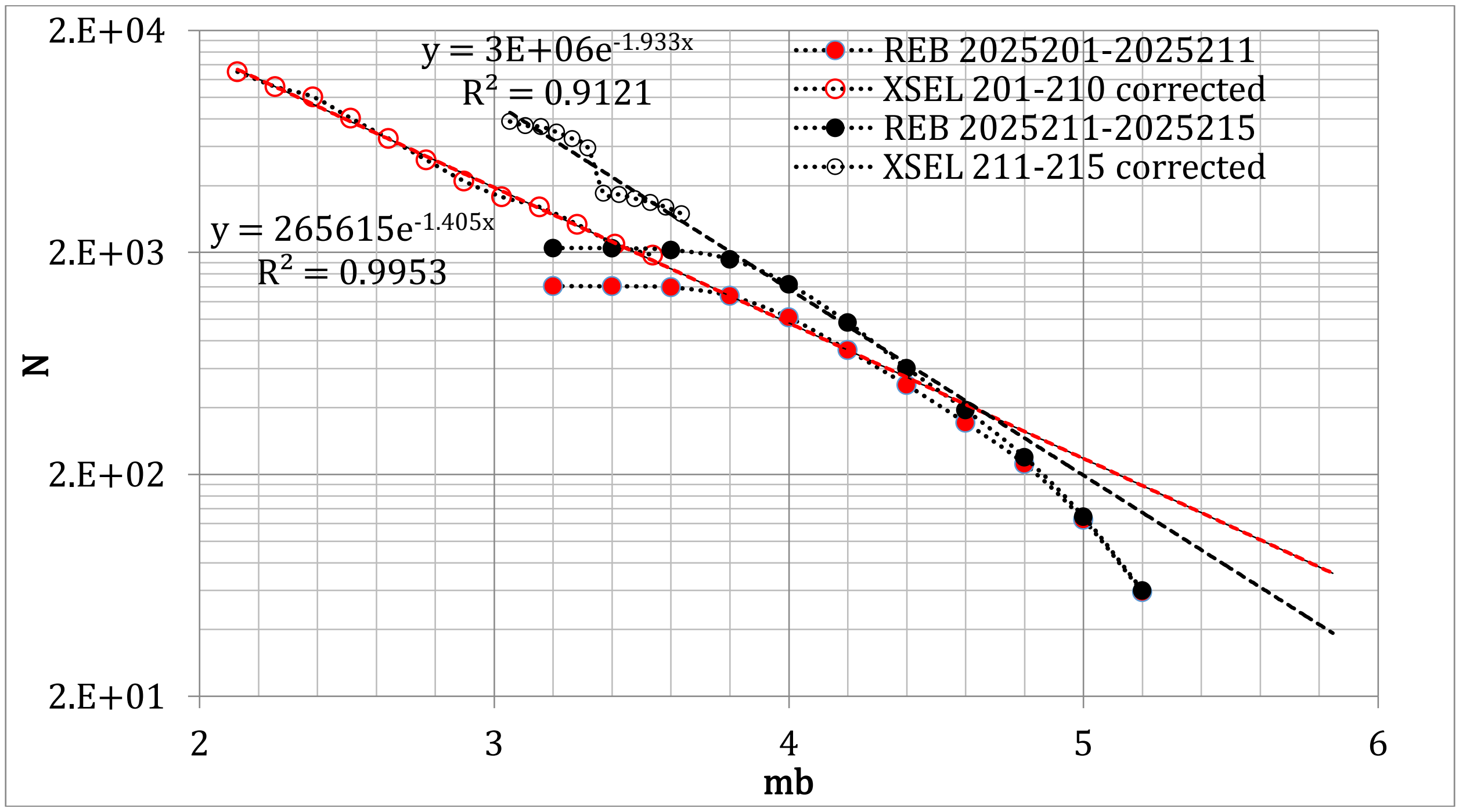

Figure 18. Scaling of the XSEL recurrence curve in Figure 16 to the REB for the same period of time.

There are presented two examples of the order number conversion into magnitudes for the XSELs related to the 12 pairs with two versions of the LA setting and 6 different origin-time tolerances. There are also different XSEL recurrence curves for the periods of lower seismic activity in Figure 16: 2025187-2025193, 2025193-2025201, and the joint period 2025193-2025201. The earthquakes from the REB for the period from 2025187 to 2025216 are shown in Figure 19. The XSEL curves cannot be directly compared to the REB recurrence curves for the same periods and then converted accordingly. Such a conversion is physically possible, because these XSELs represent actual seismicity below the corner magnitude of the local or global REB curves.

Two examples of order-number conversion into magnitudes are presented for the XSELs related to the 12 pairs with two versions of the LA setting and 6 different origin time tolerances. There are also different XSEL recurrence curves for the periods of lower seismic activity in Figure 16: 2025187-2025193, 2025193-2025201, and the joint period 2025193-2025201. The earthquakes from the REB for the period from 2025187 to 2025216 are shown in Figure 19. The XSEL curves cannot be directly compared to the REB recurrence curves for the same periods and then converted accordingly. Such a conversion is physically possible, because these XSELs represent actual seismicity below the corner magnitude of the local or global REB curves.

***XSEL recurrence curves for the period of the highest seismic activity***

The Kamchatka earthquake on July 29, 2025, generated a record number of aftershocks per day detected by the IDC. There are 760 events in the REB for July 30. This makes one event per 114 seconds. During the first 6 hours of July 30, one event per 82 s was detected. An enormous number of smaller events were hidden in the noise generated by the very high-magnitude aftershocks, as Figure 19 shows. This effect has a negative impact on the recurrence curve for the whole region. The number of events closer to the corner magnitude is heavily underestimated, with the deficit increasing with a decreasing magnitude. The smaller, but above the corner magnitude, events can be and are found during periods of lower activity, participating in defining the recurrence curve, its slope, and corner magnitude. The latter values have to be biased, however, as the mid-size and larger earthquakes are counted accurately. A larger portion of these low-magnitude events are recovered in the WCC processing. At the same time, the number of new XSEL events, *i.e.*, events not matching arrival(s) associated with any REB event at one or more IMS stations, is relatively low.

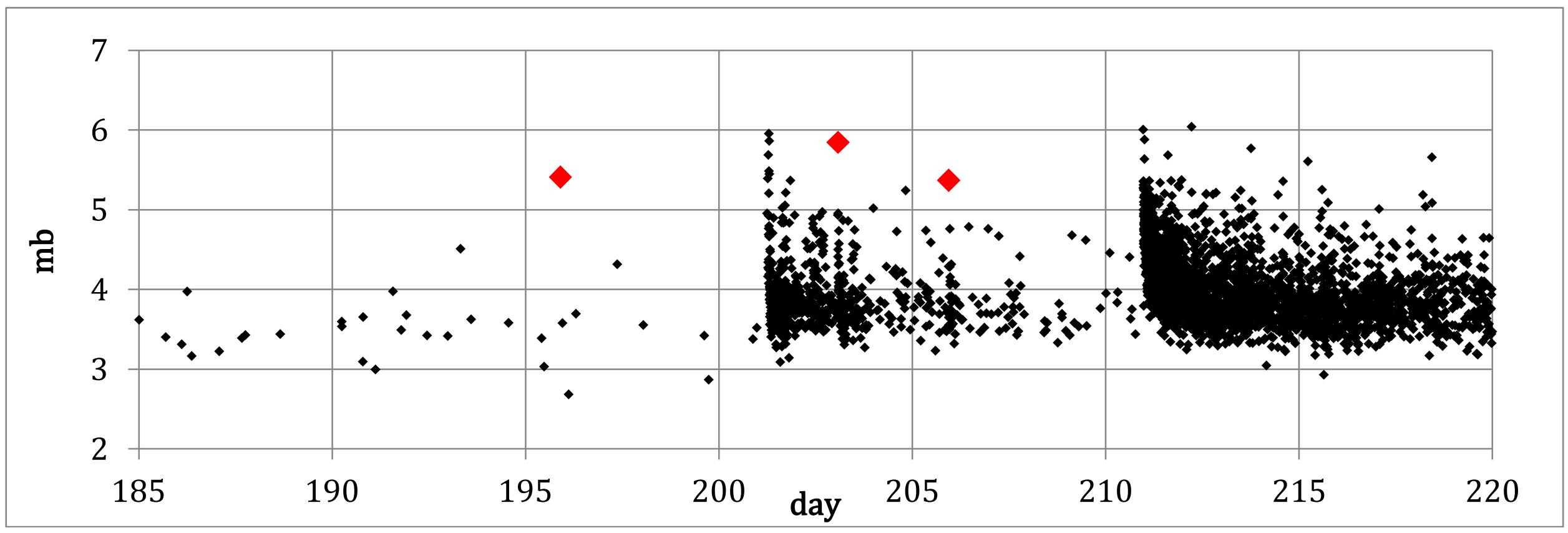

Figure 19. Magnitude of the REB events as a function of origin time for the period from July 4 to August 8, 2025. Three events with $m_b$ around 5.5 potentially affecting the XSELs are highlighted by red.

One of the peculiarities of the 2025211-2025215 curve in Figure 16 is an abrupt fall in the number of XSEL events in the transition from the weak LA version to the strict one. There is no such effect in the 2025201-2025210 XSEL curve. A more detailed analysis of this phenomenon is presented in Figure 20. In the left panel, the J29 XSELs are presented. The second (2011.375) and fourth (2011.875) 6-hour-long intervals on July 30, with the highest recorded seismic activity, demonstrate not only an abrupt drop in XSEL numbers, but also that these numbers remain constant for all six cases of the same LA version. This suggests that the threshold magnitude of the XSELs for all cases of the same version is the same because the quality of arrivals is so high (large SNR and SNRcc values) that the scattering of origin (or travel) times is less than 0.25 s for at least 3 of them at the stations with the highest weights. The LA creates the same events except may be for a small difference observed as slight changes in the XSELs. The shift in the level between the weak and strict versions indicates a change in the threshold magnitude related to the required event statistical significance defined by several quality parameters, including the minimum weight for a valid XSEL event hypothesis.

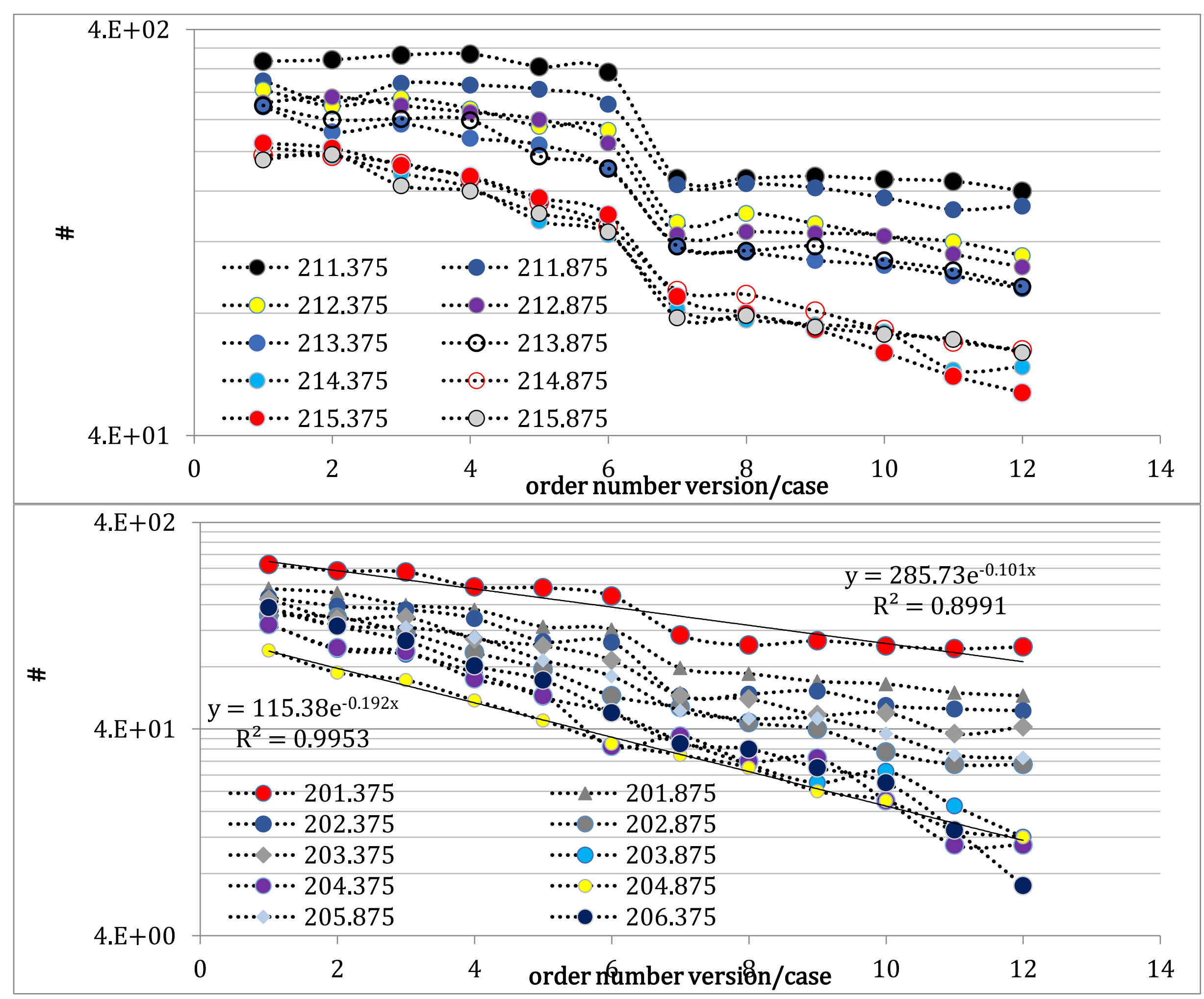

Figure 20. Comparison of the 12 pairs' performance in the periods of very high activity. Upper panel: aftershock sequence of the J29 in the even quarters (6 hours) of the day. Lower panel: Same for the J20 aftershocks. The curves are shown for every second 6-hour interval.

Since July 31, the number of XSEL events has started to decrease in correlation with the case number, eventually reaching a shape similar to those shown in the lower panel of Figure 20, where the XSEL curves are presented for the J20 aftershock sequence. Only during the first two days following the J20 event, a slight step between the weak and strict versions is observed. Then, the curve begins to resemble an exponential trend. Significant gaps in the XSEL numbers are observed between 2011.875 and 2012.375 as well as between 2014.375 and 2014.385. The difference between the curves within the groups separated by these gaps is much smaller. These gaps are interesting for understanding the process of post-seismic activity evolution. The J20 aftershock sequence does not demonstrate such gaps except for the difference between the first (201.375) and the second (201.875) curves, which is likely related to a natural decrease in seismic activity.

The curves in Figure 20 for the period between July 20 and August 3 include many XSEL hypotheses matching the REB events. Figure 21 presents the number of matched REB events in the first four 6-hour intervals on July 30, 2025. For comparison, the number of SEL3 events (left panel) and REB events (right panel) are shown. The XSEL is an automatic bulletin that should be compared to the SEL3, but its primary objective is to match the REB. During the period of peak activity, the XSEL does not match all the SEL3 events that were used as seed events for the REB. The REB for the first 6 hours of July 30 lists 278 events in the studied area, with 198 in the SEL3. Several stations, such as PETK, KURK, and MKAR, have phases associated with almost all the REB events, making around 300 arrivals in 6 hours.

The WCC detector is tuned to a lower level of seismic activity relevant to the earthquake preparation process and produces approximately 60 to 90 detections per hour. The strict version with a 2.0 s origin-time tolerance generates around 250 XSEL events, and the weak version generates ~400. This is the highest possible event rate for a given WCC configuration. It can be tuned to higher rates, however, in order to study the most active periods. This should be done to count the number of REB-ready low-magnitude events and to correct the overall recurrence curve for the region.

In the current WCC version, the XSEL events for July 30, 2025, match a large number of purely REB events that are not based on the automatic seeds. This is an important observation to support the level of statistical significance of the XSEL events not matching any REB event. Retaining the same statistical constraints used in the WCC detection, LA, and Conflict Resolution processes guarantees a reliable link between the results obtained at different levels of seismic activity.

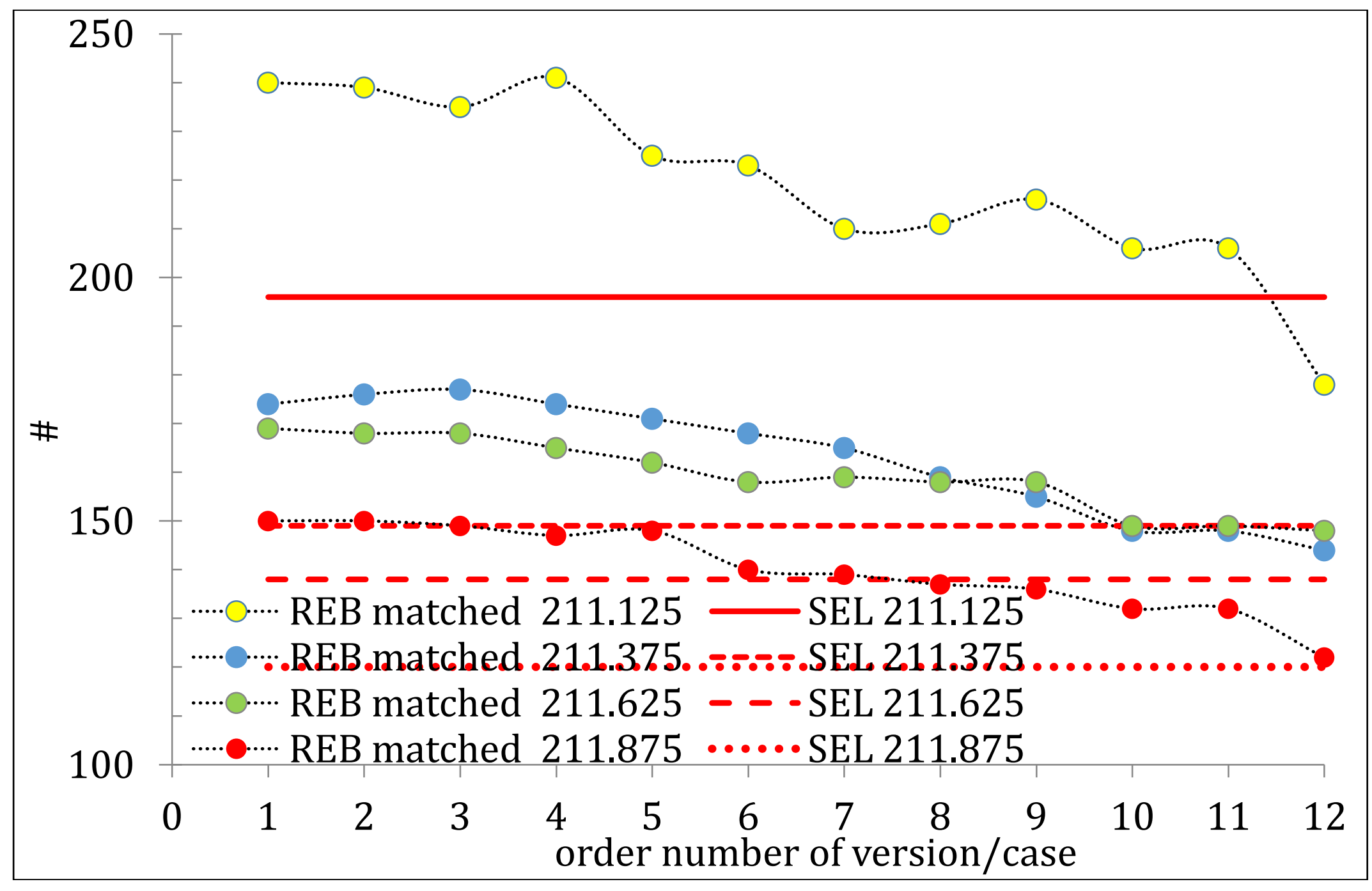

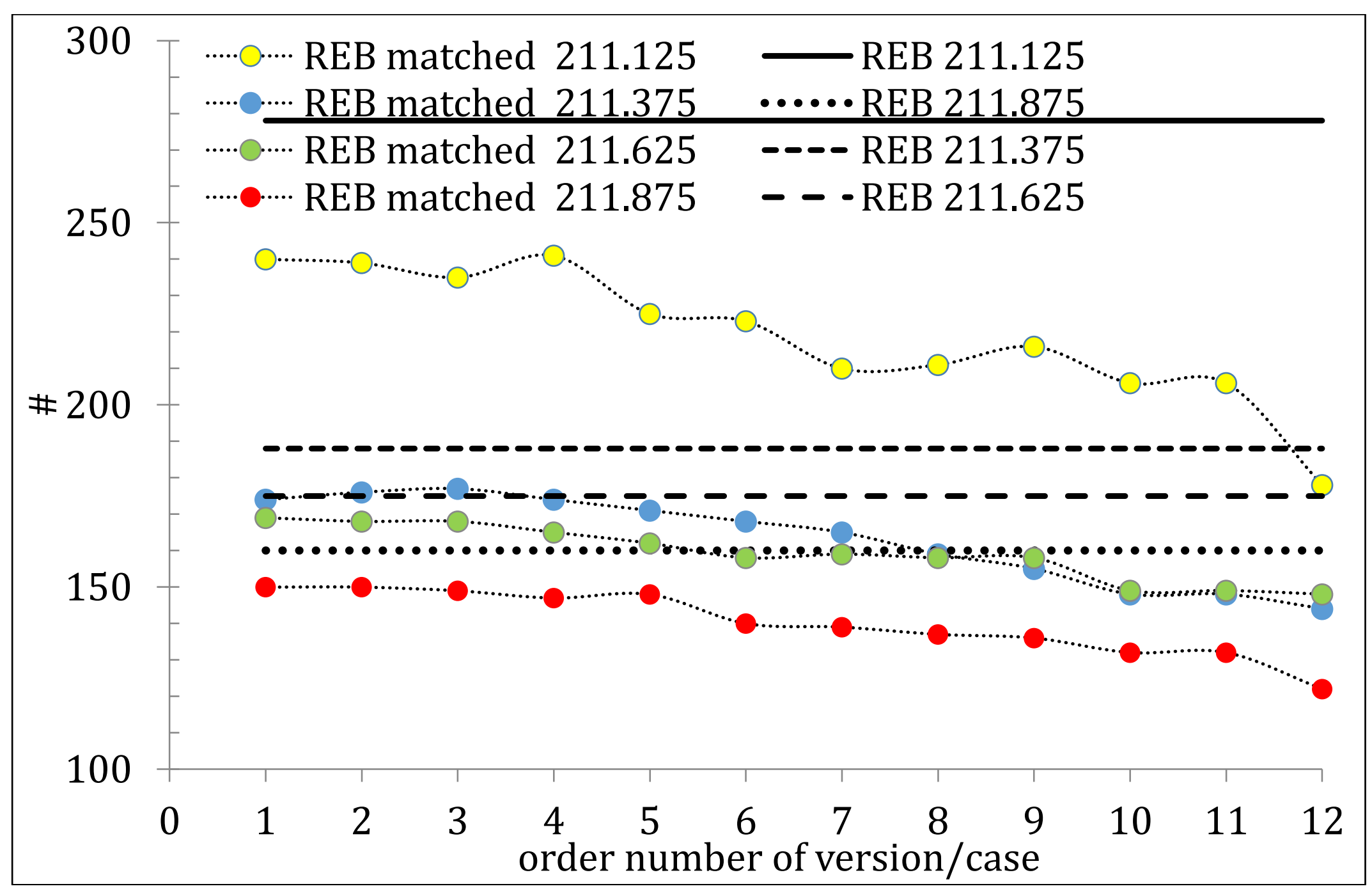

Figure 21. The number of XSEL events matching the SEL3 (upper panel) and the number of REB (lower panel) events for the same periods. The corresponding REB and SEL3 numbers are shown by red lines.

There are dozens of REB events in Figure 21 that are not matched in the XSEL. The reason behind this underperformance could be related to the WCC processing and the quality of the REB events obtained during the interactive review. Figure 22 displays the frequency distribution of SNR values for the first P-phases associated with all REB events on July 30, 2025. All SNR estimates are sourced from the IDC database. The pattern is controversial. There are 5241 arrivals with SNR>10, which are high-quality signals not shown in Figure 22 because they are well matched by the XSEL. The curve between 0 and 10 includes 15802 values in 0.2-wide bins. There are 4872 detections with SNR<3.4, with 244 of them having a default value of -1.0 and fixed at the SNR=0 point. They fall below the automatic detection thresholds at the IMS stations and were added by IDC analysts. The peak of the distribution below 3.4 is around 2.0, with 2049 arrivals having SNR <2.0, and 321 with SNR <1.0.

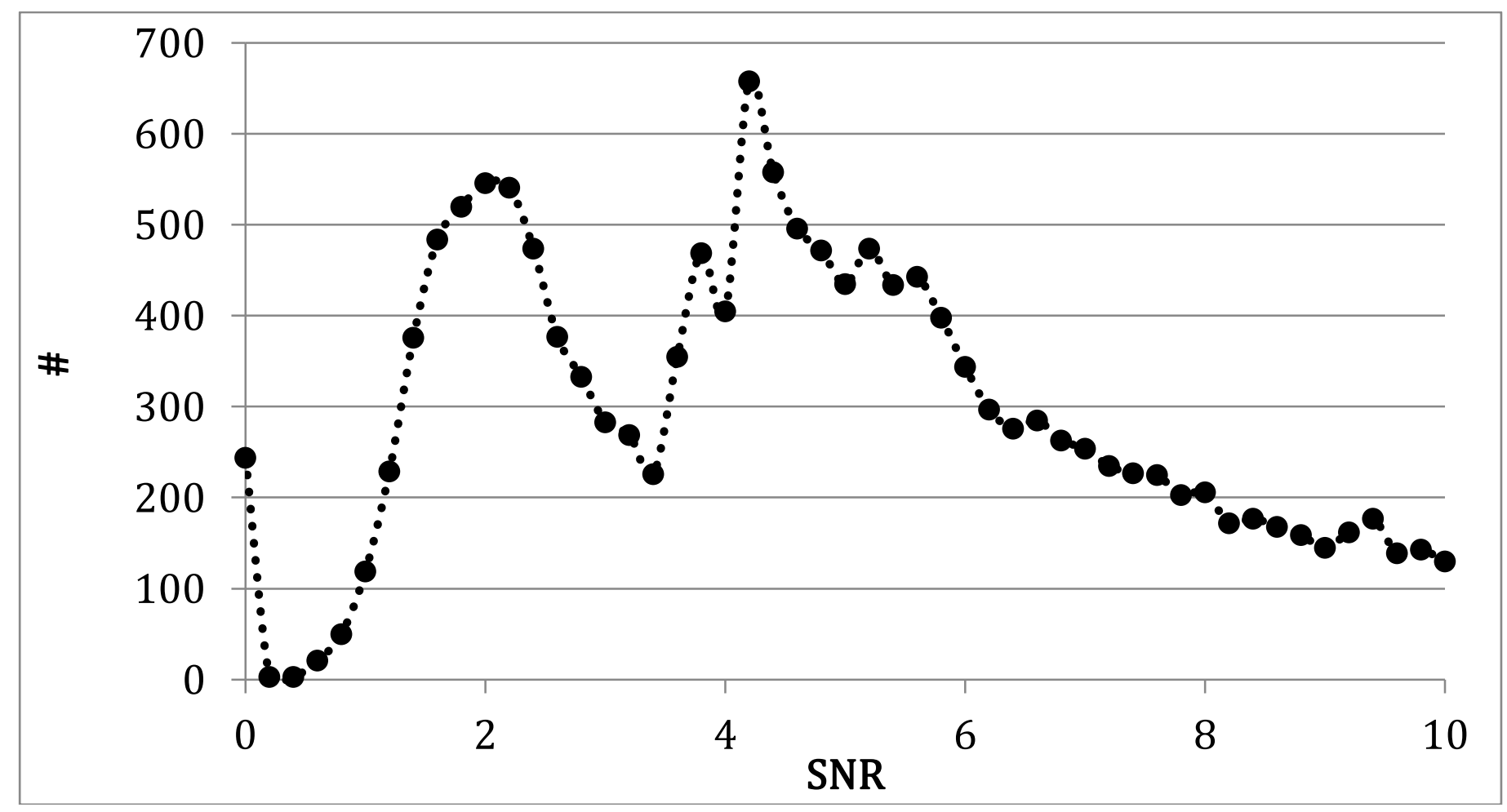

Figure 22. Frequency distribution of SNR values reported by the IDC for all P-phases associated with the REB events on July 30, 2025.

The quality of REB detections added by analysts is formally defined only by their experience and the task not to miss events of concern. The SNR estimates for the added signals, assuming they are correct, pose a challenge for any detector, including those based on the WCC. These detections were missed by the IDC automatic detector primarily due to their low quality. Furthermore, detections lacking SNR estimates should not be included in the database, as their quality is crucial for GA processing. The XSEL, when based on routine WCC processing, may encounter internal limitations in recovering events during periods of extremely high activity, which consequently affects the quality of the REB events.

***Statistical power***

The XSELs include numerous events with statistical significance controlled by the LA parameters for a given detection list. The question of their statistical power is related to the match of REB events in the XSELs with various LA versions and origin-time tolerance. The strict LA version with low tolerance has to be adjusted to create event hypotheses matching the REB and a small number of new XSEL events which could match the REB-potentially missed in the IDC processing and recovered by the WCC. Table 4 lists the match statistics for two LA versions and the origin-time tolerance of 2.0 s, as adopted by the REB. The first number in the "XSEL matched" column is for the weak version and the second for the strict version. The "REB only" and "SEL3" columns split the "REB total" into automatic and interactive production stages. "XSEL LA matched" refers to the match by XSELs for individual MEs before the CR is applied.

Table 4. Statistics of REB and SEL3 match in the XSEL for the period 2025201-2025210.

| **Day of** | **Q** | **REB** | **XSEL** | **REB** | **XSEL** | **SEL3** | **XSEL** | **XSEL LA** | **New XSEL** |
|---|---|---|---|---|---|---|---|---|---|
| **201** | Q2 | 109 | 104 / 81 | **27** | 22 / 8 | **82** | 82 /73 | 108/96 | 113 / 15 |
| | Q3 | 75 | 70 /63 | **14** | 12 / 8 | **61** | 59 / 55 | 74 / 69 | 120 / 17 |
| | Q4 | 51 | 50 / 44 | **8** | 8 / 3 | **43** | 42 /41 | 51 / 48 | 113 /27 |
| **202** | Q1 | 31 | 30 / 29 | **4** | 3 / 2 | **27** | 27 / 27 | 30 /30 | 120 / 24 |
| | Q2 | 45 | 38 / 37 | **10** | 6 / 5 | **35** | 32 / 32 | 41 / 41 | 129 / 24 |
| | Q3 | 44 | 42 / 41 | **10** | 9 / 5 | **34** | 33 / 33 | 44 /42 | 141 / 25 |
| | Q4 | 23 | 22 / 18 | **5** | 4 / 1 | **18** | 18 / 17 | 22 / 18 | 121 / 23 |
| **203** | Q1 | 55 | 52 / 46 | **20** | 17 / 13 | **35** | 35 / 33 | 55 / 50 | 102 / 16 |
| | Q2 | 31 | 30 / 30 | **4** | 3 / 4 | **27** | 27 / 26 | 30 / 30 | 116 / 23 |
| | Q3 | 20 | 20 / 18 | **8** | 8 / 6 | **12** | 12 / 12 | 20 / 18 | 118 / 24 |
| | Q4 | 11 | 10 / 10 | **2** | 1/ 1 | **9** | 9 / 9 | 11 / 11 | 112 /18 |
| **204** | Q1 | 4 | 4 / 4 | **0** | 0 | **4** | 4 / 4 | 4 / 4 | 101 / 24 |
| | Q2 | 5 | 5 / 5 | **1** | 1 / 1 | **4** | 4 / 4 | 5 / 5 | 123 / 33 |
| | Q3 | 11 | 11 / 11 | **0** | 0 | **11** | 11 / 11 | 11 /11 | 134 / 47 |
| | Q4 | 9 | 9 / 9 | **2** | 2 / 2 | **7** | 7 / 7 | 9 / 9 | 91 / 19 |
| **205** | Q1 | 7 | 7 / 7 | **1** | 1 / 1 | **6** | 6 / 6 | 7 / 7 | 99 / 24 |
| | Q2 | 10 | 10 / 10 | **3** | 3 / 3 | **7** | 7 / 7 | 10 / 10 | 105 / 17 |
| | Q3 | 6 | 6 / 5 | **3** | 3 /2 | **3** | 3 / 3 | 6 /6 | 133 / 37 |
| | Q4 | 22 | 22 / 22 | **5** | 5 / 5 | **17** | 17 / 17 | 22 /22 | 131 / 28 |
| **206** | Q1 | 16 | 16 / 16 | **4** | 4 / 4 | **12** | 12 / 12 | 16 / 16 | 164 / 45 |
| | Q2 | 4 | 4 / 4 | **1** | 1 / 1 | **3** | 3 / 3 | 4 / 4 | 137 / 36 |
| | Q3 | 2 | 2 / 2 | **0** | 0 | **2** | 2 / 2 | 2 / 2 | 129 / 33 |
| | Q4 | 6 | 6 / 6 | **0** | 0 | **6** | 6 / 6 | 6 / 6 | 122 / 23 |
| **207** | Q1 | 3 | 3 / 3 | **0** | 0 | **3** | 3 / 3 | 3 / 3 | 87 / 15 |
| | Q2 | 4 | 4 / 4 | **1** | 1 / 1 | **3** | 3 / 3 | 4 / 4 | 107 / 28 |
| | Q3 | 9 | 9 / 9 | **1** | 1 / 1 | **8** | 8 / 8 | 9 / 9 | 90 / 19 |
| | Q4 | 4 | 4 / 4 | **0** | 0 | **4** | 4 / 4 | 4 / 4 | 107 / 37 |
| **208** | Q1 | 0 | 0 | **0** | 0 | **0** | 0 | 0 | 110 / 24 |
| | Q2 | 4 | 4 / 4 | **1** | 1 / 1 | **3** | 3 / 3 | 4 /4 | 89 / 16 |
| | Q3 | 0 | 0 | **0** | 0 | **0** | 0 | 0 | 93 / 27 |
| | Q4 | 5 | 5 / 4 | **2** | 2 / 1 | **3** | 3 / 3 | 5 / 5 | 89 / 24 |
| **209** | Q1 | 3 | 2 / 3 | **2** | 1 / 2 | **1** | 1 /1 | 3 / 3 | 131 / 29 |
| | Q2 | 2 | 2 / 2 | **0** | 0 | **2** | 2 / 2 | 2 /2 | 84 / 16 |
| | Q3 | 1 | 1 / 1 | **0** | 0 | **1** | 1 / 1 | 1 /1 | 80 / 13 |

|  | Q4 | 1 | 1 / 1 | **0** | 0 | **1** | 1 / 1 | 1 / 1 | 135 / 38 |
|---|---|---|---|---|---|---|---|---|---|
| **210** | Q1 | 2 | 2 /2 | **0** | 0 | **2** | 2 / 2 | 2 / 2 | 127 / 33 |
|  |  |  |  |  |  |  |  |  |  |
| **Total** |  | 635 | 604 / 555 | 139 | 119 / 81 | 496 | 472 / 454 | 626 /593 | 4103 / 921 |

The period between the J20 and J29 earthquakes is likely the best time to assess the statistical power of the XSEL events. Two days following the J20 are characterized by rates ranging from 100 to 30 REB events per 6 hours. The second quarter-day (Q2) of 2025201 includes 109 REB events. The XSEL for the weak LA version matched 104 out of 109, and the strict version matched 81. This difference between the weak and strict versions is relatively large, but it is decreasing with time from the J20. Since the second half of 2023203, there is no difference in the matches between the weak and strict versions for SEL3 events or for the pure REB. The watershed is likely 30 events per quarter-day. The larger the number of REB events the lower the match rate, and the reason for this deviation is discussed in the previous section.

The final column in Table 4 lists the number of new XSEL events. The weak LA version produces 114±19 events per 6 hours, whereas the strict version produces 26±9 events per 6 hours for the studied period. This difference is expected, given to the design of the detection threshold, which results in a decreasing number of XSEL events as the pair order number or the corresponding magnitude threshold increases. The number of new events in the strict LA version is slightly larger than 50% to 100% of the number of REB events with the expectation of approximate equality, as related to the exercises with interactive review of XSEL bulletins. For the 0.25-second tolerance case, the number of XSEL events after the 2025203 ranges between 2 and 11. This pair of the LA version and tolerance case is designated as the transition to the REB. The statistical power of the XSELs used in this study is demonstrated by a perfect alignment with the REB, encompassing both actual and potential REB-ready events.

The importance of the results in Table 4 lies in the equal statistical significance of XSEL events that match the REB and those that do not. This implies that a complete match of the REB events during the period of relatively high activity between July 22 and July 29, 2025, can be extended to other XSEL events, thereby assigning them the same quality as the REB. In turn, the XSELs for the period between 201 and 210 of 2025 were obtained using the same procedure as before the J20 earthquake. The WCC processing introduced in [Kitov, 2026] and further developed in this study can be applied to IMS data for earthquakes across various regions. This approach guarantees both statistical significance and statistical power for the XSEL event hypotheses.

## Results

General aspects of WCC processing and its specific characteristics across different levels of seismic activity have been discussed. The seismicity within the studied region before and after the J20 and J29 earthquakes was also examined. The selection procedure and structure of the ME set were customized to fulfill the processing objectives. All calculations in this study were performed using identical WCC settings, ensuring direct compatibility throughout the entire processing period and across specific areas. The results include XSELs for individual MEs as well as the final XSELs obtained via the CR process. The full processing period spans from July 12 to July 31. Additionally, the best 100 MEs were subjected to an extended processing period from July 6 to August 3. Finally, two 3-day periods without REB events were processed using this 100 best ME set, providing a baseline reference for XSEL interpretation.

### *XSEL seismicity during low seismicity and periods without REB events*

The two 3-day periods without REB events allow for the estimation of recurrence curves exclusively for XSEL events. The first period, spanning November 29 to December 1, 2023 (jdate = 2023333-2023335), can be extended into the second quarter (Q2) of day 2023332. Additional calculations were performed for December 2 (5 REB events) and December 3 (1 event) to analyze the transition from a quiet period to the activity captured in the REB events. Similarly, the quiet period from February 24 (2024055) to February 26 (2024057), 2024, was also extended by two days exhibiting REB activity. Because this study focuses on rapid fluctuations in seismic activity below the IDC detection level, the data are processed in quarter-days intervals (Q1-Q4). Given that the activity levels can be exceptionally low, with zero XSEL events across several consecutive quarters, a four-quarter moving sum, denoted as MS(4), is calculated with a one-quarter step to ensure robust event statistics.

The results of the WCC processing using the best 100 MEs are displayed in Figure 23. The upper panel shows the recurrence curves, where the x-axis represents the order numbers. These XSEL recurrence curves demonstrate the feasibility of converting these order numbers into magnitudes; furthermore, deviations from the regression lines are potentially related to fluctuations in seismic activity near a specific magnitude level. The curves between 333.125 (Q1/333) and 335.875 (Q4/335) follow exponential trends, exhibiting deviations from these trends that increase in amplitude as the order number increases. The summary curve for the entire 15-quarter quiet period yields an $R^2$=0.98. On day 2023336, 5 REB events occurred, and two curves 336.875 and 337.125 - show a dramatic 3- to 4-fold increase in seismic activity throughout all the curves. This predominant growth is concentrated within the mid-range tolerance values of the strict LA version, peaking between 1.0 s and 2.0 s. A similar behavior is observed in the curves for some, but not all, quiet days. The most prominent deviations occur toward the end of the quiet period, while the summary curve shows a low-amplitude deviation for a 1-second tolerance.

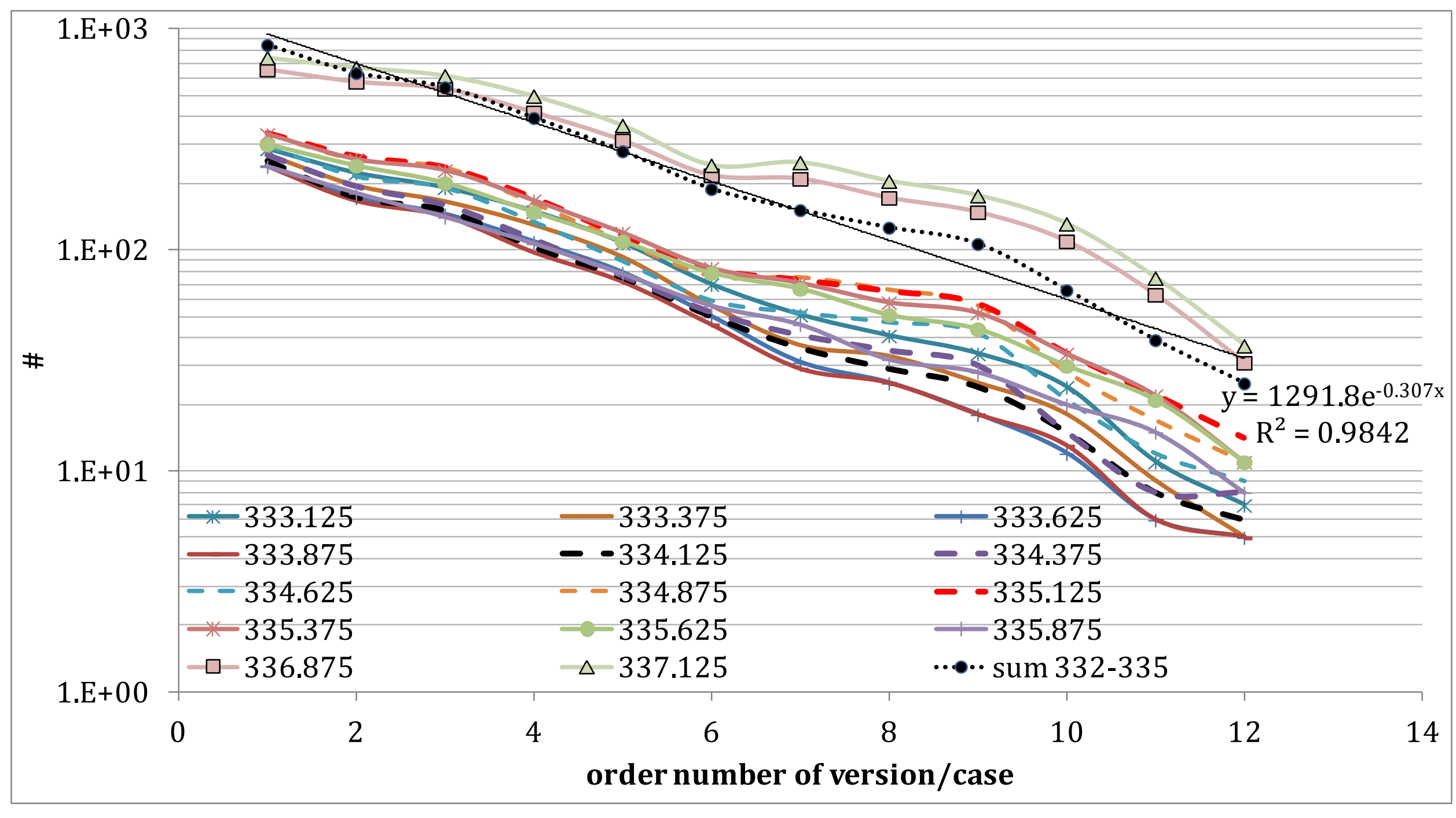

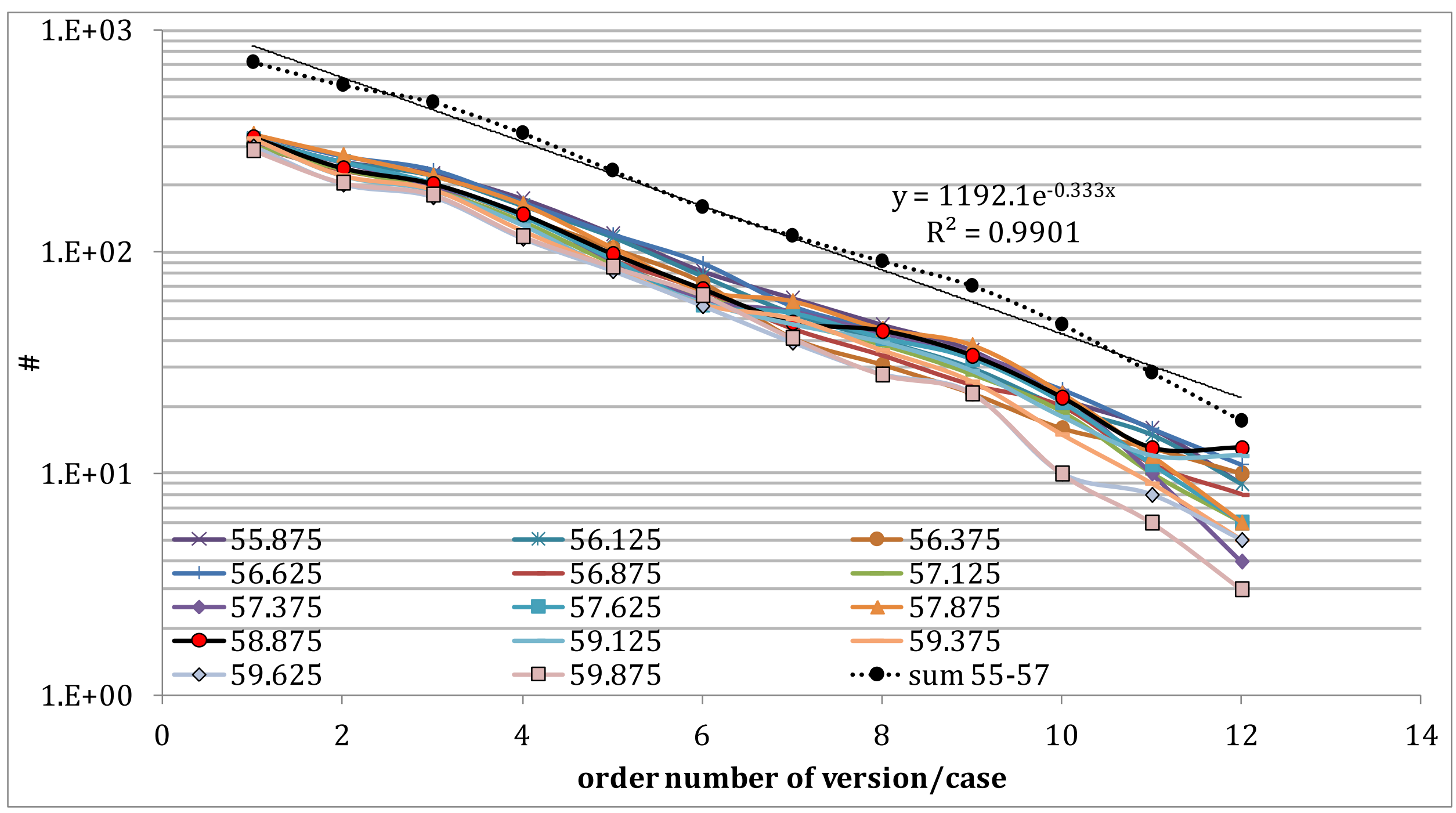

Figure 23. The recurrence curves of the moving sum of four consecutive 6-hours intervals for 12 XSELs from $v_1c_1$ to $v_2c_6$ with order number from 1 to 12. Upper panel: days between Q2/2023332 to Q4/2023337. Each curve is marked by its end quarter: curve 333.175 is the sum of 4 quarter-days between Q2/2023332 and Q1/2023333. Lower panel: the days between 2024055 and 2024059.

The recurrence curves illustrate the integral characteristics of the XSEL event distribution over a linear scale, such as a magnitude-like one. This representation is important as it proves the quality of the observational tool and the unambiguous conversion of the order number of the version/case into a magnitude scale. The scale is not standard in terms of the measurement procedure underlying the values. There are many magnitude scales, each with slightly different characteristics, taken from different parts of the signal wavetrain from the first P-wave to Love waves. An inverse representation of these scales is also possible similar to that described for Figure 18.

The time evolution of the seismic process for a set of magnitude thresholds is crucial for understanding the earthquake preparation process. The numbers of XSEL events for 12 thresholds are shown in Figure 24 for the quiet period 2023332-2023335 with an extension into the days 2023336 and 2023337 with REB events. The XSEL curves, which represent the MS(4) of the previous four quarter-days, look smooth and synchronized for each LA version over the entire period including the quiet and active seismicity periods.

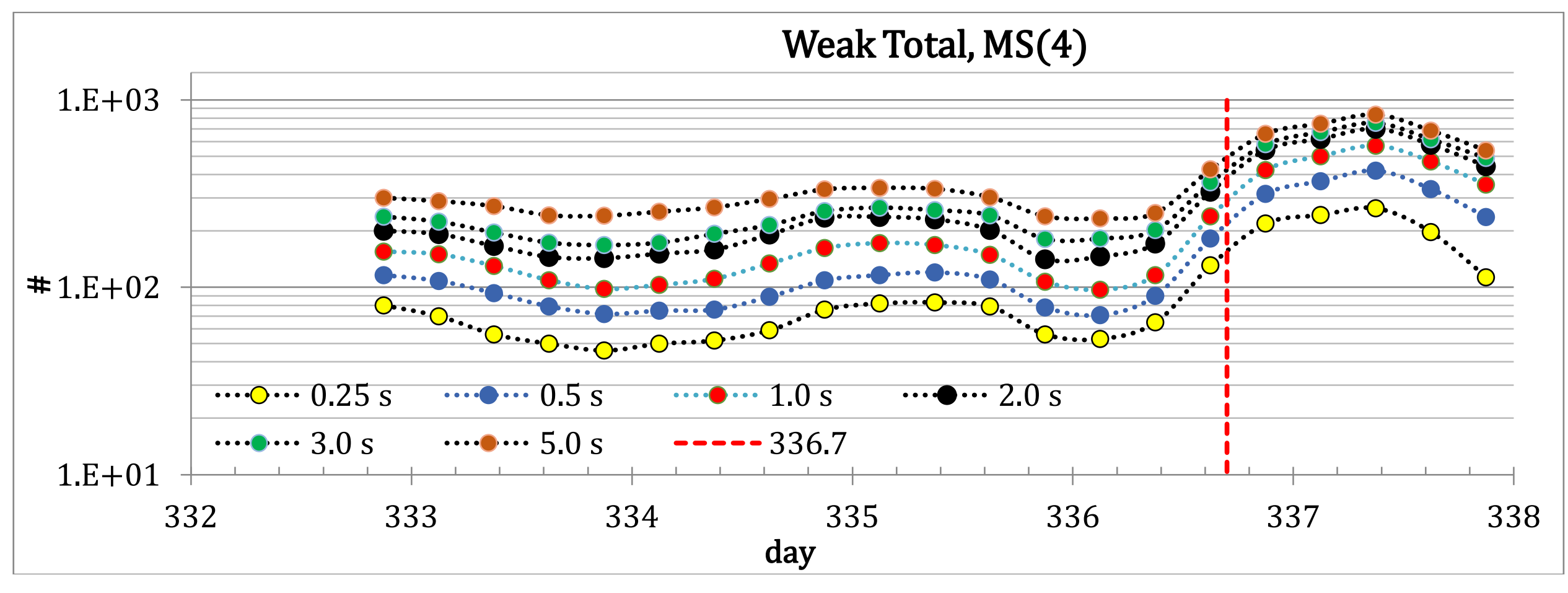

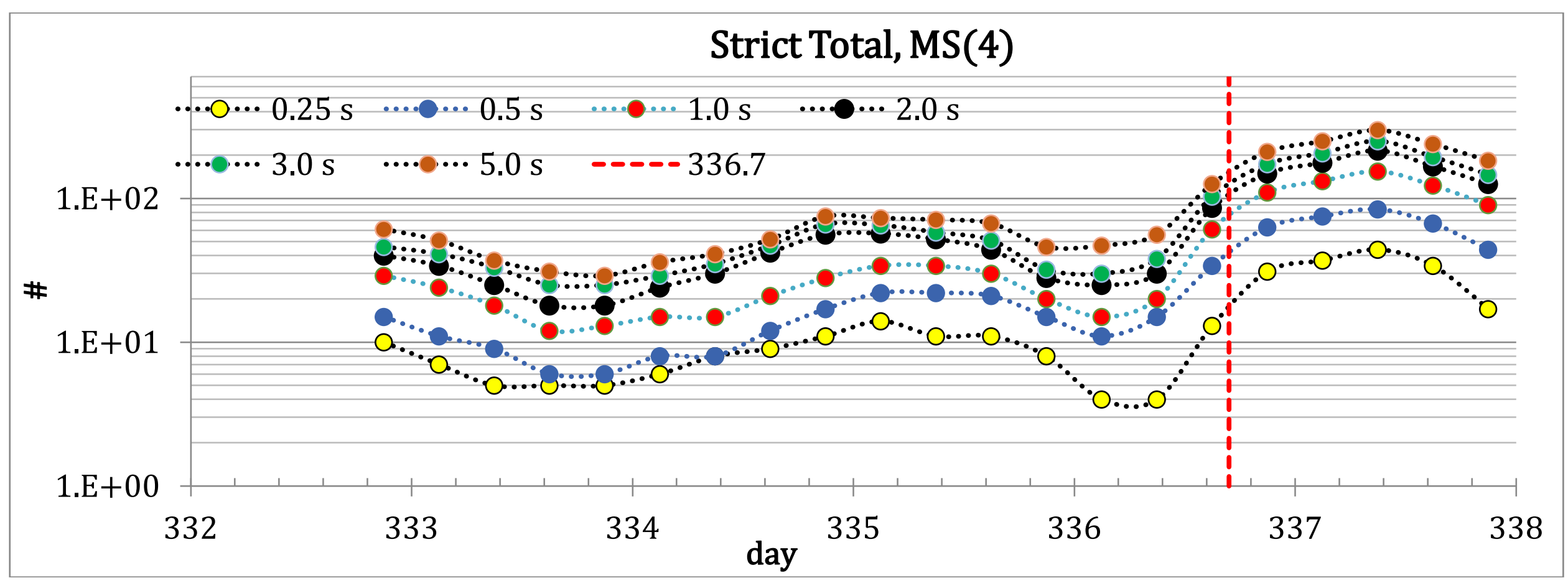

Figure 24. The evolution of the number of XSEL events in 6 weak cases (upper panel) and 6 strict cases (lower panel). Vertical dashed line – the origin time of the $m_b$=3.95 event occurred at 17:22:35 on December 2, 2023. Hypocenter is ~100 km due south of the J29 earthquake: 51.60ºN, 159.86ºE, d=0 km.

The origin time of the largest event on 2023336 with $m_b$=3.95 is indicated by a vertical red line. There is a noticeable increase in the XSEL numbers starting two quarter-days before this event. The only deviation is the drop in the 0.25 s curve for the strict LA version, which began approximately 20 hours before the earthquake. The ratios of the weak and strict LA versions for all six cases are displayed in Figure 25. These curves highlight the variances in seismic activity above the respective thresholds. The 0.25 s ratio curve begins to rise three quarter-days before the earthquake, peaks roughly 8 hours prior to the earthquake, and then declines as it approaches the red line. A similar pattern was observed in the 0.25 s curve for the May 24, 2013, Sea of Okhotsk earthquake, while the other five curves exhibit different behaviors. The resemblance in the peaks in the 0.25 s curve is likely coincidental, but similar patterns before other major earthquakes are the objects of this study. For the 2023336 earthquake, the increase in the 0.25 s curve may be observable due to the absence of larger-magnitude seismic activity during the previous 96 hours.

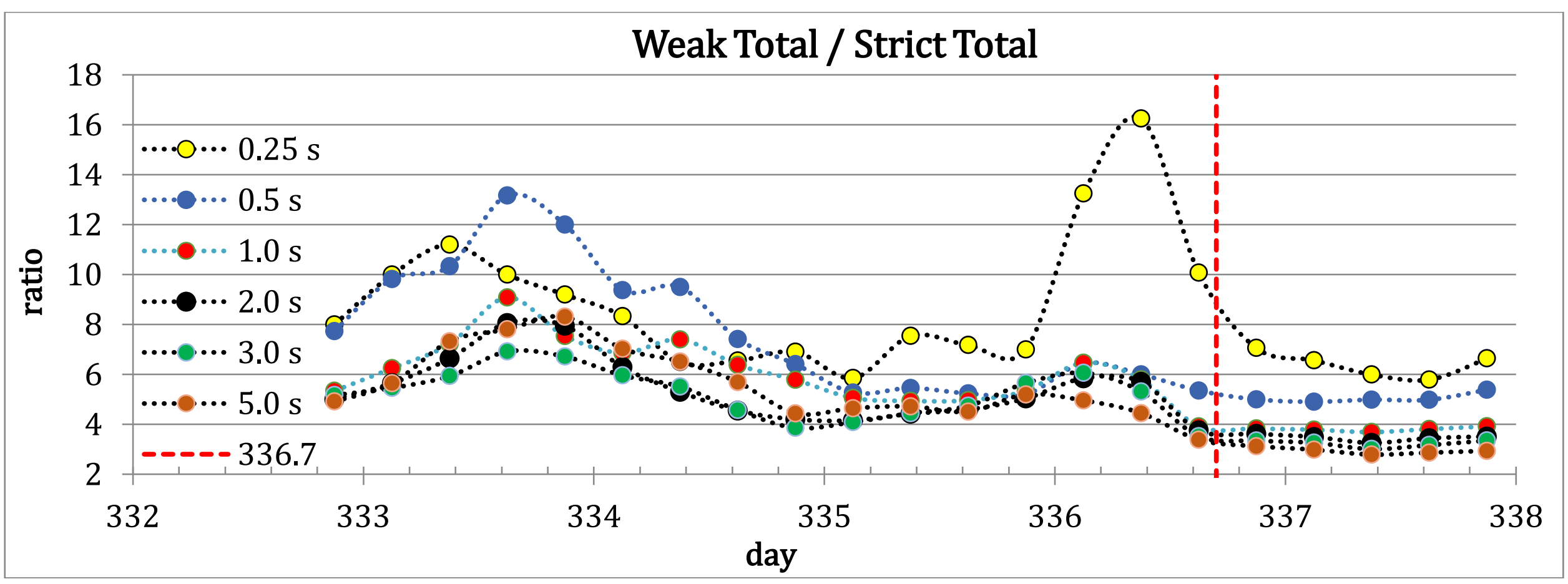

Figure 25. The ratios of the XSEL for the weak and strict LA versions and the same origin time tolerance case. The curve for the 0.25 s tolerance has a peak approximately 10 hours before and then falls just before the earthquake at 17:22:25. This peak corresponds to the fall in the XSEL for strict LA version in Figure 23.

The 3-day quiet period in 2024 is shown in Figure 26. It displays a pattern similar to that observed in 2023, with a peak in the 0.25 s ratio curve. Figure 27 depicts this curve starting to rise 5 quarter-days before peaking a day before the event on February 28 at 11:11:31. The epicenter coordinates are 53.93°N, 160.34°E, with a depth of 75.4 km, and $m_b$=3.85. This peak is observed long before the event origin time, and the event itself is not large. Therefore, there may be no causal link between the peak and the earthquake. Nevertheless, the peak and the events are measured independently and are where they are.

There were no significant changes observed in the XSEL numbers throughout the entire period between February 24 and February 29. However, the strict LA version demonstrated larger variations. The day of February 29 was added to the initial interval due to a sudden rise in the 0.25 s curve right up to the end of February 28, as observed in Figure 27. This rise is related to the decreasing 0.25 s curve for the strict LA version in the lower panel. The only REB event on February 29 occurred at 19:37:35 at a depth of 499 km, far to the west of the J29 epicenter (51.45°N, 151.28°E). The magnitude was 3.47, large enough for the event to be detected by almost all IMS stations worldwide.

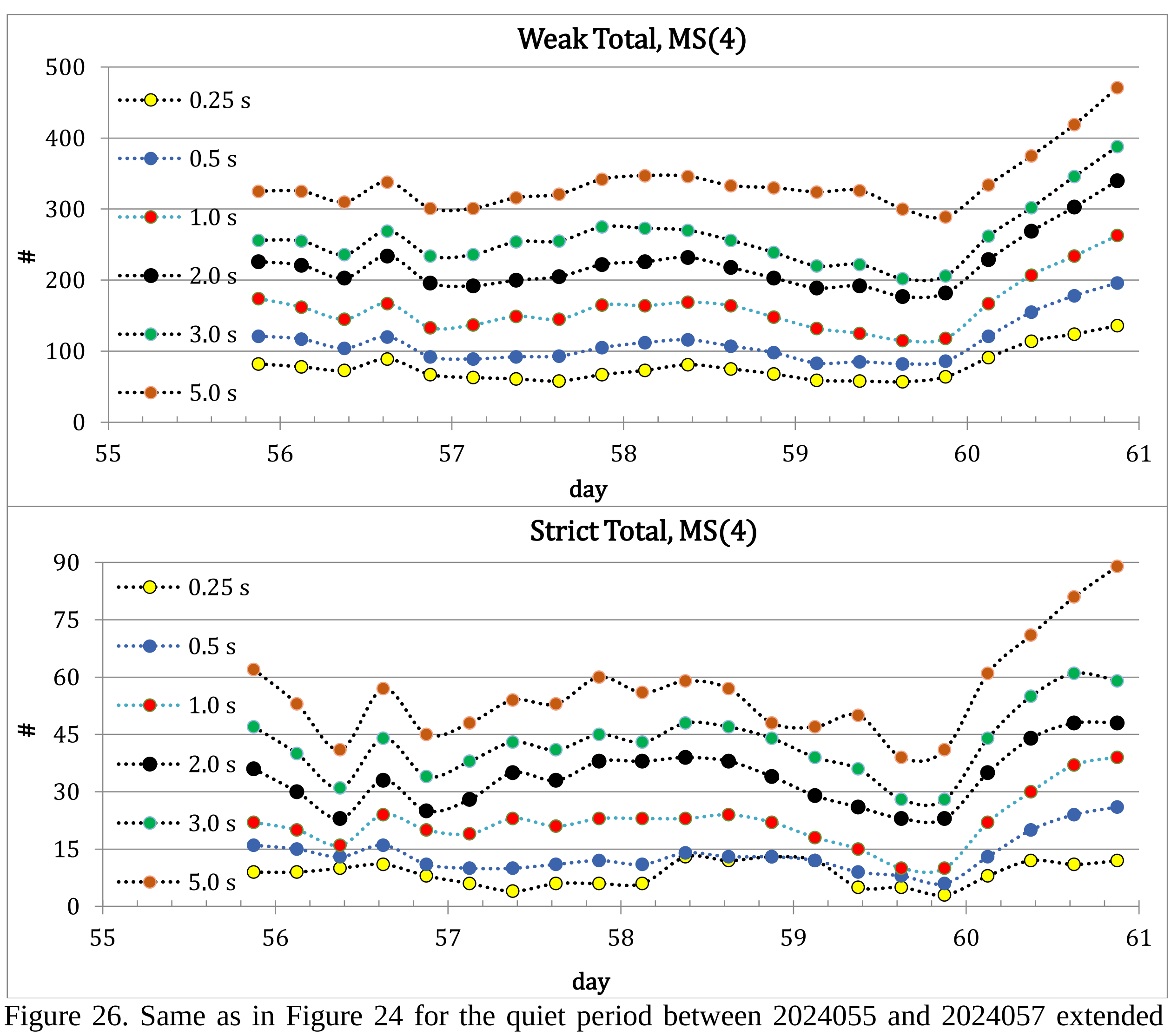

Figure 26. Same as in Figure 24 for the quiet period between 2024055 and 2024057 extended into 2024058-061.

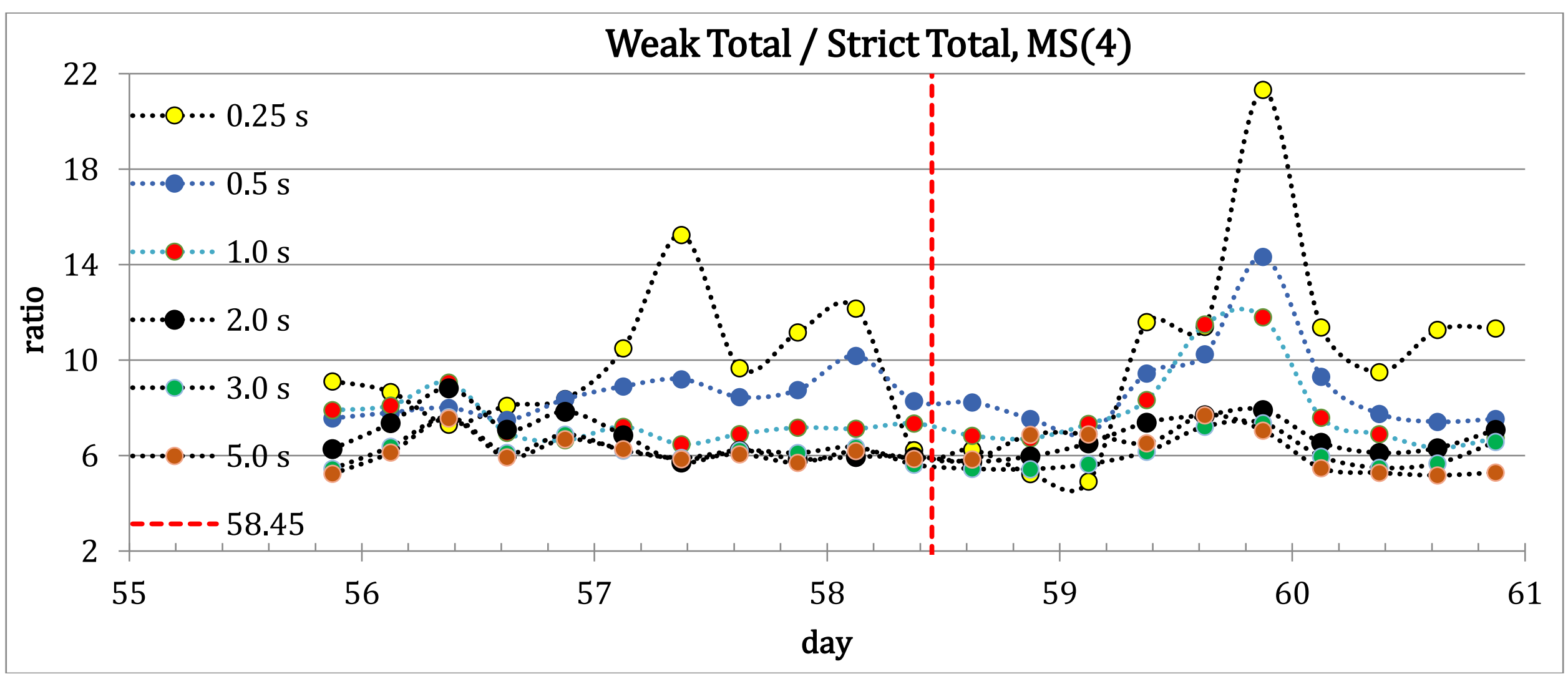

Figure 27. Same as in Figure 25 for the quiet period between 2024055 and 2024057 extended into 2024058-059.

***Seismic activity prior to the J20 earthquake***

During the quiet periods, the seismic activity detected by the WCC is nearly identical to that observed during the periods of low activity periods preceding the J20 earthquake. Figure 28 displays the trajectories of the running four-quarter moving sum, MS(4), of the quarter-day-long XSELs between July 6 and July 20, 2025. This period was fully processed using the best 100 MEs and presents a baseline reference for relatively low seismicity prior to the J20 M7.4 earthquake (see Figure 19), which featured only one large REB event on July 14 (origin time: 21:38:01, epicenter: 46.66°N, 151.30°E, depth: 87 km, Mw~5.7-5.8). Results for both the weak (upper panel) and strict (lower panel) LA versions, each including six origin-time tolerance cases from 5.0 s to 0.25 s, are shown.

Several intervals with significant variations in the MS(4) are marked by ovals. Between July 10 and 12, there was deep through, with all curves for both the weak and strict version changing almost synchronously. The strict/0.25 s curve dropped down to 1 XSEL per day during three consecutive quarter-days. There were four quarter-days without XSEL events in a row and only one XSEL event during a day and a half. There were also no REB events on 2025189. Given that the final REB event on 2025188 occurred at approximately 16:00 and the first event on 2025191 appeared at 06:00, the total event-free period lasted roughly 38 hours. The strict LA version with a 0.25 s origin time tolerance exhibits approximately the same sensitivity as the REB.

This serves as an interesting example of a rapid 2- to 3-fold decline in low-magnitude seismicity within a very large geographical region, especially when compared to the subsequent 38-hour REB-event-free period between the morning of 2025197 and the late evening of 2025198. During this latter period, all curves for the weak LA version rise and then fall back to their initial levels. In this case, the strict/0.25 s curve also demonstrates growth, albeit emerging from a trough at the end of 2025196.

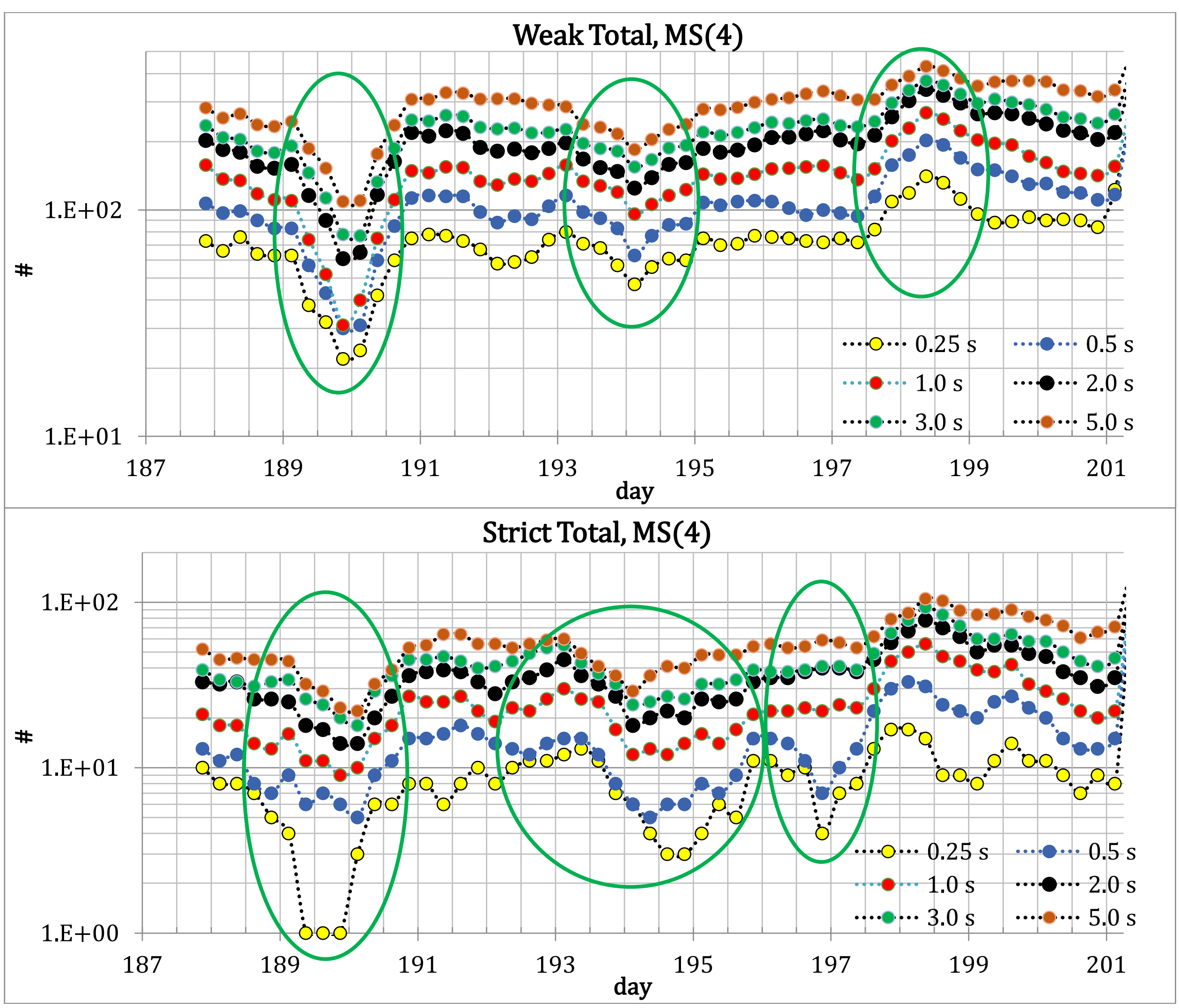

Figure 28. The evolution of the running MS(4) of the quarter-day number of XSEL events between 2025187 and 2025201. The weak (upper panel) and strict (lower panel) with six origin time tolerance cases (from 5.0 s to 0.25 s) each are shown. Several intervals of with significant variations are marked by ovals.

Figure 29 compares two LA versions and two cases, highlighting the periods of synchronous and asynchronous behaviors across various version/case permutations. When seismicity fluctuates simultaneously at all magnitude levels, the corresponding ratios evolve in sync. Conversely, the asynchronous onset of growth and decline may be attributed to waves of low-magnitude seismicity passing through the magnitude thresholds that correspond to the respective version/case order numbers. The weak LA curves are nearly synchronous, with the exception of the final quarter-day (201.125, representing the first six hours of July 20). Specifically, the weak/0.25 s curve, which remains systematically below the weak/0.5 s curve, exhibited a sharp increase to align with the 0.5 s curve level. The strict curves did not mimic this behavior; consequently, the weak-to-strict ratio for the 0.25 s tolerance increased, as illustrated in Figure 30. This is an example of low-seismicity progress through the magnitude range confined between the weak and strict versions.

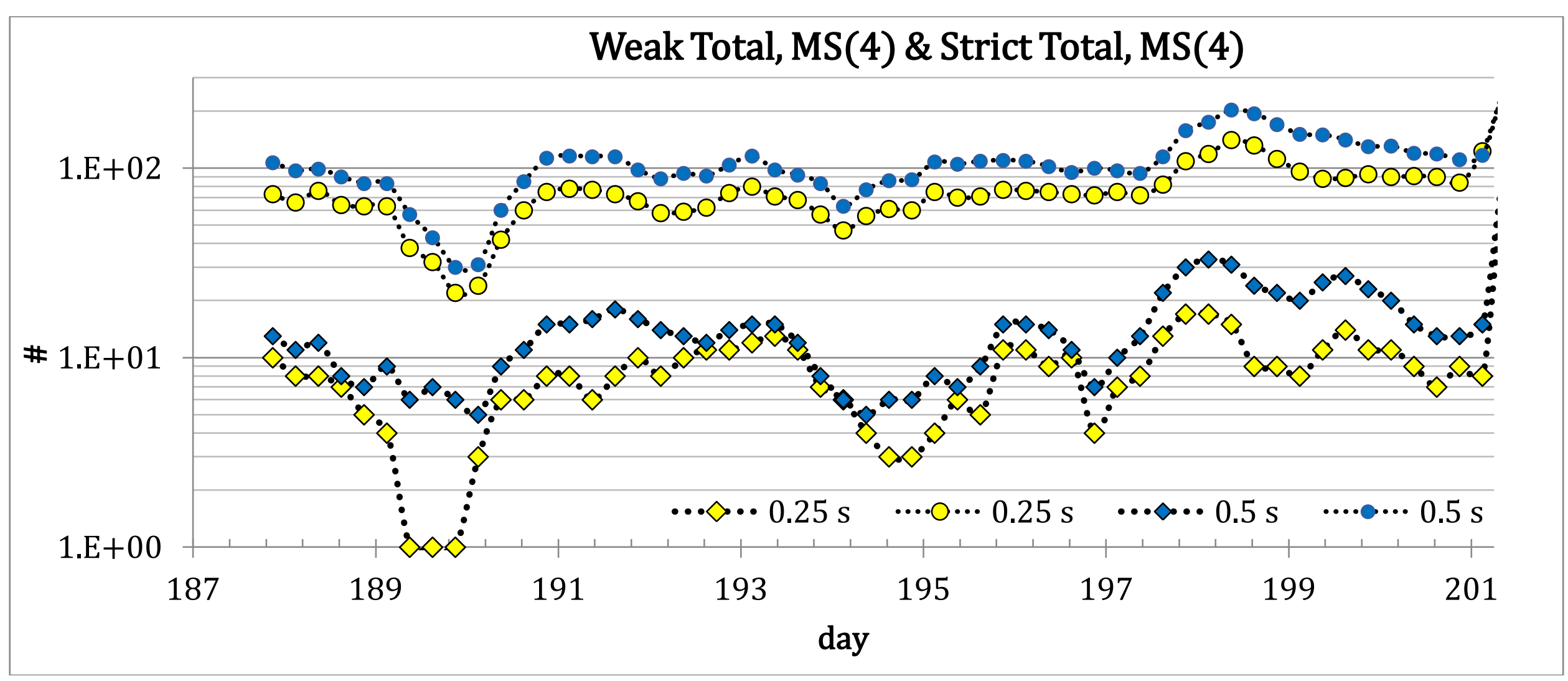

Figure 29. Comparison of the 0.25 s and 0.5 s cases illustrating the synchronous in asynchronous behavior of the two LA versions and two cases.

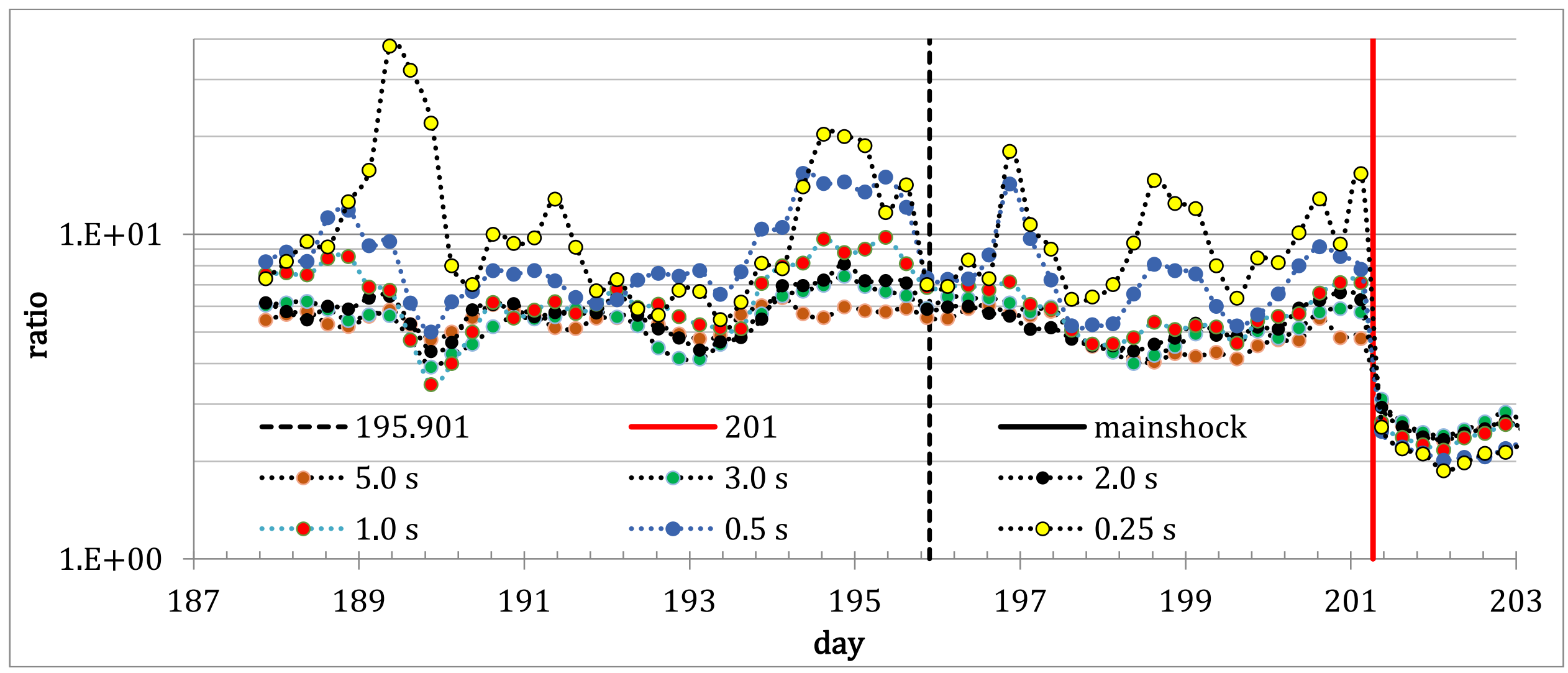

Figure 30. The ratio of weak-to-strict LA versions for six origin tolerance cases between 2025187 and 2025203. The event at 21:38:01 on July 14 is marked by a black vertical dashed line.

Periods of significant decline in the strict/0.25 s curve compared to the weak curve were observed several times during the period between 2025187 and 2025201. The weak-to-strict ratio rose to high values, albeit against the background of decaying seismic activity. A prominent spike reaching a value of 38 occurred in the 0.25 s ratio curve in the second quarter-day of 2025189, when the XSEL count for the strict version fell to 1. A broader peak was observed on 2025195, when the strict curve dropped to a level of 3. Similarly, a sharp peak in the fourth quarter-day of 2025196 was induced by an abrupt decline in the strict LA version. This specific point marked the onset of growth in the strict version at a faster rate than in the weak version. This corresponding peak likely indicated the transition to the event on July 20.

The two weeks preceding the J20 earthquake demonstrate a variety of peaks and troughs in the MS(4) of quarter-day XSELs and in the ratio of the weak and strict MS(4) XSELs. These peaks are most prominent in the curves for the 0.25 s origin-time tolerance, which defines the magnitude range closest to the corner magnitude of the REB recurrence curve and generates the most statistically significant XSEL event hypotheses. The XSEL ratio for this specific case may reveal the preparation for an impending earthquake, provided it is induced by an increasing number of events in the weak LA version, as was previously observed for the Sea of Okhotsk earthquake.

An *ex post* analysis allows for the differentiation of two distinct classes of peaks in the ratio curves. The first class is associated with a rapid drop in the number of events in the strict XSELs, accompanied by a much slower decline or no decrease at all in the weak XSEL version. The resulting peaks are sharp and exhibit large amplitudes, which are likely associated with periods of extremely low seismic activity. Statistically, recording only one to three events per day within a highly seismic region of a size of approximately 2000 km × ~1400 km is inherently characterized by statistical uncertainty. In contrast, the peaks in the weak-to-strict ratio curves observed prior to the J20 earthquake are supported by more robust statistics within the strict LA version.

The 0.5 s ratio curve also demonstrates peaks in nearly the same time intervals; however, these peaks are systematically lower, whereas the underlying MS(4) XSEL curves remain higher than the 0.25 s curves. The 0.5 s ratio curve exhibits a peak along the growth trend two quarter-days prior to the J20 earthquake and then decreases slightly. This behavior differs from that observed before the Sea of Okhotsk earthquake, where the 0.5 s curve began to rise a few quarter-days before the 0.25 s curve, after which both curves peaked synchronously right before the mainshock. This sequence was interpreted as a wave of events with an increasing magnitude progressing through a set of magnitude thresholds defined by the increasing order number of the LA version/origin-time tolerance case.

The four other curves also exhibit minor shifts toward the peaks of the 0.25 s curve. There is an episode of countermotion near 2025189, where the other five curves form a trough while the 0.25 s curve approaches its peak value. The significance of XSEL bulletins with the largest origin time tolerances depends on the sensitivity of the observational networks and the magnitude of the earthquake under study. For the Sea of Okhotsk earthquake, all six XSEL ratio curves contributed specifically to estimating the level and tracking the progression of the seismic activity wave from the lowest magnitudes to the mainshock. In contrast, for the study of the Kamchatka earthquakes, they likely serve only as a baseline reference, assisting in the reconstruction of the recurrence curve, such as those presented in Figure 18.

***Seismic activity between the J20 and J29 earthquakes***

The period between the J20 and J29 earthquakes is of special interest because it involves seismic activity that is significantly higher than that observed prior to the July 20 earthquake. Although the J20 aftershock sequence decays rapidly toward the J29, as illustrated in Figure 19, it could pose a potential challenge for the WCC processing to distinguish between the REB matching and new XSEL events before the J29 mainshock. While the REB matching statistics in Table 4 illustrate the efficacy of the WCC in detecting REB events, it remains a question whether the WCC can differentiate between the mixture of low-magnitude J20 aftershocks and J29 foreshocks, or if this mixture represents a single, continuous process of earthquake preparation. Up to this point, the XSELs included all events, encompassing both the REB-matching and newly detected ones. The latter subset of XSEL events does not differ significantly from the total version for quiet days, and could therefore be utilized instead of the total number for earthquake prediction purposes.

The period between the J20 and J29 events, however, is not characterized by seismic quietness. Figure 31 compares the total number of events with the number of new XSEL events for both the weak and strict LA versions under the 0.25 s and 5.0 s origin-time tolerance cases. The difference between these curves represents the number of REB-matching XSEL events. It rises across both versions and for both cases. Within the strict version, the gap between the 0.25 s and 5.0 s cases is small on the first day after the J20 event and practically disappears after the J29 mainshock. For the weak version, the 5.0 s total and new curves virtually coincide after 2025207, whereas the 0.25 s curves converge only two days prior to J29. In contrast, the 0.25 s strict total and new curves retain the gap until the final minute prior to the J29. Overall, because the total XSEL counts inherently include J20 aftershocks up to the onset of the J29, these aftershocks and the J29 foreshocks cannot be distinguished. Nevertheless, the J20 earthquake failed to follow the J29 evolution steps, suggesting that the overall preparation for the mega-earthquake may not be disrupted by a relatively small loss of the total elastic energy accumulated in the entire region in J20 mainshock and its aftershocks.

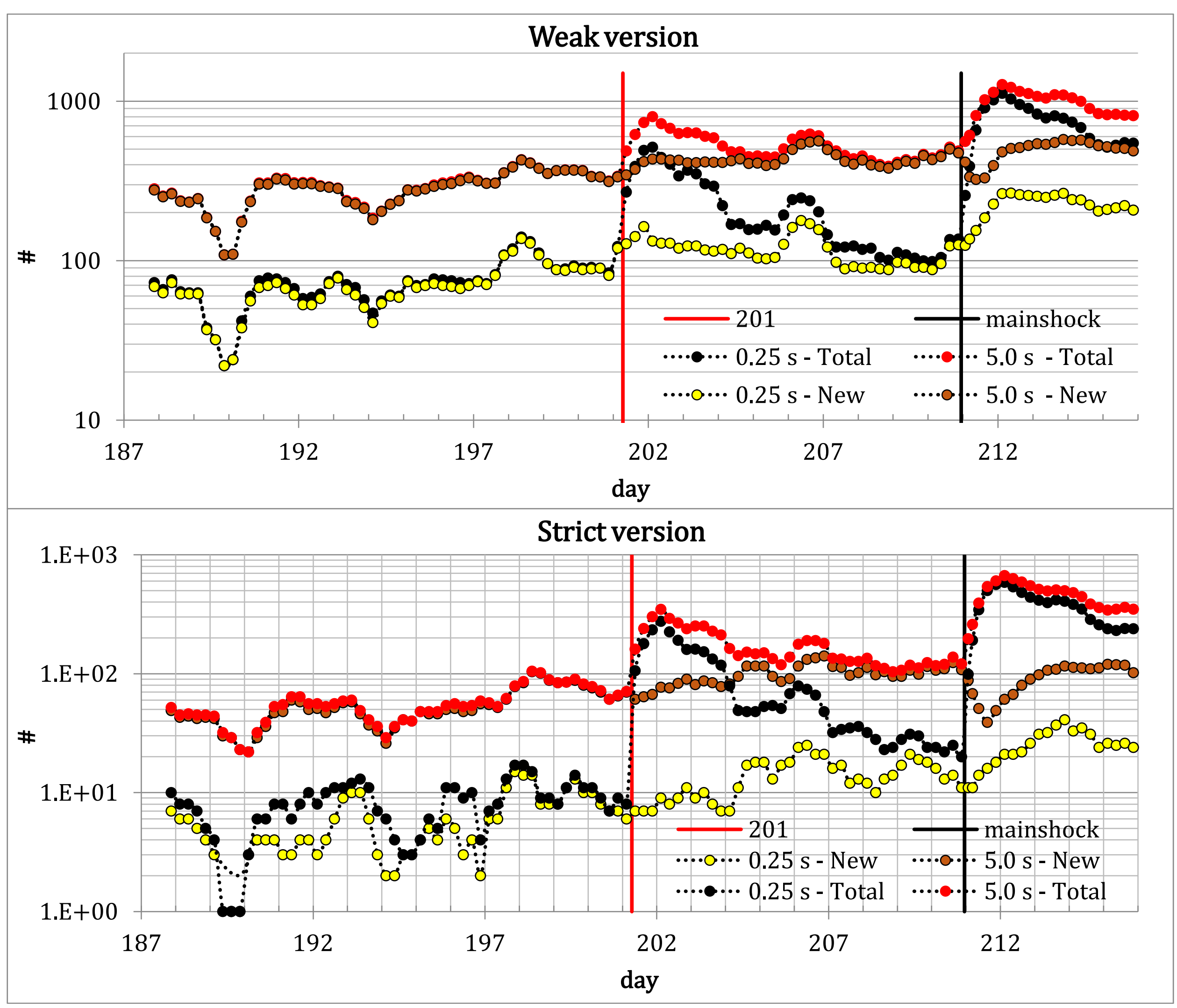

Figure 31. Evolution of weak (upper panel) and strict (lower panel) MS(4) numbers in the total XSEL vs. only the new events XSEL. Two origin time tolerances are used: 0.25 s and 5.0 s.

The evolution of the MS(4) for the total XSEL curves between 2025200 and 2025211 is presented in Figure 32. Six curves representing all cases within the same versions move nearly synchronously, displaying only a few minor deviations. Furthermore, curve pairs sharing the same origin-time tolerance case appear highly similar across both LA versions. The principal divergence among the curves in Figure 32 is observed during the day preceding the J29 mainshock. The 0.25 s curve for the weak version demonstrates a clear upward trend, whereas the strict version exhibits a downward trend common to all six cases, punctuated by a minor peak two quarter-days prior to the J29 rupture.

The weak-to-strict ratios for the six origin time tolerance cases in Figure 33 illustrate a synchronous evolution of the MS(4) metric, characterized by an overall positive trend from 2025200 to 2025211. The most pronounced variations in these ratios are confined to the 0.25 s and 0.5 s cases. This pattern is common to all events analyzed thus far: the Sea of Okhotsk, the J20, and the J29 events, as well as the two quiet periods. Following the J29 earthquake, all six ratio curves drop sharply to a baseline level of 2, a feature that is directly linked to the abrupt drop in the MS(4) levels illustrated in Figure 20. Because of the smoothing effect inherent to the running sum window, the first three quarter-days following the mainshock still incorporate MS(4) values from the pre-seismic interval, thereby smoothing the transition to the lower post-seismic baseline. A ratio value of 2 corresponds to a regime where the number of newly detected XSEL events is negligibly small compared to those matching the REB bulletin. In other words, the magnitude threshold for the WCC processing is shifted very close to the corner magnitude of the total recurrence curves in Figure 15 or 17, depending on the respective depth estimates in the REB and XSEL.

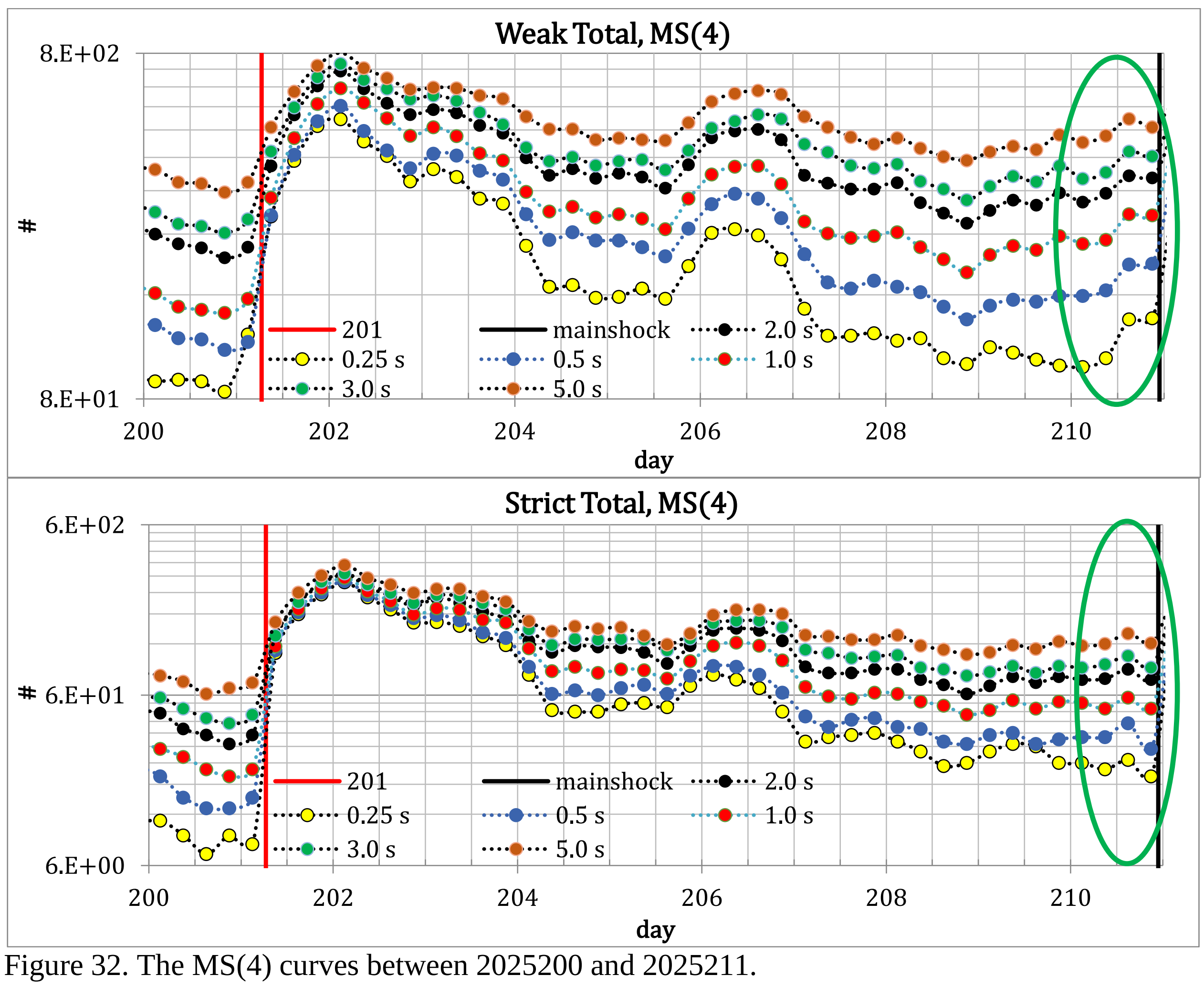

Figure 32. The MS(4) curves between 2025200 and 2025211.

The weak-to-strict ratios for the 0.25 s and 0.5 s tolerance cases in Figure 33 rise from a level of 4.0-4.5 up to 7.0 during the last two to three quarter-days prior to the J29. This peak is observed within the last 5 hours preceding the mainshock, which occurred at 23:24:48 (IDC origin time). The last hour of 2025210 was excluded from the pre-event WCC processing. This last hour was also omitted from the post-seismic processing because both the REB and XSEL are severely biased by extraordinary noise levels. There is a study devoted to the recovery of aftershocks in the initial minutes following the J29 [Kitov *et al.*, 2026]. The exclusion of this hour should not introduce any measurable bias into the weak-to-strict version ratio, as it affects both versions proportionally and exhibits no discernible influence on the three cases ranging from 2.0 s to 5.0 s. The 1.0 s case demonstrates an increase, peaking near 5.5.

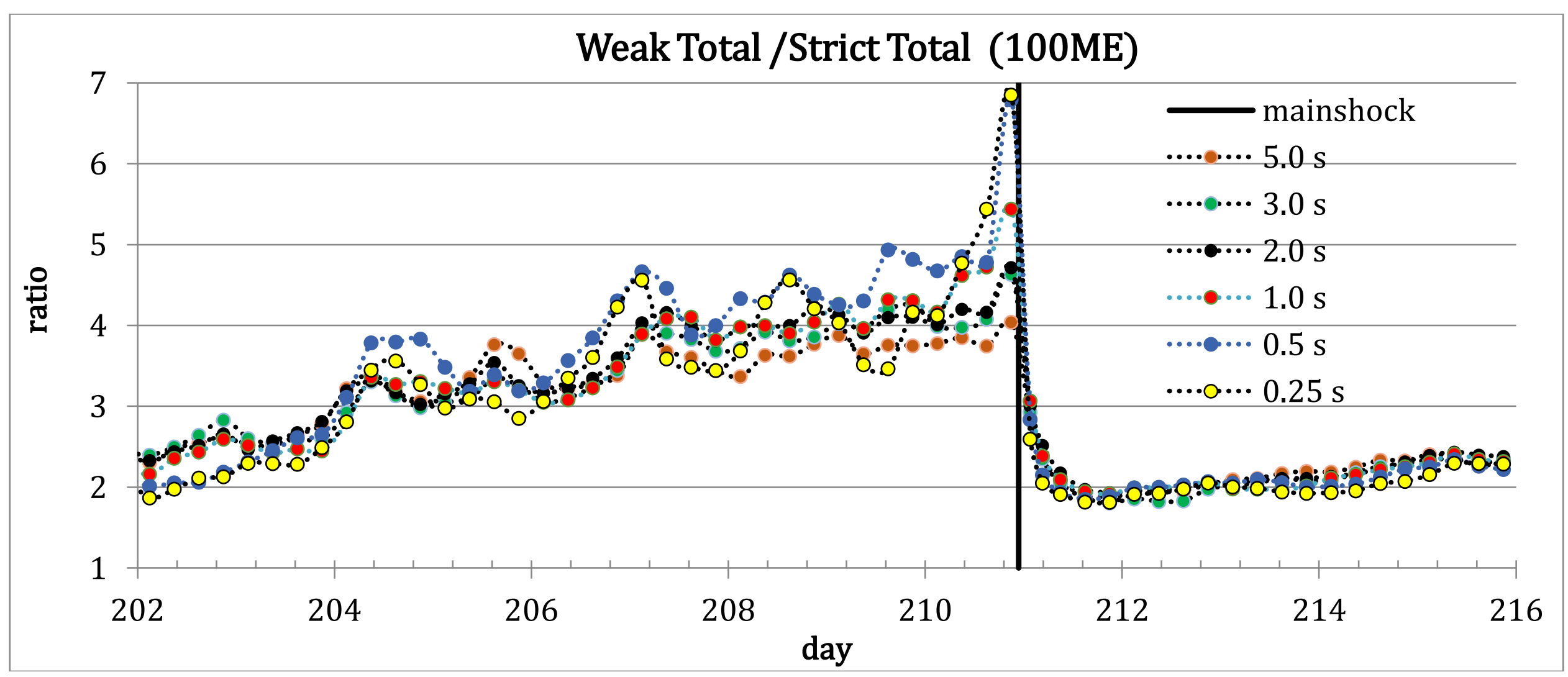

Figure 33. The ratios of the number of XSEL events in the weak and strict LA versions for the period of time between 2025200 and 2025215.

The peaks in the 0.25 s and 0.5 s curves immediately preceding the J29 mainshock, with the depth fixed at the free surface, represent a potential short-term earthquake prediction trigger. A similar precursor was previously observed prior to the Sea of Okhotsk M8.3 earthquake. The consistent behavior of low-magnitude seismicity in both cases provides compelling evidence in favor of such a hypothesis.

***Recurrence curves for the ME sets***

The best 100 MEs constituted the initial configuration covering the entire studied region. Thus far, this specific set has been utilized in the calculation of the XSEL bulletins, as well as the corresponding time series of the MS(4) metrics and the weak-to-strict ratios. There is another ME set consisting of 195 REB events, each featuring a mandatory associated P-wave arrival at station PDYAR. This latter set also covers the same geographical area and depth range but was selected from the REB after 2023, following the upgrade of this station to an operational array at the end of 2022. A cumulative total of 295 MEs were utilized to create the XSEL events; both constituent sets are illustrated in the left panel of Figure 13.

The expanded set allows for the investigation of several critical problems related to the precursory peaks observed in the 0.25 s and 0.5 s weak-to-strict ratio curves in Figures 30 and 33. First, it is necessary to determine which of the two configurations exhibits greater efficiency in generating XSEL events, given their subtle variations in spatial distribution and sample size. The top 100 MEs are likely characterized by superior sensitivity and resolution, a feature supported by their selection procedure, which relies on a comprehensive pairwise WCC analysis with neighboring REB events. The ME set anchored by station PDYAR is larger and directly benefits from the high sensitivity of this array to seismic events within Flinn–Engdahl region 19 (Kuril–Kamchatka).

The subsequent question involves resolving the geographical provenance of the XSEL events that drive the precursory trigger shown in Figure 33. The earthquake preparation processes may structurally encompasse the entire regional volume bounded by coordinates 45°N - 65°N, 145°N - 165°N and a depth range from 0 km to 700 km. Alternatively, the preparation zone can be restricted to a small spot, such as the J20 aftershock sequence in Figure 12, concentrated in the lithosphere, or a broader spatial envelope corresponding to the J29 aftershock distribution. To test these hypotheses, the total 295 ME set is partitioned into four distinct geographic domains in the right panel of Figure 13:

- The entire region including the subducting Pacific plate.

- The localized zone surrounding the potential asperity that arrested the slip propagation of the J20 event along the eventual J29 rupture line (hereafter referred to as the "Asperity" zone, which excludes a single deep-focus event within its boundary and incorporates 49 MEs).
- The remaining spatial domain outside this core zone, designated as "Out of Asperity".
- The southwestern periphery situated beyond the immediate J29 aftershock zone, designated as the "Rim" region, which utilizes 79 MEs from the “PDYAR” configuration.

This spatial subdivision is baseline-exploratory, aimed at identifying potential regional accents in the precursory signal. If necessary, this tentative division can be optimized during a subsequent technical calibration procedure. One such optimization parameter involves the recurrence curve, which characterizes the performance of the WCC pipeline in terms of how accurately the order numbers of specific version-case pairs correspond to a uniform sequence of magnitude thresholds. The upper panel of Figure 34 illustrates this parameter computed for the period between 2025193 and 2025201. The curve for the best 100 MEs is identical to that presented in Figure 16, whereas the «JOINT» curve represents the complete set of 295 MEs. The «JOINT» bulletin is not a simple cumulative sum of the best 100 and 195 “PDYAR” XSELs; rather, it is the direct product of joint processing incorporating a Conflict Resolution (CR) algorithm those cross-rejects redundant XSEL events between the two constituent bulletins. For instance, the «JOINT» XSEL yields 3752 events in the $v_1c_1$ pair (weak version, 5.0 s tolerance), whereas the standalone best 100 MEs generate 2625 events and the “PDYAR” configuration produces 2642 events. For the $v_2c_6$ pair (strict, 0.25 s), the event counts are 140, 74, and 92 events, respectively. Thus, the value added to the «JOINT» XSEL bulletin scales from 50% for the $v_1c_1$ pair to nearly 100% for the $v_2c_6$ pair.

The respective curves in the upper panel of Figure 34 evolve synchronously, displaying similar low-amplitude deviations within the weak LA version. However, the strict LA version for the “PDYAR” subset generates larger numbers of XSEL events across all six tolerance cases. This behavior is likely attributed to the larger number of MEs physically encompassing a broader spatial domain during the high-resolution grid search, which operates with a maximum radius of 43.2 km in the strict LA version. For the weak version, the grid radius is 81 km, and the neighboring MEs have overlapping search areas leading to the CR process. This phenomenon is important for the future ME setting – the number of MEs should be sufficient to cover the whole region without detection-creation blind zones. The “PDYAR” master subset was partially introduced to address this issue, and the «JOINT» set is likely an important step on the way to an optimal ME number and distribution. The increasing number of MEs can yield additional XSEL events, but it would require higher computing power or longer processing time.

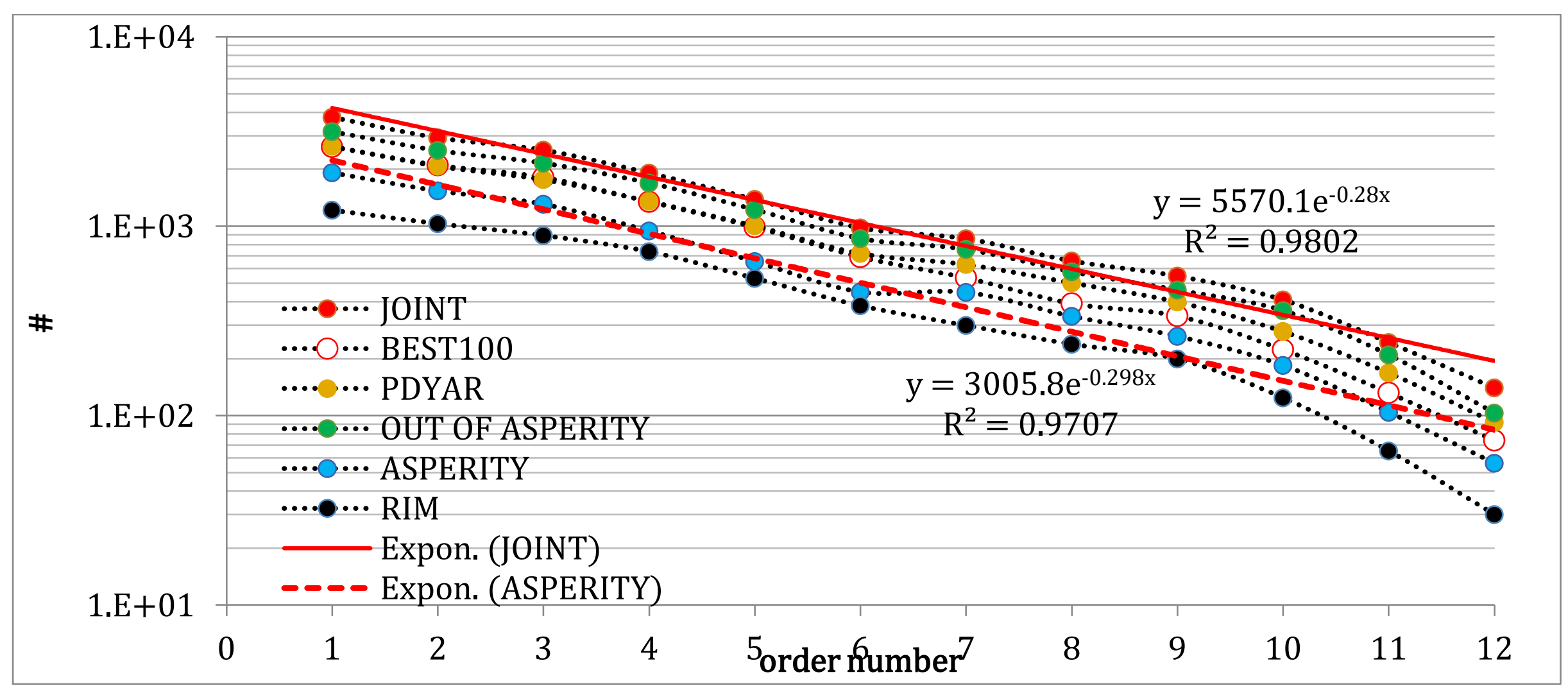

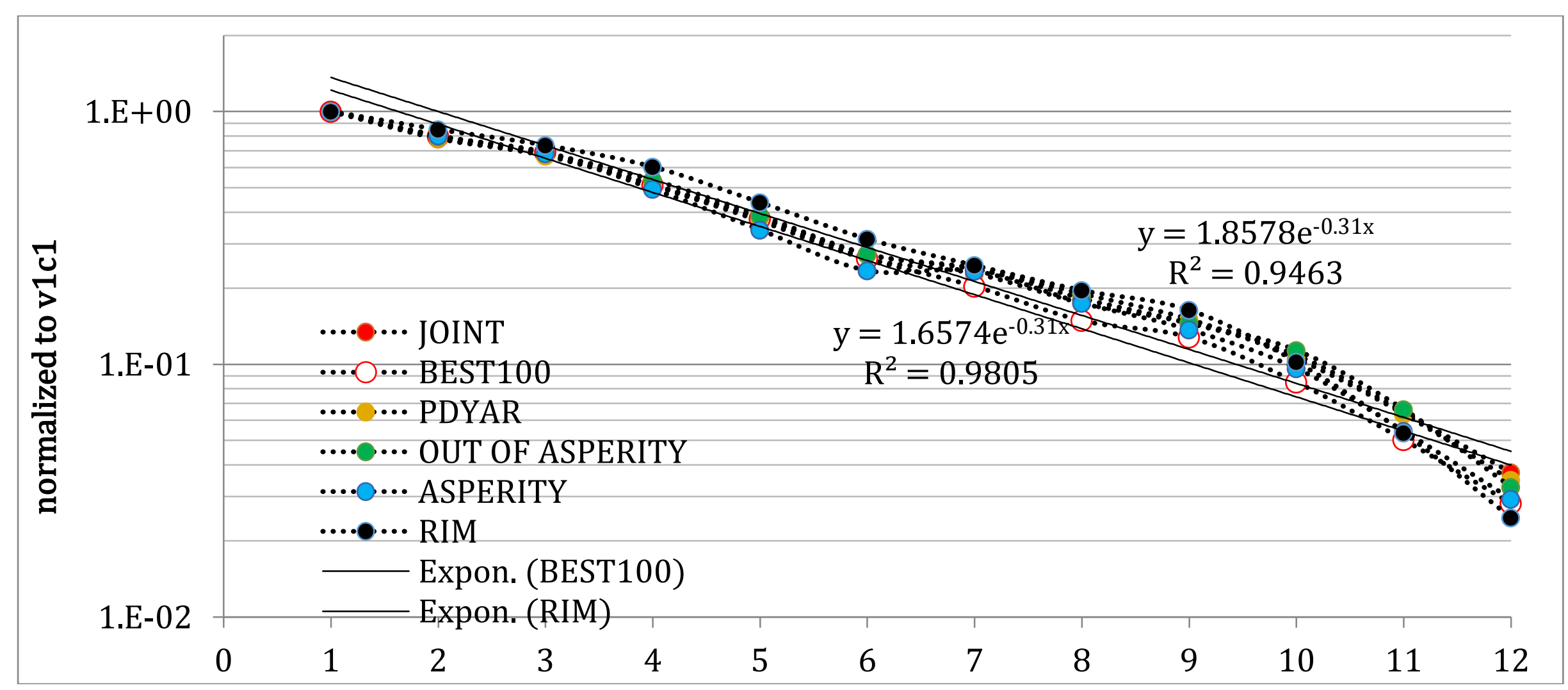

Figure 34. Recurrence curve characteristics evaluated across distinct Master Event (ME) sets. Upper panel: The recurrence curves for the 12 LA version/tolerance case pairs computed for different ME configurations. Lower panel: The corresponding recurrence curves from the upper panel, normalized to their respective absolute levels in the v1c1 configuration to highlight relative morphological variations and outliers.

There are minor deviations from the «JOINT» XSEL exponential regression line within the strict version XSELs. The lower panel of Figure 34 presents the recurrence curves from the upper panel, normalized to their respective maximum values for the $v_1c_1$ pairs. These relative variations effectively highlight the outliers across the configurations. The highest coefficient of determination, $R^2$=0.9805, is yielded by the top 100 MEs set. This value is statistically significant enough to validate the correlation between the order numbers and the uniform sequence of magnitude thresholds. This robust correspondence represents a key feature underlying the reliability of the weak-to-strict version ratio as a predictive parameter. The largest fluctuations are demonstrated by the "Rim" curve with $R^2$=0.94. This specific curve exhibits a progressive deficit in XSEL events when approaching the highest order numbers. Both regression lines have the same slope of -0.31.

***Evolution of the number of XSEL events***

Because the weak-to-strict version ratio represents a measurable parameter designed to serve as an operational trigger for a major earthquake alert, the fundamental properties of the underlying calculations become crucial. For each monitoring interval, six distinct ratio curves are evaluated, corresponding to the six origin-time tolerance figures ranging from 0.25 s to 5.0 s. These metrics are derived from the respective XSEL bulletins generated by the WCC detection and LA procedures for individual MEs. The aggregation of these individual XSELs into a consolidated final XSEL is governed by a CR procedure, which is finely calibrated against the baseline daily event rates observed during the periods between major earthquakes. The backbone of this entire WCC processing pipeline comprises a spatial network of MEs combined with waveform templates of the first P-wave recorded across all stations associated with each respective ME.

A standalone ME creates event hypotheses strictly within the spatial grid search radius relative to its own hypocentral position reported in the REB. While the probability of detecting an event situated outside this grid boundary is non-zero, its reconstructed location will be artificially shifted inside the grid. This artifact can introduce significant errors in the absolute location, a limitation that is highly common in automatic bulletins. The REB events are also inherently mislocated relative to their true positions; however, the confidence ellipses derived from travel time residuals provide an estimate of the 95% probability contour containing the actual source. In general, a larger number of associated stations reduce the dimensions of this confidence ellipse. These structural parameters, alongside potential phase misassociation, render the choice of MEs highly vulnerable to location errors and template quality.

The set of 295 MEs includes REB events of varying quality regarding absolute location accuracy, the number of associated stations, and the availability of valid templates of the first P-phase. The operational efficacy of individual MEs can vary significantly across the studied region. This variability can introduce measurable fluctuations in both the resolution and sensitivity of the WCC processing, directly affecting the final XSEL bulletins. Smaller ME subsets assigned to specific geographical areas may exhibit localized underperformance due to poor ME quality, which could be misinterpreted as low seismic activity. Genuine spatial fluctuations in seismic activity across the selected could be erroneously interpreted as instrumental or algorithmic underperformance.

The partitioning into these geographical subsets is accomplished using simple straight lines that delineate quasi-rectangular areas. The probability space within which an individual ME can generate a valid event hypothesis can readily extend into adjacent zones. A discrete XSEL bulletin can incorporate events that geographically belong to a different ME subset. Within the complete ME network, however, these overlapping MEs would generate competing hypotheses that are ultimately resolved by the CR procedure. For example, the "Out of Asperity" domain contains numerous MEs situated immediately along the boundary of the "Asperity" core zone. This spatial overlapping introduces a non-zero bias in the independent XSEL estimates computed for each discrete subset. Within the integrated «JOINT» master set, this specific spatial bias is successfully suppressed by the presence of the MEs that explicitly belong to the "Asperity" set.

The operational performance of all individual ME subsets can be systematically compared to the «JOINT» set as an aggregated output of all MEs in all subsets. Panel a) in Figure 35 depicts the six weak-to-strict ratios between 2025193 and 2025201 for the «JOINT» set. There are notable differences observed in the behavior of the 0.25 s and 0.5 s curves when compared to the best 100 MEs subset in Figure 30. The precursory peak amplitude in the «JOINT» 0.25 s curve decreases to 13.1, compared to 15.4. The double-peak structure preceding the J20 event in Figure 30 is smoothed, and the highest peak moved to the first place in their sequence. The peak at 2015199 lost its relative amplitude to 8.8 from 14.7 for the best MEs 100 set. The third peak near 2025197 is the highest for the presented interval. The period before 2025194 is not shown as the calculations for the «JOINT» set were conducted from 2025193 to 20252012. The 0.5 s curves are nearly the same for both sets with slightly lower amplitude relative to the highest peak. The remaining four tolerance curves are almost identical across both sets.

Overall, the 0.25 s curve has not lost its predictive capability, despite the fact that the weak MS(4) curve in Figure 36a exhibits only a single upward shift prior to the J20 event, while the strict MS(4) curve declines at a faster rate than the weak curve, followed by a sharp increase along the final leg preceding the J20 rupture, which is marked by the vertical red line. For the «JOINT» curve, the broader morphology of the precursory peak is driven primarily by an accelerated decline in the strict curve, mimicking the behavior observed during previous peak intervals. This underlying mechanism differs significantly from the forcing factors that govern the standalone best 100 MEs set.

The “PDYAR” subset in Figure 35b is characterized by a broader precursory peak that begins to grow three days before the J20 mainshock. The prominent trough near 2025200, which is clearly observed in the best 100 and «JOINT» ME sets, loses its depth in the “PDYAR” subset, appearing instead as a smooth transition between two individual maxima. Both of these peaks are driven by a faster decline in the strict curve relative to the weak LA version. The final quarter-day preceding the J20 event is characterized by a rapid amplitude rise in the strict curve – signaling an increasing activity of the XSEL events closer to the corner magnitude of the REB recurrence curve – which occurs alongside a more gradual increase in the weak curve. The peak level of the 0.25 s curve is 16.8, strongly supporting its precursory character. The 0.5 s curve begins to decline three quarter-days prior to the J20 earthquake, consistent with the behaviors documented in the other two ME sets. This synchronous signature further reinforces the hypothesis of an impending major earthquake.

The 0.25 s ratio curve for the "Asperity" ME subset in Figure 35c exhibits features similar to those of the "PDYAR" subset, albeit with a few distinct deviations. The "Asperity" curve also starts to rise on 2025198, but reaches its peak value of 14.5 five quarter-days prior to the J20. Unlike in the «JOINT» and "PDYAR" subsets, there is a peak near 2025195 induced by a deep trough in the strict curve. There is a similar peak in the best 100 MEs curve. This empirical observation suggests that the sharp declines in the strict curves are driven by local conditions and depend on individual ME quality and spatial distribution, rather than encompassing the entire region. Such local peaks are effectively smoothed by the cumulative activity in the «JOINT» set. Only the largest anomalies remain visible in the curves that cover the entire region. For instance, the prominent peak near 2025197 is clear for the best 100, "PDYAR", and «JOINT» sets, yet it remains statistically significant within the "Asperity" curve. Furthermore, the final precursory peak preceding the J20 event is significantly smoothed and displays a lower amplitude relative to the other maxima analyzed above.

a)

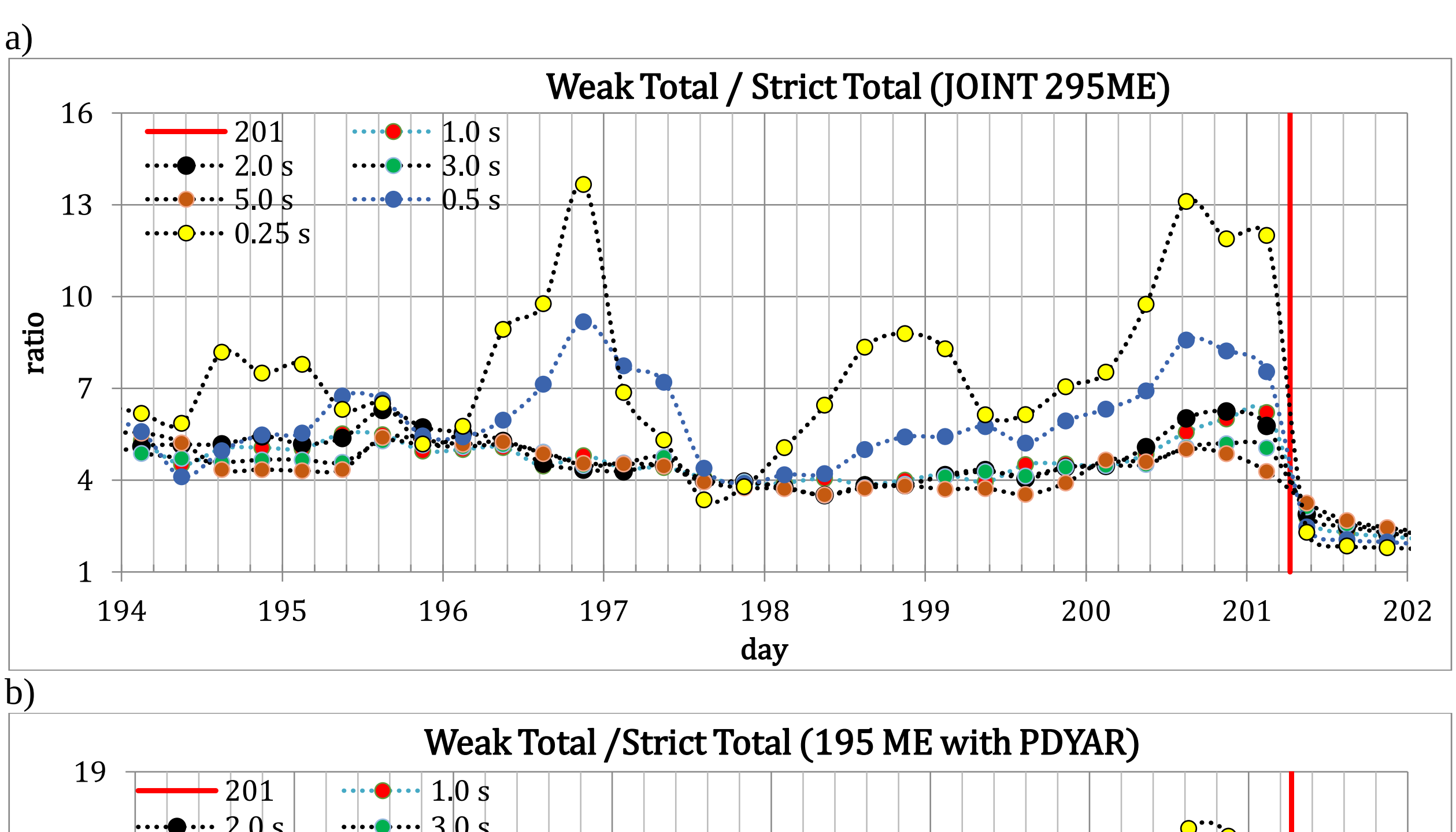

b)

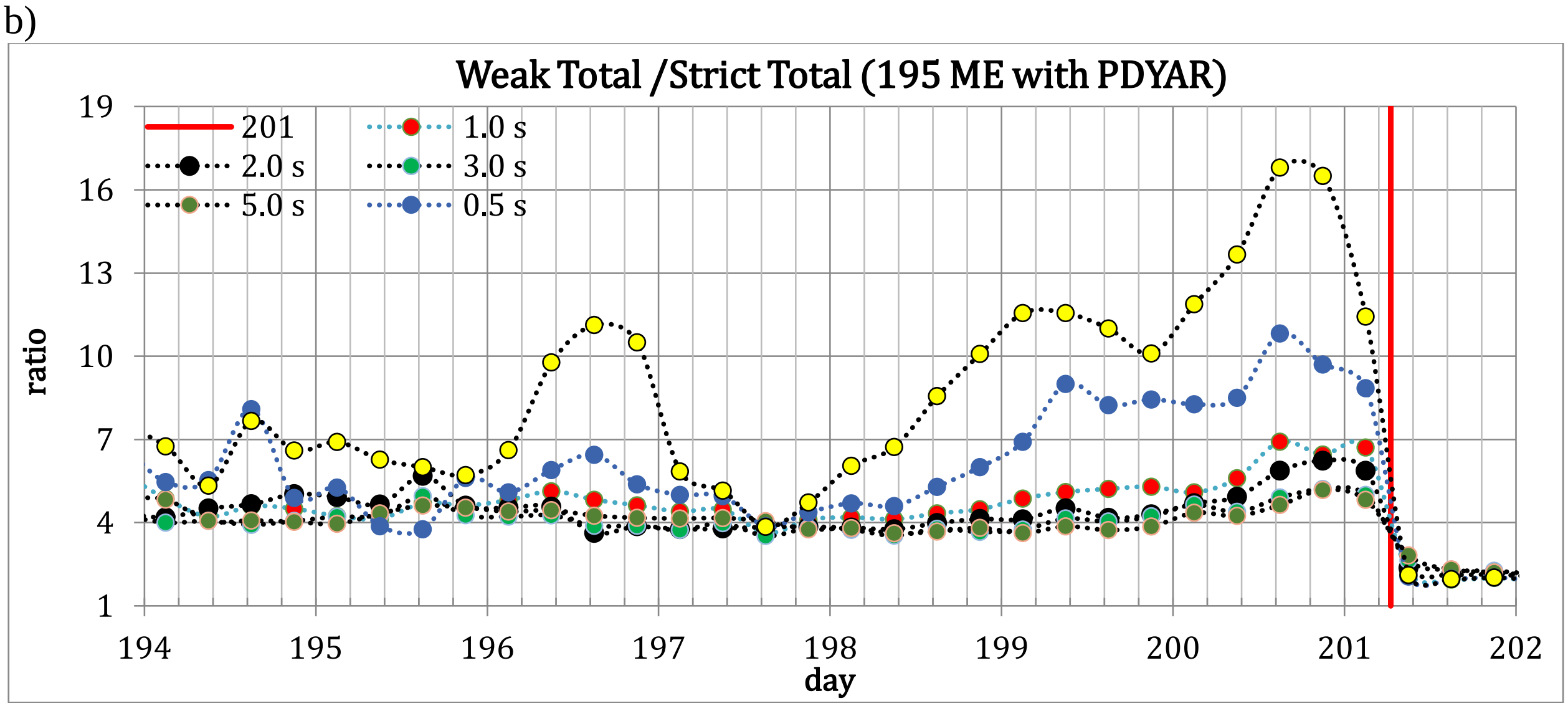

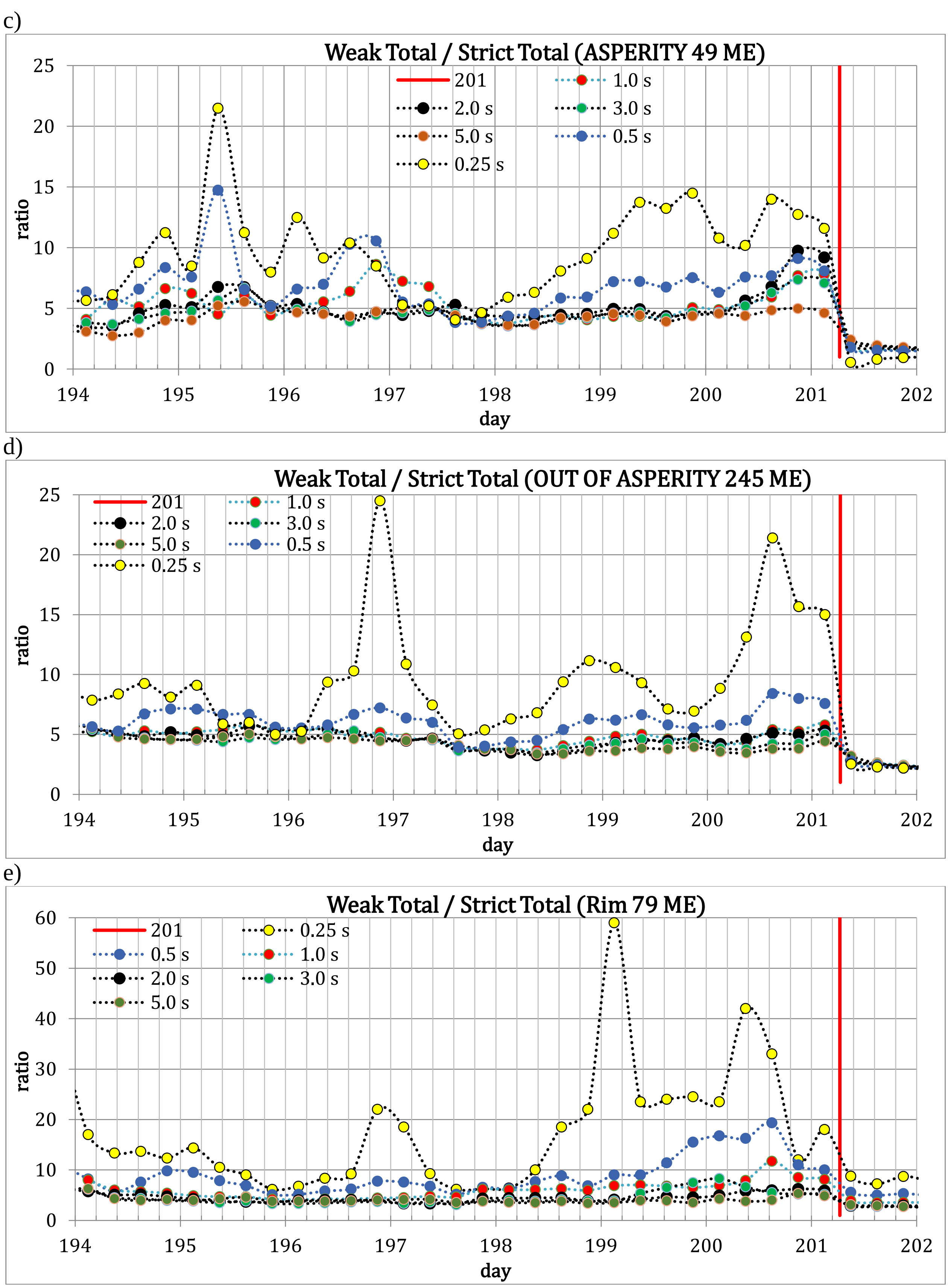

Figure 35. Evolution of the weak-to-strict LA version ratios calculated across six origin time tolerance levels preceding the J20 earthquake. Five distinct master event (ME) configurations are presented to resolve the spatial and geographical dependence of the precursory ratios: a) «JOINT» configuration comprising the complete macro-set of 295 MEs; b) Standalone “PDYAR” master subset incorporating 195 MEs with a mandatory associated arrival at station

PDYAR; c) "Asperity" subset consisting of 49 MEs (excluding one deep-focus event located within the horizontal boundaries of the asperity area) clustered within the immediate epicentral zone of the J20 and J29 events, capturing the stable tectonic asperity that arrested the J20 rupture propagation; d) "Out of Asperity" subset comprising 245 MEs situated outside the core zone; e) "Rim" subset incorporating 79 MEs from the “PDYAR” configuration located along the southwestern periphery of the analyzed domain.

The “Out of Asperity” 0.25 s ratio curve in Figure 35d demonstrates two distinct high-amplitude peaks: one near 2025197 and another three quarter-days prior to the J20. The latter maximum reaches an absolute value of 21.4. It could serve as an optimal short-term precursor to the J20 earthquake, especially given that its culmination is immediately followed by a systematic two-quarter-day decline leading directly to the J20 origin time. This peak is driven by a fast decline in the strict curve in Figure 36d, followed by an equally accelerated recovery that initiates three quarter-days before the J20. The growing number of events in the strict version XSEL is a feature potentially related to earthquake preparation. The geographical distribution of MEs within the “Out of Asperity” set suggests that the seismic activity prior to the coseismic phase was concentrated outside the zone of the “Asperity” subset.

The “Rim” subset is presented in Figure 35e. It was designed to cover the area beyond the aftershock sequence of the J29 mega-earthquake. It should inherently remain insensitive to any seismic activity related to the preparation of the J20 and J29 earthquakes. It has an extremely high peak near 2025199. While similar peaks are observed across the sets covering the entire region, they are much lower in relative amplitude, thereby confirming the strictly localized character of this anomaly. The “Rim” set in the southwestern periphery of the region undergoes its own independent seismic activity on 2025197, which should not be factored into the core J20/J29 nucleation process. There is another sharp peak with an amplitude of 42.0 four quarter-days before the J20. Its amplitude is directly related to the drop in the strict curve to the level of 1 to 2 per day, as illustrated by the corresponding MS(4) curves in Figure 36e. Despite its high amplitude, this peak is statistically insignificant due to the extremely small sample size. Mechanically, this anomalous spike can be attributed to the over-the-border spatial sensitivity of MEs positioned near the primary preparation zone of the J29 earthquake.

a)

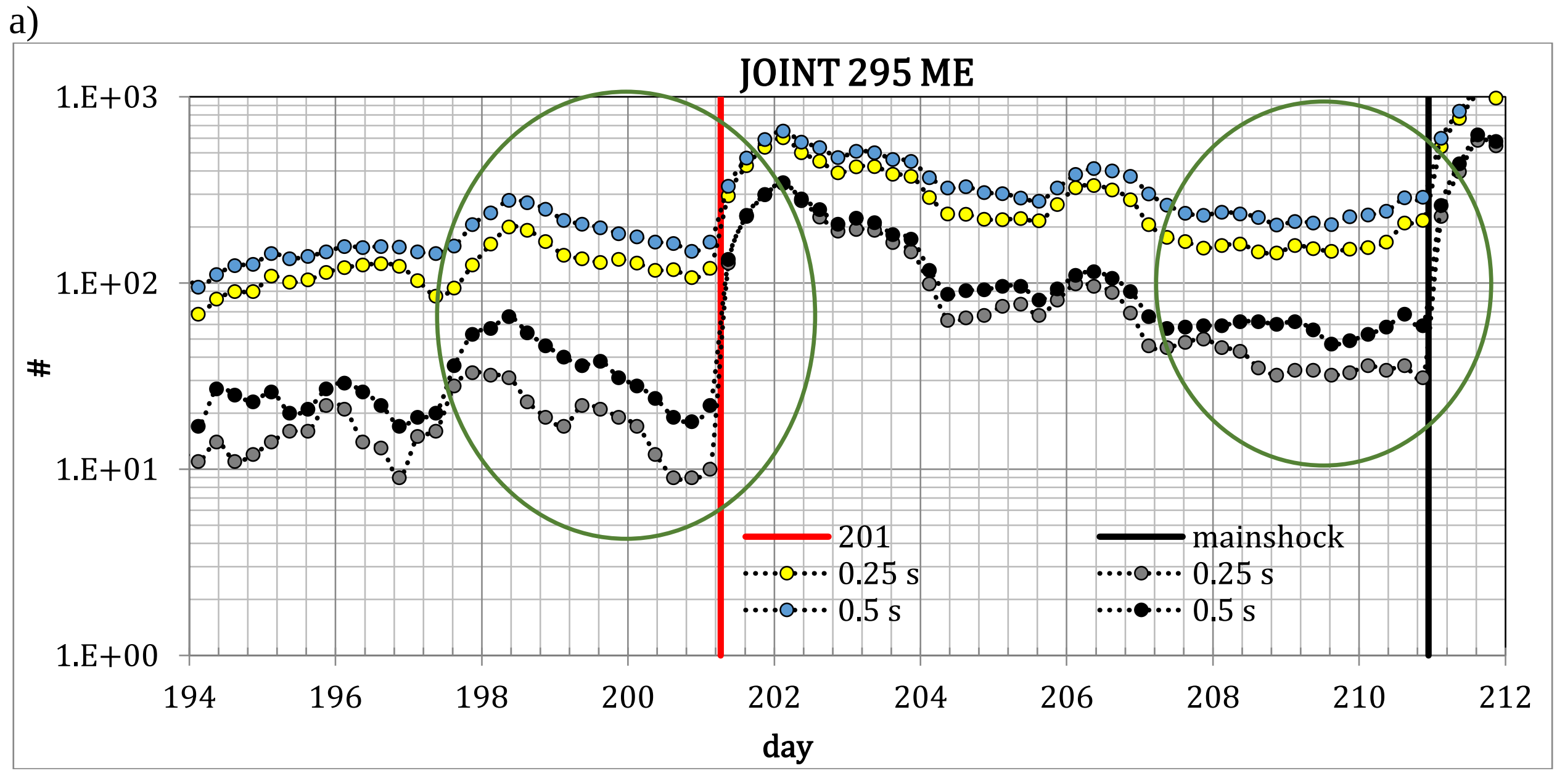

b)

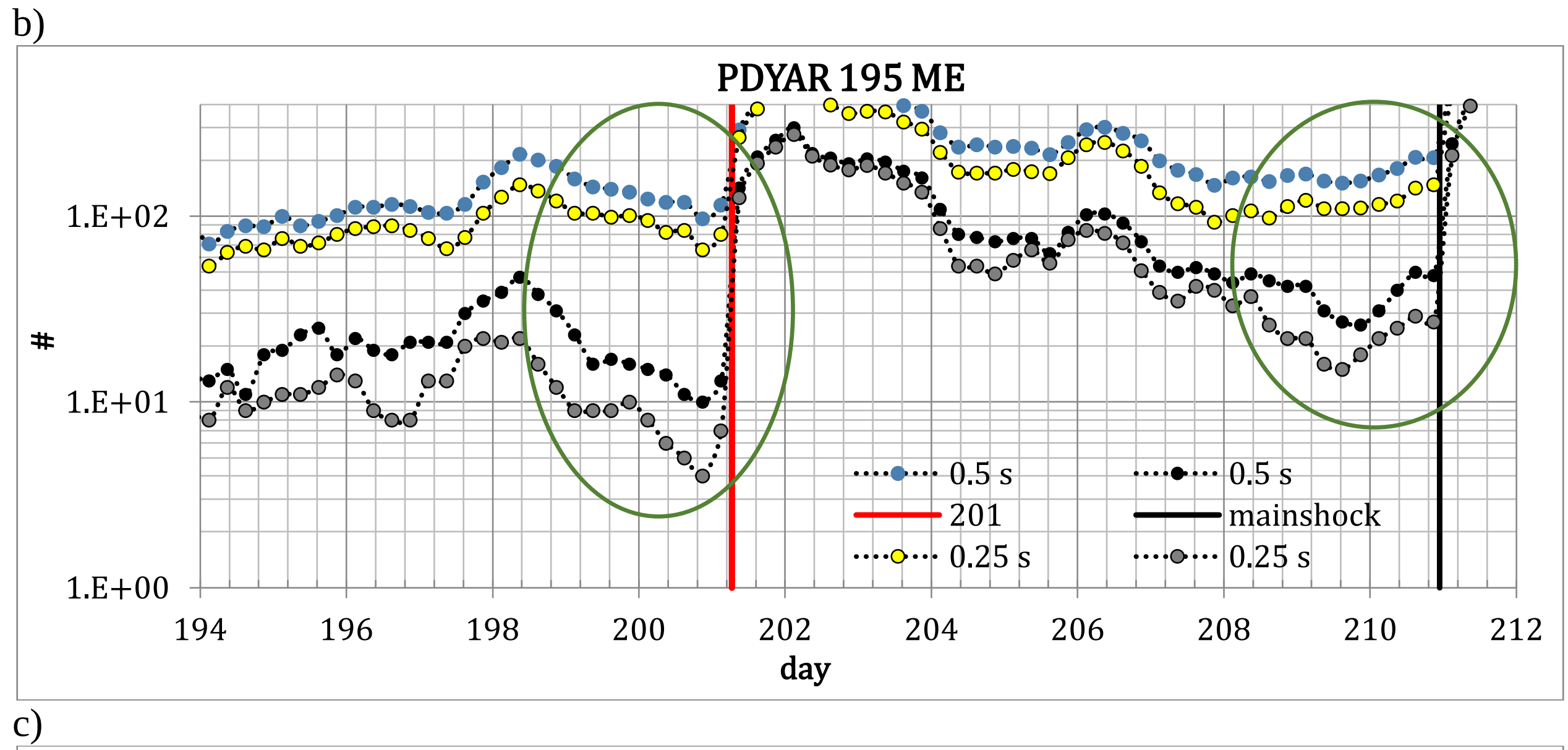
PDYAR 195 ME
1.E+02
1.E+01
1.E+00
#
0.5 s
201
0.25 s
0.5 s
mainshock
0.25 s
194
196
198
200
202
204
206
208
210
212
day

c)

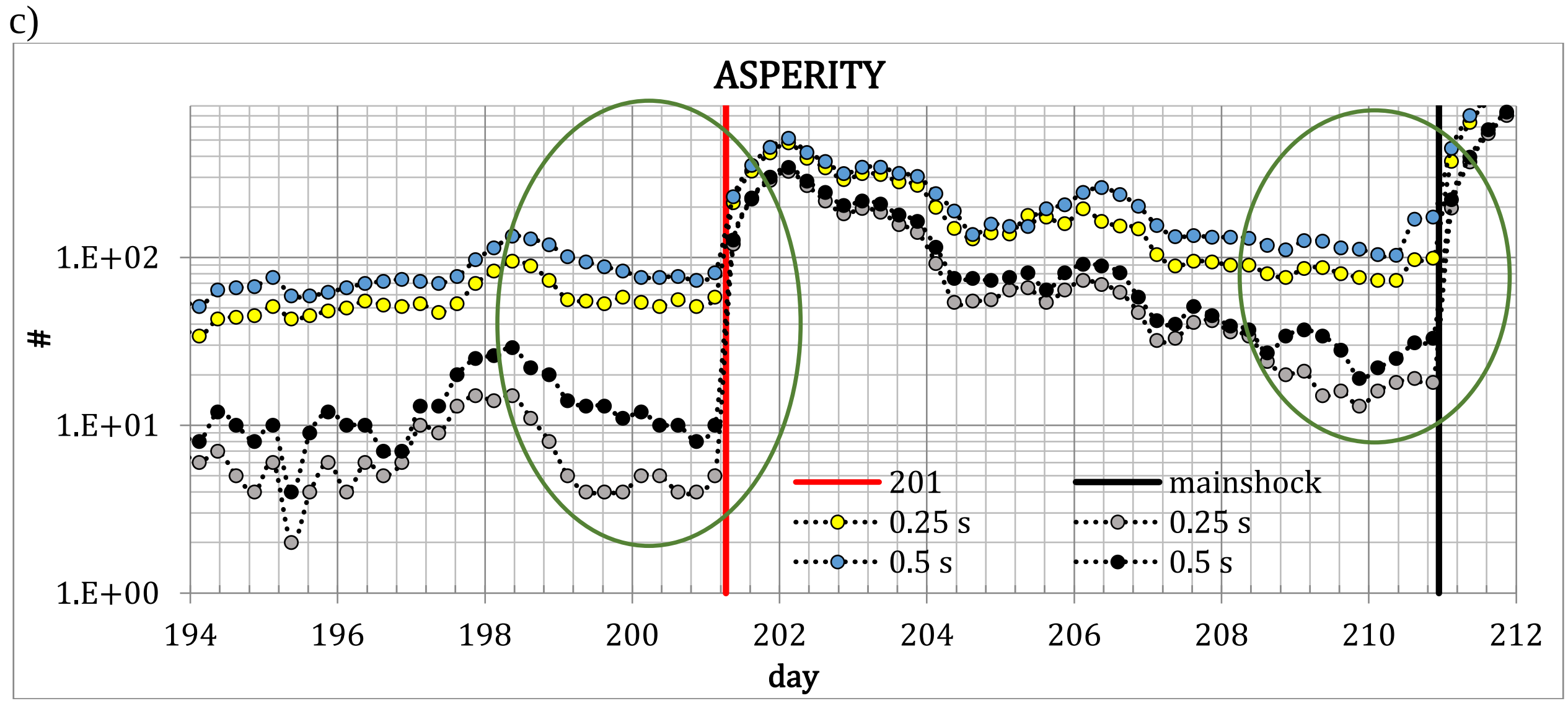
ASPERITY
1.E+02
1.E+01
1.E+00
#
201
0.25 s
0.5 s
mainshock
0.25 s
0.5 s
194
196
198
200
202
204
206
208
210
212
day

d)

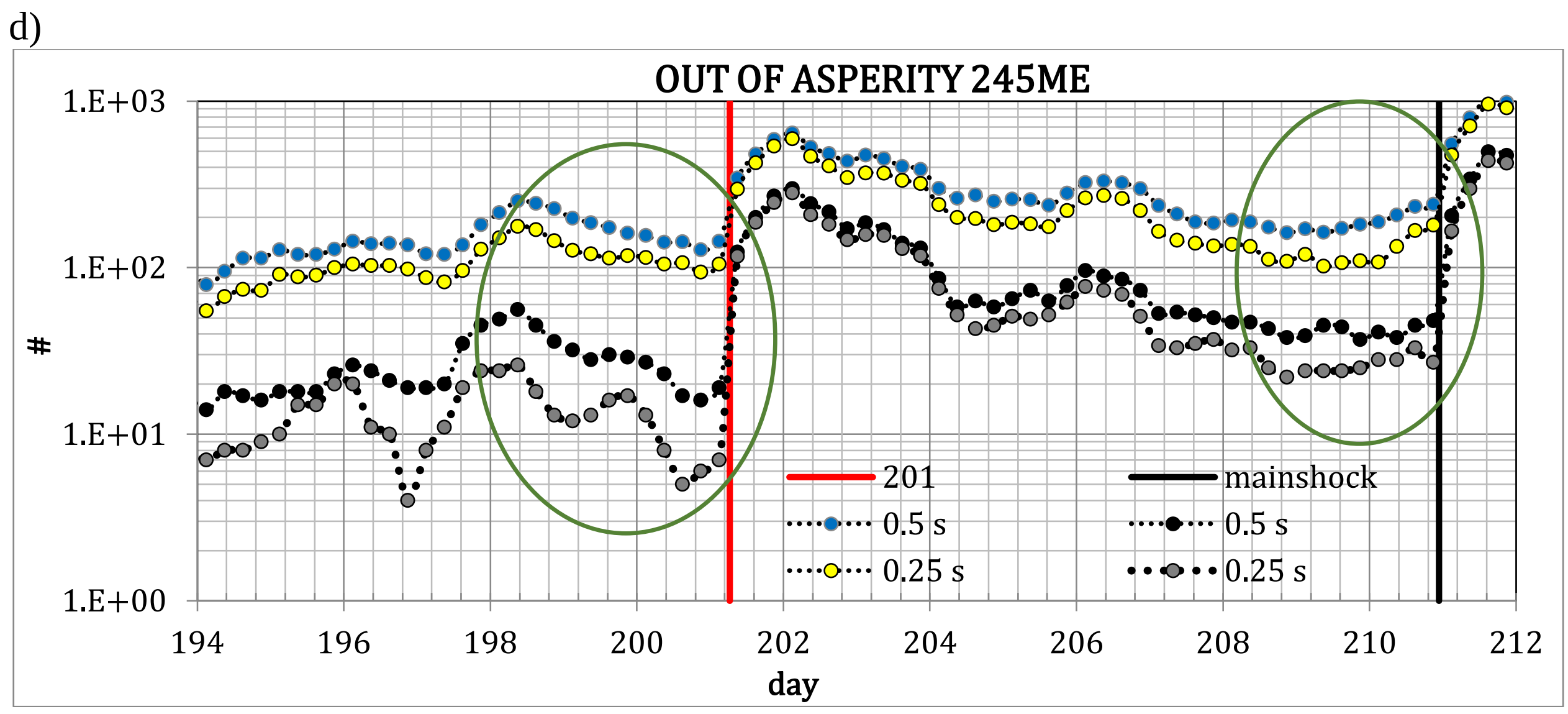
OUT OF ASPERITY 245ME
1.E+03
1.E+02
1.E+01
1.E+00
#
201
0.5 s
0.25 s
mainshock
0.5 s
0.25 s
194
196
198
200
202
204
206
208
210
212
day

e)

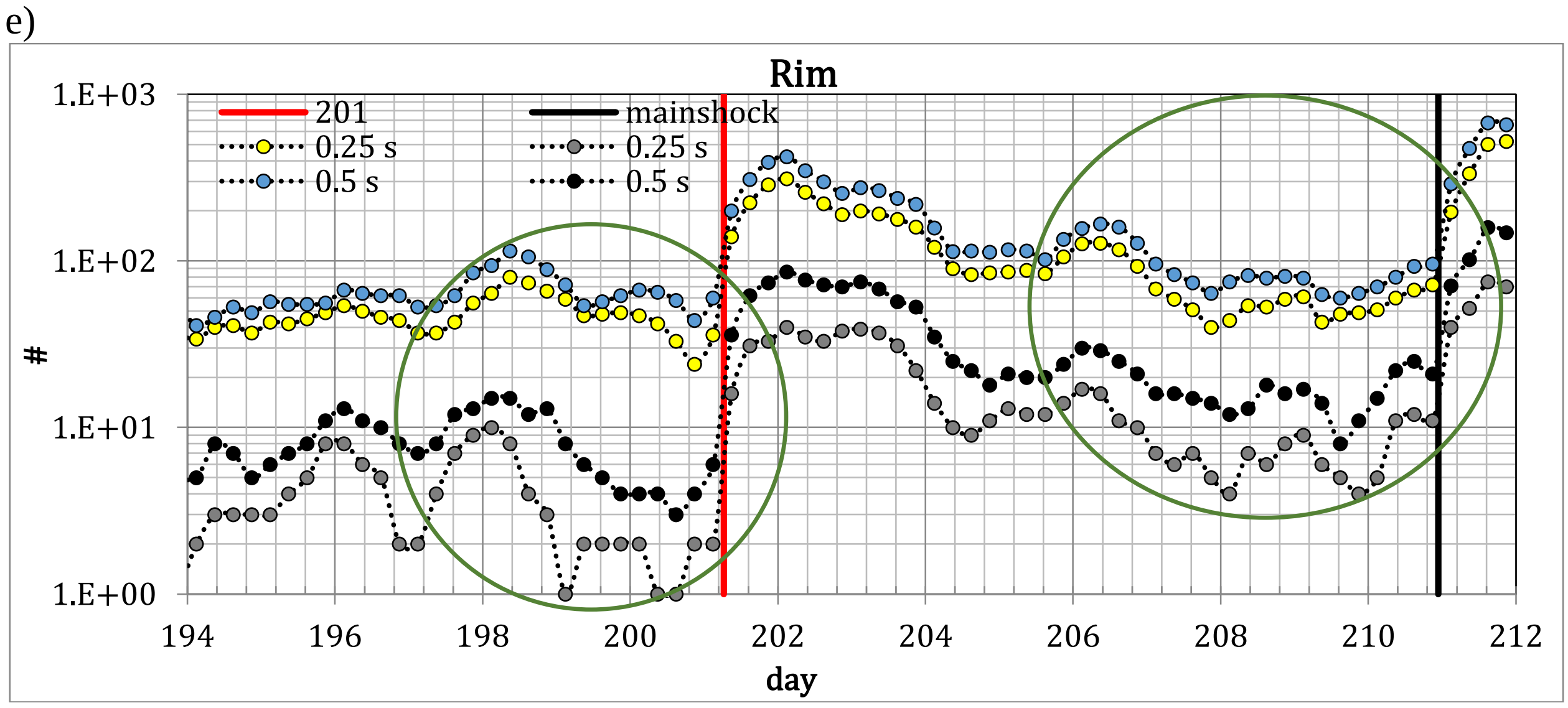

Figure 36. Comparison of the curves for the weak and strict LA versions and the origin-time tolerances of 0.25 s and 0.5 s for the same five ME configurations as in Figure 35.

Initially, the J20 earthquake was not the primary target of this study, with the original objective to analyze the J29 mega-earthquake. However, the fundamental physical mechanisms underlying major earthquake preparation remain consistent within a given seismogenic zone; consequently, the nucleation phase of a catastrophic rupture can overlap with and mask the preparation process of an intermediate event. As a result, the parameters used in the analysis of the J29 earthquake preparation may introduce a systematic bias when applied to the J20 process. Nevertheless, the observations presented in Figures 35 and 36 suggest that the weak-to-strict ratio curve for the 0.25 s origin time tolerance preserves its validity as an operational trigger for short-term earthquake early warning systems. The density and geographical distribution of MEs required to ensure the optimal resolution of this precursor can be systematically calibrated using the REB as a robust source. Furthermore, the core parameters of the WCC pipeline – ranging from individual station detection thresholds to the conflict resolution algorithm – remain fully adjustable. Such a comprehensive network optimization is highly feasible given sufficient computational resources, and thus lies beyond the scope of the current study.

For the J29 mega-earthquake, the weak-to-strict versions ratio for the 0.25 s origin time tolerance exhibits a sharp peak preceding the rupture onset time. The 0.5 s curve in Figure 33 is nearly synchronous with this 0.25 s curve, forming a more robust potential trigger for an earthquake early warning alert. Taken together, these two individual peaks would be sufficient to achieve the objective of the J29 earthquake forecasting. The Sea of Okhotsk earthquake also exhibits two similar peaks right before the mainshock. Their statistical significance is validated by a progressive increase in the number of XSEL events detected within the strict LA version for both the 0.25 s and 0.5 s cases during three consecutive quarter-days preceding the mainshock. The stochastic creation of these strict hypotheses is practically excluded, as the LA defining parameters are calibrated to suppress not only random phase alignments but also extraneous detections related to the side-sensitivity of the highest-weight IMS array stations.

The «JOINT» set of MEs generated XSELs for all 12 LA version-case pairs computed for the period between 2025193 and 2025212. The corresponding weak-to-strict ratios of the MS(4) curves for the six distinct tolerance cases are presented in Figure 37a. The 0.25 s ratio curve exhibits a sharp peak in the fourth quarter of July 29, immediately preceding the mainshock, which is marked by the vertical black line. The curve undergoes a rapid acceleration over two consecutive quarter-days, reaching a maximum amplitude of 7.0. This value is marginally higher than the respective peak in the best 100 MEs subset illustrated in Figure 33. This peak is driven by the rise in the weak curve, as observed in Figure 36a. The strict curve retains a practically constant level during the two days prior to the J29 earthquake. This behavior is nearly identical to that demonstrated by the respective curves of the 100 best MEs subset, despite the difference in the XSELs total numbers of events, as shown in Figure 34.

The 0.5 s curve peaks at a level of ~5 in the «JOINT» configuration, likely smoothed out by the competing MEs from the "PDYAR" subset. The geographical position of MEs generating extra XSEL events for the 0.5 s weak-to-strict ratio is not clear yet. Otherwise, the «JOINT» and the best 100 MEs curves are characterized by highly synchronous evolution between 2025202 and 2025212, punctuated by two distinct low-amplitude peaks on 2025204 and 2025207. These peaks are likely driven by the increased seismic activity after two earthquakes on 2025203 and 2025206, both featuring a body-wave magnitude ($m_b$) of approximately 5.5, as illustrated in Figure 19.

The «JOINT» set and the best 100 MEs subset can both serve as effective predictive tools, displaying only minor differences. In contrast, the "PDYAR" subset in Figure 37b demonstrates a markedly different behavior. The 0.25 s ratio curve is characterized by a broader peak, with a maximum amplitude of 7.3 that is shifted chronologically to Q3/2025209. This peak is driven by a steady decline in the strict curve, which reaches its lowest point in Q3/2025209, as illustrated in Figure 36b. The strict curve then rises faster than the weak curve, causing the weak-to-strict ratio to decline continuously up to Q3/2025210. In Q4/2025210, the strict curve drops by 2 events relative to the preceding quarter-day, while the weak version retains its level. The Q4 lasts only 5 hours, as the J29 origin time is approximately at 23:25. A secondary, lower peak with an amplitude of 5.5 is observed immediately preceding the J29 mainshock, likely corresponding to the primary precursory maxima identified in the «JOINT» set and the best 100 subset. The 0.5 s ratio curve exhibits an isolated peak reaching an amplitude of 6.2 in Q4/2025209, followed by a gradual decline to the level of the remaining four tolerance curves right before the mainshock. As for the 0.25 s curve, this 0.5 s peak can be explained by the decrease in the strict curve between Q1 and Q4 of 2025209.

The predictive capability of both the 0.25-s and 0.5-s curves appears partially suppressed by the high ambient seismicity within the boundaries of the "Asperity" zone. This core domain represents a relatively compact area characterized by the highest level of historical seismicity and the nucleation of several extremely large historical earthquakes (Figures 12 and 13). While the primary mechanical source of the precursory indicator marking an impending rupture is clearly situated within this focal zone, the low-magnitude seismic activity recovered by the WCC pipeline likely incorporates events generated by alternative, long-term tectonic processes that operate concurrently with the immediate mainshock preparation cycle.

a)

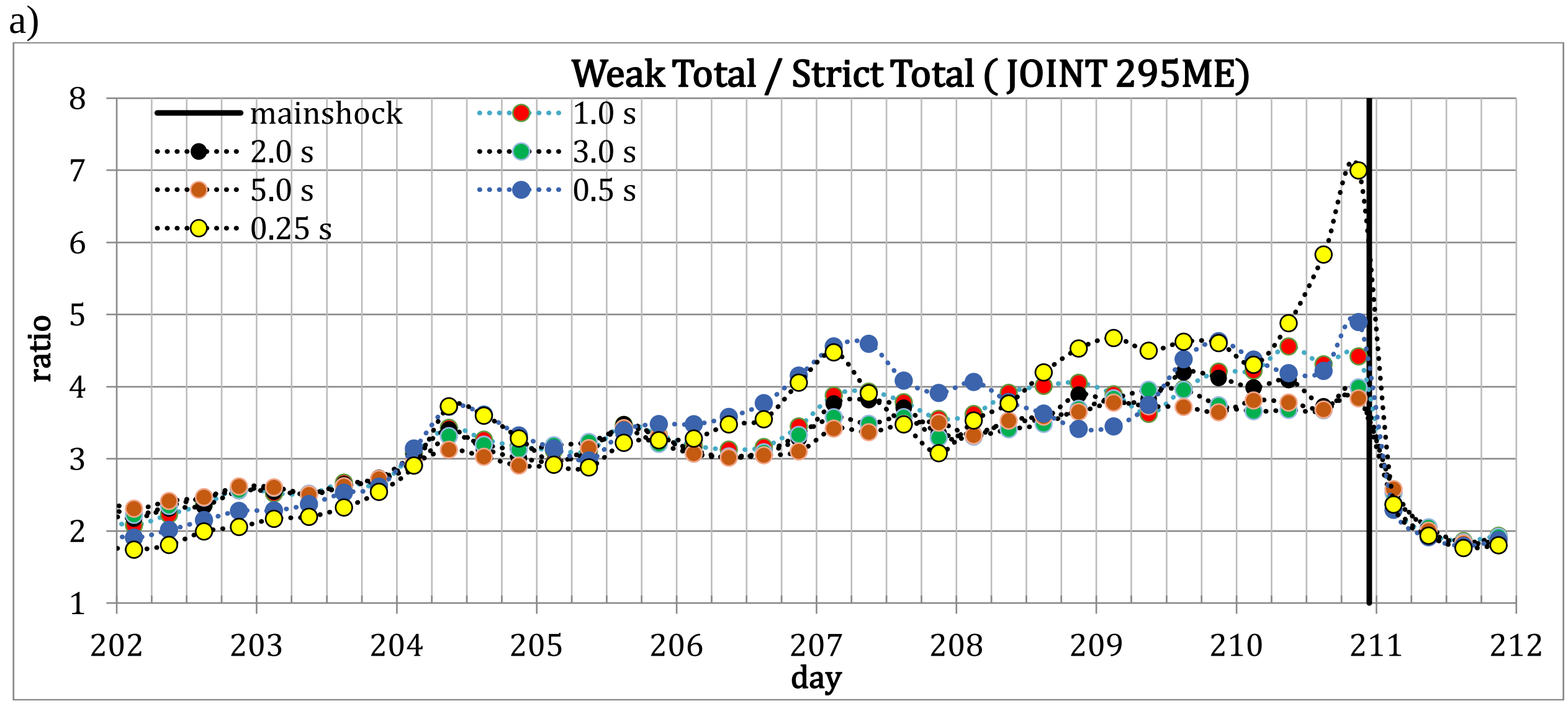

b)

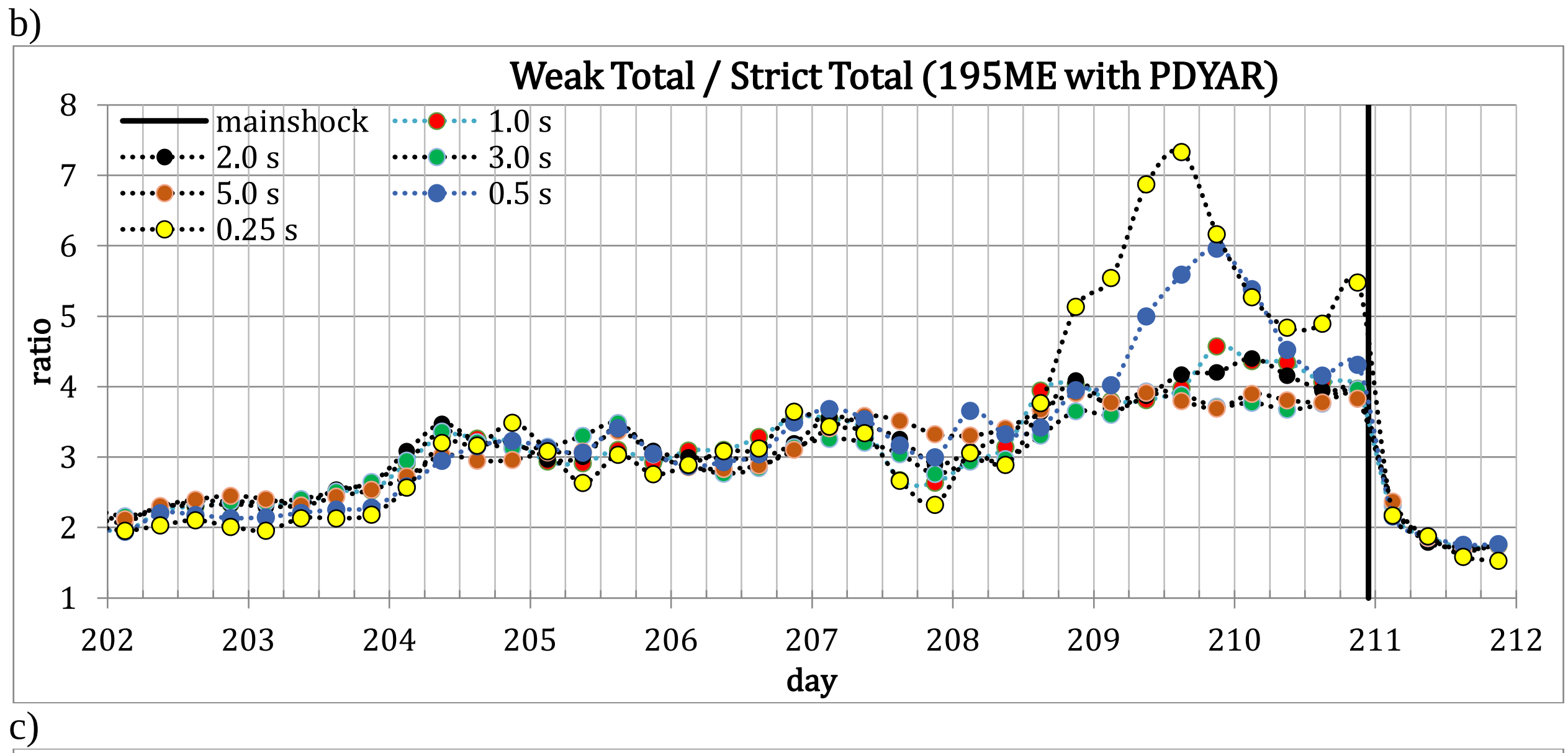

Weak Total / Strict Total (195ME with PDYAR)
mainshock
2.0 s
5.0 s
0.25 s
1.0 s
3.0 s
0.5 s
ratio
day
8
7
6
5
4
3
2
1
202
203
204
205
206
207
208
209
210
211
212

c)

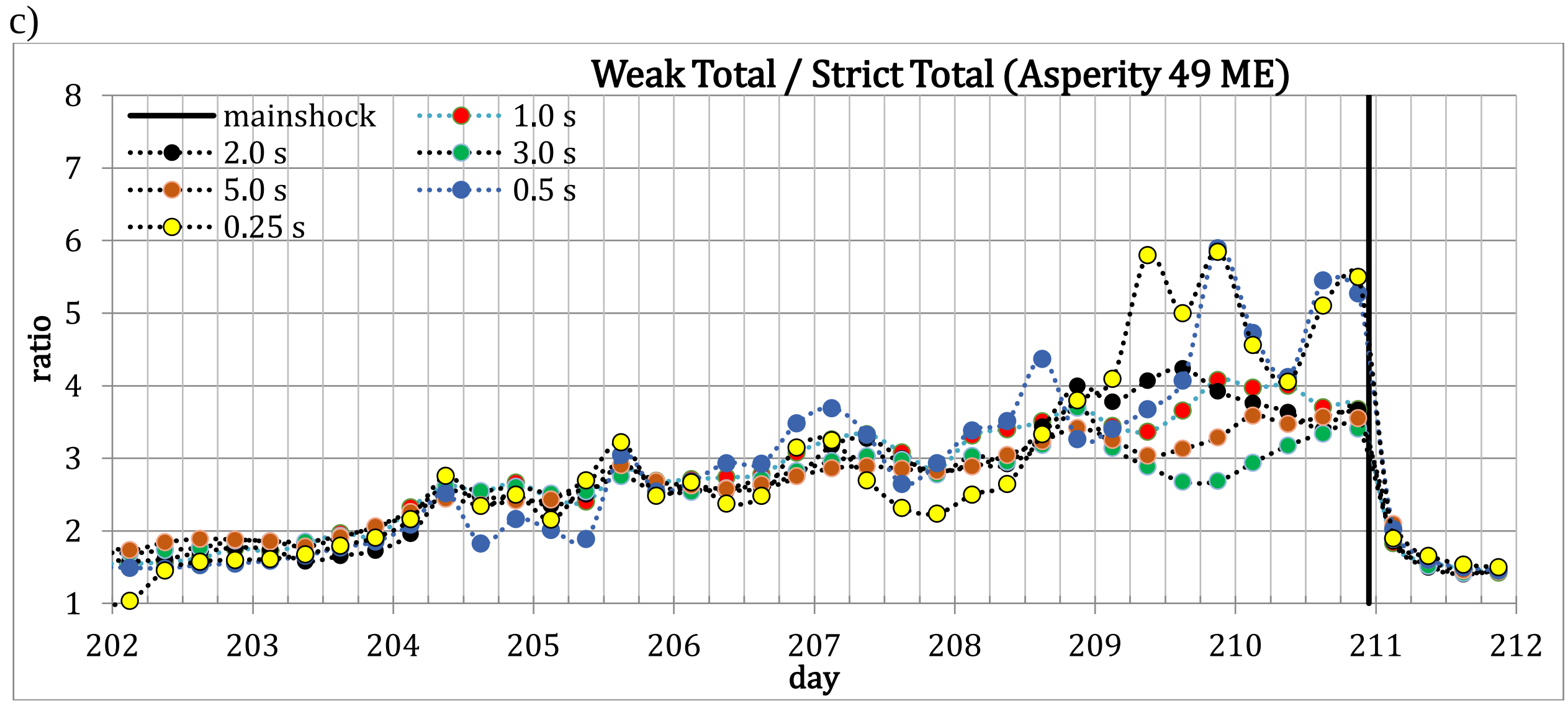

Weak Total / Strict Total (Asperity 49 ME)
mainshock
2.0 s
5.0 s
0.25 s
1.0 s
3.0 s
0.5 s
ratio
day
8
7
6
5
4
3
2
1
202
203
204
205
206
207
208
209
210
211
212

d)

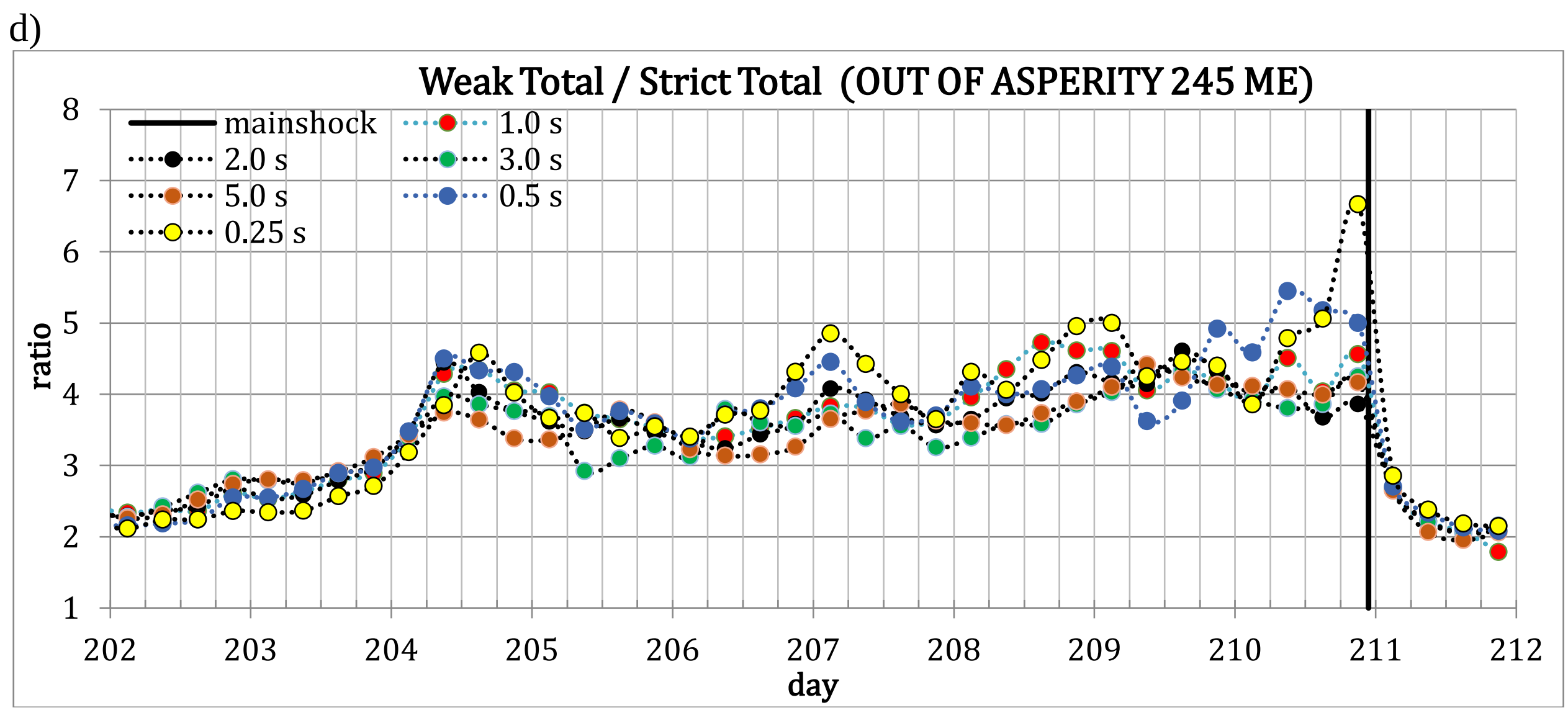

Weak Total / Strict Total (OUT OF ASPERITY 245 ME)
mainshock
2.0 s
5.0 s
0.25 s
1.0 s
3.0 s
0.5 s
ratio
day
8
7
6
5
4
3
2
1
202
203
204
205
206
207
208
209
210
211
212

e)

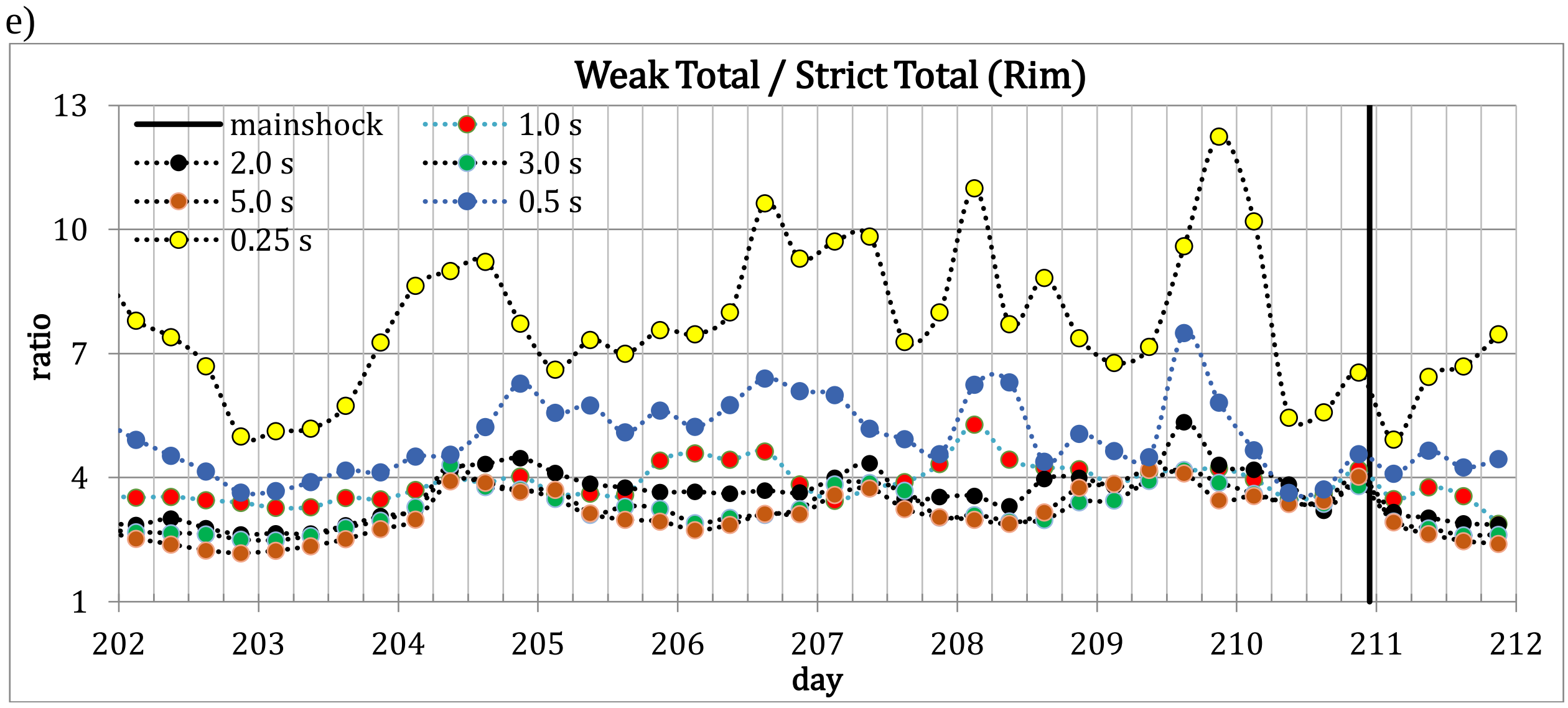

Figure 37. Same as in Figure 35 for the earthquake on July 29, 2025

The "Out of Asperity" ME subset yields results that are highly similar to those of the «JOINT» master set. Specifically, the 0.25-s version ratio curve exhibits a sharp peak immediately preceding the J29 mainshock, reaching maximum amplitude of approximately 7.0. Prior to the J29 event, the "Out of Asperity" 0.5-s curve reaches a level marginally higher than that recorded by the integrated «JOINT» configuration. This close similarity is primarily attributed to the absolute sample size of this subset; the "Out of Asperity" network incorporates 245 out of the total 295 MEs deployed in the «JOINT» configuration. Crucially, the operational performance and predictive capability of this subset do not suffer any significant degradation when the "Asperity" core events are entirely omitted from the processing stream. This empirical outcome demonstrates that the critical short-term earthquake preparation processes operate effectively across the broader spatial domain encompassed by these 245 master events, rather than being strictly restricted to the focal asperity itself.

The "Rim" subset serves as a critical counterpoint to establish the statistical significance of precursory anomalies. The 0.25-s version ratio curve presented in Figure 37e demonstrates high-frequency fluctuations in amplitude, with its maximum peak emerging during Q4 of day 2025209. During the entire day preceding the J29 mainshock, this 0.25 s curve drops to a flat baseline level between 5.0 and 6.0, displaying no anomalous peak that could be interpreted as a precursor. This behavior demonstrates that the crustal volume encompassed by the "Rim" configuration remains structurally independent of the seismogenic and mechanical processes operating across the core preparation zone of the studied region. Figure 36e illustrates this conclusion by the effect of the weak (and strict) LA version curves for all origin-time tolerance cases fail to reach the post-seismic level observed after the J29 mega-earthquake in Figure 20. The other four configurations in Figure 36 demonstrate this essential feature of the XSEL recurrence curves.

The 79 MEs defining this southwestern “PDYAR” subset can be safely excluded from the direct investigation of the seismic process leading to the J29 rupture. This spatial selectivity is robustly supported by the corresponding findings related to the preceding J20 earthquake. Mechanically, the largest peak observed in the MS(4) ratio curve during the fourth quarter of 2025209 is artificially induced by a sharp drop in the strict curve down to a baseline of only 4 events per day, as illustrated by the running sum trajectory in Figure 36e.

## Discussion

The ultimate objective of this study is the recovery of low-magnitude seismicity prior to a major earthquake. The IMS array stations represent the best global seismic network, providing lower detection thresholds for remote regions. In the standard beamforming detection technique, this is achieved by the destructive interference of ambient noise accompanied by constructive interference of the sought signals. The WCC-based processing creates a pipeline that provides additionally reduced detection thresholds, effective suppression of detections not relevant to the studied localized areas, and accurate local association of the preselected detections with the event hypotheses created by high-quality master events. The conflicts for the same physical signals associated with independent hypotheses created by two or more neighboring master events are resolved in a Conflict Resolution procedure, which optimizes the resulting bulletin to the actual event rate – from very low to that observed during the first hours after the mega-earthquakes such as Sumatra in 2004, Tohoku 2011, and Kamchatka 2025.

The WCC-based approach was advocated by Schaff and co-authors [2025] as a technique to obtain information on the seismic processes not detectable by the standard methods. The WCC processing has been widely used to detect low-magnitude events at regional and teleseismic distances. One of the problems associated with low-magnitude events is that one cannot accurately estimate their standard magnitudes from amplitude and period measurements, as the detected signals are hidden in the ambient noise. In this study, we further developed the technique based on a set of threshold defined by the statistical significance of valid events introduced by Kitov [2026ab]. For the Kamchatka 2025 major earthquakes and their aftershocks, these thresholds were directly linked to standard magnitude estimates. This finding allowed us to follow the evolution of low-magnitude seismicity across a set of increasing magnitude thresholds toward the onset times of the J20 and J29 earthquakes. The measurable parameter potentially useful as a mega-earthquake precursor is based on the ratio of the numbers of XSEL events above two thresholds. This parameter signals that the rate of such events increases in the corridor between the thresholds. Alternatively, a decreasing rate above the upper threshold also leads to a rise in the ratio, but does not indicate an approaching large event.

The absolute duration of the investigated interval was limited to approximately two weeks, primarily due to available computing power. However, this period was sufficient to resolve highly specific precursory features preceding the J29 mainshock and its potential foreshock, the J20 mainshock. A few days prior to the J29 megathrust failure, the acceleration in seismic activity progressed into the most statistically significant $v_1c_1$-$v_2c_1$ bin configuration – the parameter space situated closest to the historical corner magnitude of the Kamchatka recurrence curve. An identical manifestation was previously documented before the Sea of Okhotsk earthquake.

There are several important findings in this study made possible by the IMS array stations and the sensitivity and flexibility of the WCC processing. Firstly, the 12 independent version-case configurations, based on different Local Association criteria and varying origin time tolerances, can be mathematically converted into a standard body-wave magnitude scale. This transformation yields robust recurrence curves that successfully resolve actual seismicity down to a baseline of approximately magnitude 2.0, extending the catalog completeness into the sub-noise register where no official REB event hypotheses are available. Secondly, synchronous, high-amplitude peaks in the 0.25 s and 0.5 s weak-to-strict version ratio curves are consistently observed immediately preceding major seismic ruptures. The stability and predictive power of this dual-window operational trigger have been robustly verified across distinct geodynamic settings, encompassing the 2013 deep-focus Sea of Okhotsk earthquake as well as the 2025 Kamchatka major earthquakes. Thirdly, the partitioning of the comprehensive «JOINT» master event catalogue into discrete geographic subareas exposes the underlying spatial sensitivity of the WCC pipeline. The interaction between overlapping grid search radii and the automated Conflict Resolution cross-rejection process successfully suppresses spatial cross-talk and boundaries blurring between adjacent zones. Fourthly, the multi-parametric spatial test effectively divides the investigated subduction region into zones that are highly significant and zones that are insignificant for precursory signal localization. The analysis demonstrates that while the core tectonic asperity acts as the primary mechanical driver of the impending rupture, the accelerated micro-seismic preparation process is heavily concentrated outside this immediate core, mapping out a wide precursory activation envelope on the periphery of the fault plane. Conversely, remote reference configurations, such as the "Rim" subset, display a flat background variance, confirming the absolute spatial selectivity and statistical significance of the developed method.

There were several precursor-like peaks found in the data before the J20 and J29 earthquakes. They were all statistically insignificant and caused by a severe drop in the XSEL number of the strict LA version. However, it is not excluded that some statistically significant precursory signals might not end in a catastrophic earthquake. Further studies are needed to understand the abundance and properties of false positive related to the precursor-like signals. This methodological projection involves several critical aspects. First, a high rate of false alarms does not constitute a viable alternative to the unmitigated impact of a catastrophic event. Comprehensive false-alarm statistics must be systematically gathered over a multi-year baseline, encompassing both high-activity clusters and prolonged seismically quiet intervals. While this verification requires significant computational resources and time, isolating signatures similar to those reported before the 2013 Sea of Okhotsk and 2025 Kamchatka events would not only precisely constrain the empirical false-alarm probability but also unlock invaluable insights into the aggregate mechanical behavior of the subducting Pacific plate.

Tectonic preparation processes that fail to reach the final co-seismic rupture stage, alongside their precise spatial positions and hypocentral depths, are directly governed by the internal architecture of the slab and the space-time evolution of stresses under regional driving forces. On a macro-scale horizon, these observations may hold greater significance for structural geophysics than the isolated estimation of false-alarm rates. For instance, the J29 mega-earthquake was preceded by an almost collocated J20 event that failed to overcome the stable local asperity, terminating as a localized $m_b$=5.39 (IDC) event instead of immediately triggering the final M8.8 rupture that materialized ten days later. A highly analogous precursory sequence was documented two days prior to the 2011 Tohoku earthquake, whereas no such distinct foreshock activation was resolved before the catastrophic December 26, 2004, Sumatra-Andaman megathrust failure.

For future investigations of the Kuril–Kamchatka subduction zone, it will be highly informative to systematically assess the large aftershock that occurred on September 18, 2025. Furthermore, as a historic alternative to the modern, well-instrumented J29 case, the major earthquake of November 15, 2006, will be analyzed. That historical event nucleated within the southwestern boundary of the active domain, explicitly belonging to the "Rim" master event configuration. To achieve maximum operational efficiency, this localized "Rim" template database will be strategically extended using core anchors selected from the top 100 MEs configuration. In 2006, a significantly reduced IMS network topology was operational, completely lacking the critical near-regional array PETK (Petropavlovsk-Kamchatskiy, Russia; 53.108°N, 157.699°E) and the highly sensitive regional array PDYAR (Peleduy, Russia; 59.655°N, 112.441°E). Additionally, the regional events of May 19, 2013, and December 20, 2018, will be investigated to complete a comprehensive re-evaluation of all major earthquakes within this domain since 2001.

The Kamchatka Peninsula represents one of the most seismically active and well-instrumented subduction zones globally; however, some other regions capable of generating catastrophic earthquakes suffer from deficient instrumentation. For instance, the IMS network across South America comprises exclusively three-component (3-C) stations, a network that catches significantly fewer phases, elevates the baseline WCC detection threshold, and prevents the high-resolution recovery of low-magnitude sub-noise seismicity. Because Chilean megathrust earthquakes exhibit structural mechanisms highly analogous to those of the Kamchatka Peninsula, they warrant thorough investigation once the regional observational assets are upgraded to match the strict array configuration requirements of the WCC pipeline.

The March 11, 2011, Tohoku earthquake commands both a catastrophic magnitude and an exceptional instrumentation baseline, captured by multiple IMS arrays operating at regional distances. Conversely, the December 26, 2004, Sumatra earthquake exhibited a catastrophic rupture volume but was recorded by relatively poor network instrumentation during the early operational years of the IMS network. Both of these historical mega-earthquakes are currently under active investigation using the optimized WCC pipeline. The compelling structural divergence between them resides in the drastically different levels of ambient micro-seismic activity operating immediately prior to their respective mainshocks.